\documentclass[twocolumn,showpacs,preprintnumbers,amsmath,amssymb]{revtex4}

\usepackage{graphicx}
\usepackage{dcolumn}
\usepackage{bm}
\usepackage{url}
\usepackage{rotating}

\begin{document}

\title{The International-Migration Network\footnote{Thanks to Matteo Chinazzi, Marco Due\~{n}as, Giuseppe Mangioni, and Tiziano Squartini for their invaluable help in providing codes and assistance on network visualization, the econometrics of gravity exercises, community detection, and null models, respectively. All remaining errors are ours. G.F. gratefully acknowledges support received by the research project ``The international trade network: empirical analyses and theoretical models'' (www.tradenetworks.it) funded by the Italian Ministry of Education, University and Research (Scientific Research Programs of National Relevance 2009)}}

\author{Giorgio Fagiolo}%
\affiliation{LEM, Sant'Anna School of Advanced Studies, Pisa (Italy)}
\author{Marina Mastrorillo}
\affiliation{University of Foggia, Foggia (Italy)}%

\date{\today}

\begin{abstract}
This paper studies international migration from a complex-network perspective. We define the international-migration network (IMN) as the weighted-directed graph where nodes are world countries and links account for the stock of migrants originated in a given country and living in another country at a given point in time. We characterize the binary and weighted architecture of the network and its evolution over time in the period 1960-2000. We find that the IMN is organized around a modular structure characterized by a small-world pattern displaying disassortativity and high clustering, with power-law distributed weighted-network statistics. We also show that a parsimonious gravity model of migration can account for most of observed IMN topological structure. Overall, our results suggest that socio-economic, geographical and political factors are more important than local-network properties in shaping the structure of the IMN.
\end{abstract}



\maketitle

\section{Introduction\label{Sec:Introduction}}
This paper studies the evolution of human international-migration patterns in the period 1960-2000 using a complex-network perspective \cite{AlbertBarabasi2002,guidosbook,Newman_Book}. 

International migration is increasingly perceived as a fundamental feature of human life for its huge impacts on the global economy and the potential to shape the world where we live by changing the demographic structure of towns, cities and nations \cite{PRB_2008,OECD_2008,world2006global}. Driven by the desire for better job opportunities, a more secure environment, or simply forced to move from home, an increasing number of migrants have contributed to unprecedented levels of cross-border flows in the last years, leading to an overall migrant world population of about 190 million in 2010 \cite{wmr_2010}.

Despite migration is not a recent phenomenon \cite{Manning2005}, and people have always left their home either involuntarily or in search of better conditions, recent predictions suggest that international migration is likely to become bigger and more complex in the next decades, due to the combined impact of population aging and increasing demographic differences, economic inequality, climate change, environmental disasters, wars and famines, new political and economic scenarios, and technological change favoring social contacts \cite{UNDP_2009}. Therefore, the expected strength and reach of the impacts of migration are likely to dramatically increase in the next years, and so do the complexity of related policies. 

International migration affects many intertwined spheres of the demographic, social and economic fabric of both origin and destination countries \cite{Ma98,Ha02}. Beside the obvious impact on country demographic composition, living conditions, cultural and social integration, national security, and socio-economic globalization, migration has strong distributional effects due to international-remittance flows \cite{Ozden_2006}, which can contribute to poverty reduction and economic growth in origin countries \cite{Or09}. Furthermore, it may unevenly affect labor-market structure, wages and development in destination countries through brain-drain \cite{St91,Be01}, and bilateral trade via a decrease in transaction costs and the emergence of new consumption-preference patterns \cite{Gould_1994,Egger_etal_2012}. 

The controversial net outcome of such a wealth of effects has led to policies fostering an increase in cross-border barriers to the movement of people, which contrast with the correspondent fall observed for cross-border trade and financial transactions. In the near future, it is expected that migration-related issues will become more and more relevant in the political arena \cite{world2006global}.

Given its massive positive and normative implications, it is imperative to enhance our empirical knowledge of international migration patterns and try to capture trends and determinants from a global perspective. So far, the lack of coordinated and homogeneous data on migration has forced researchers to focus either on country-specific data or to address global migration issues using very small samples of reporting countries \cite{GlobBook,GastonNelson_2011}. In all these studies \cite{Massey_etal_1993}, international migration has been always considered essentially as a bilateral phenomenon, where migration flows or stocks between any two countries are taken as independent of any other seemingly unrelated link \footnote{This parallels what has happened in the study of international trade, see \cite{Fagiolo2010jeic} for a discussion.}. 

More recently, however, a reliable and homogeneous bilateral-migration database covering most of world countries in the period 1960-2000 has been made available \cite{Ozden_etal_data_2011}. This eventually allows one to study the evolution of international-migration patterns from a complex-network perspective. Building on this idea, this paper introduces the International Migration Network (IMN), defined as the weighted-directed human-migration graph where nodes represent world countries and directed (weighted) links describe international migration corridors (and related migration intensities) between any two countries. Such an approach allows one to explore the evolution of migration patterns using a systemic perspective \cite{Le96}, where one can study the embeddedness of countries into the network, their centrality and the often-complicated local clusters that can be formed due to the interplay of social, economic, geographic and political factors, in line with what has been recently done for international-trade \cite{Garla2004,Fagiolo2009pre,BariFagiGarla2010} and financial \cite{SRF_2010_QF} macro-economic networks. Borrowing the interpretation suggested in this stream of literature, weighted directed links in the IMN can be interpreted as an additional interaction channel between countries, beside e.g. bilateral trade, foreign-direct investment and financial linkages. Indeed, the number and intensity of inward migration flows reflect the potential of a country to attract people from other countries, and therefore indirectly signal the existence of some social, economic and political imbalance or, in the migration jargon, the existence of significant push-and-pull factors \cite{Lee_1966}. Such imbalances, similarly to what happens in trade, investment and financial networks, may be important to explain the diffusion of economic shocks originating in a given country and possibly spreading to seemingly unrelated countries \cite{Forbes2002}.

Despite migration data can be naturally described as a network, only a few studies have tried to recast this phenomenon using a graph-theoretic perspective \cite{Slater_2008} and the topological properties of global migration networks still remain poorly understood. A network perspective has been instead heavily used to study human-mobility issues in general \cite{Br08,Wo11,Si12}; air, cargo and maritime transportation networks\cite{Ba04,Hu09,Ka10,Wo11,Ka11}; and inter-urban traffic \cite{De07}. These works show that understanding human-mobility networks can be important to predict the properties of diffusion processes occurring on them, e.g. global epidemics \cite{Colizza_etal_2006,Ba09}. 

In this paper we perform a thorough analysis of the topological properties of the IMN and their evolution in the period 1960 to 2000. Furthermore, we explore geographical, social, political, and economic determinants of IMN properties. We show that the IMN displays a small-world structure where a few hubs and a lot of local structures coexist, generating high clustering and disassortative patterns. Geography, population size and country language seem to explain a large part of the observed modularity in the IMN. Furthermore, the number of identifiable clusters has decreased through the years, hinting to some globalization of migration patterns. The reorganization of migration links across time has occurred mostly at the weighted level. The topology of migration corridors has instead remained quite stable and homogeneously distributed across countries. Conversely, the total number of migrants of each country (as well as many other weighted-network statistics) are characterized by power-law distributions, hinting to      
a sort of rich-get-richer effect at the weighted level. We also find that, contrary to what happens in the international-trade network, the local structure of the network cannot replicate its higher-level properties such as assortativity and clustering and that instead geographical distance, country-level characteristics, and other bilateral country variables are necessary to provide a good characterization of the observed IMN architecture. We show that geographical distance not only negatively affects migration as expected, but it strongly shapes the observed assortativity and clustering patterns in unexpected ways. Finally, we show that a gravity model of migration can reproduce very nicely most of local and global IMN observed topological properties, hinting to a preponderance of country specific and bilateral effects in prediction of the complex structure of the global migration network.    

The rest of the paper is organized as follows. Section \ref{Sec:Data} presents the dataset that we employ in our analysis. A brief overview of what we know at the aggregate level about migration patterns is contained in Section \ref{Sec:AggregatePatterns}. We then introduce the International Migration Network (Section \ref{Sec:IMN}) and we present results about its topological properties and their determinants in Sections \ref{Sec:Topology} and \ref{sec:Determinants}. Finally, Section \ref{sec:Conclusions} concludes.   

\section{Data\label{Sec:Data}}
Finding detailed data on international migration patterns and migrant characteristics is extremely difficult. Data problems are especially acute when compared to higher-frequency data on international trade and finance flows. Nevertheless, thanks to the combined efforts by the United Nations Population Division, the Statistics Division of the United Nations, the World Bank and the University of Sussex, a reliable source about bilateral international migration compiled using the United Nations Global Migration Database has been made available to the community of researchers \cite{Ozden_etal_data_2011}. 

Starting from about 3500 individual census and population register records from more than 230 destination countries and territories from across the globe, the final database comprises 5 origin-destination 226$\times$226 matrices for each decade in the period 1960-2000  \footnote{Data are available at http://data.worldbank.org/data-catalog/global-bilateral-migration-database}. For each year $t=1960,\dots,2000$, the generic element $(i,j)$ of each matrix records the stock of migrants (corresponding to the last completed census round) originating in country $i$ and present in destination $j$.

The database partially solves three main hurdles facing migration data. First, the very definition of migrants is often contradictory. In the database, instead, country of birth (and in its absence nationality) is consistently employed to define the status of a migrant. Second, political turmoils usually lead to emergence of new countries and disappearance of old ones, thus generating artificially large flows of new migrants. The database avoids this problem by disaggregating migrants among a constant set of 226 countries consistently over the whole 1960-2000 period \footnote{For example, countries created after the break up of the Soviet Union or Yugoslavia were always considered within the sample of countries from 1960. Therefore, people moving across the former Soviet Union were considered as international migrants.}. Finally, issues due to missing or omitted data are solved in a homogeneous way using interpolation, imputation using existing bilateral data, and when data are not sufficient, alternative external sources. 

Bilateral data in the database refer to stocks and not to flows. Generally, immigrant stock statistics are easier to interpret than data on migrant flows \cite{Parsons_etal_2000data_2007}. The latter are typically characterized by a higher degree of heterogeneity. This is because flow data are typically gathered through surveys, which may differ according to the definition of total population (\textit{de facto} vs \textit{de jure}), timing, categories of people considered, alternative country coding used to record the responses and lack of standardization between the questions asked during the census. Furthermore, complete bilateral flows are only available for a limited set of countries (typically a subset of OECD ones), thus substantially limiting world coverage \cite{OECD_2008}. As a result, data on migrants stocks are the only available covering a large set of countries and years. 

Some remarks are in order. To begin with, care must be taken in interpreting the change of bilateral figures across two consecutive decades, as the stock at time $t$ will be obtained from the stock at time $t-10$ after having considered all possible in and out (unobserved) flows during the decade. Throughout the paper we will then interpret the bilateral stock $(i,j)$ at $t$ in two ways. First, as the sheer number of migrants born in $i$ and living in $j$ at time $t$. Second, as a proxy of the potential that country $j$ has in attracting people from $i$, i.e. of the strength of the interaction channel from country $i$ to $j$ at $t$. Notice however that, given the definition of a migrant in terms of foreign birth, directed paths $\{i_1,i_2,\dots,i_{k-1},i_k\}$ of order $k\geq 2$ between any non-directly connected pair of countries $(i_1,i_k)$ do not necessarily entail a corresponding flow of people born in $i_1$ and ended up in $i_k$ after a number of migration flows through intermediate countries (otherwise one would have had a direct link $i_1\rightarrow i_k$ in the IMN). Rather, directed paths between unconnected countries convey information about the strength of migration chains, i.e. existence of migration corridors that channeled a lot of migration flows in the past.          

Notice also that our data do not consider two important facets of migration, namely illegal and (within-border) internal migration. Whereas the former is almost impossible to measure on a consistent basis, internal migration is estimated to be of an order of four times international migration (even using conservative definitions).

Finally, year coverage does not allow us to evaluate the impact of the global recession on migration. More recent data \cite{wmr_2010} indicate that, overall, the total stock of migrants has not substantially decreased, even if flows have been weakening due to reduced opportunities or more restrictive policies. There was no evidence either about mass return of migrants to countries of origin, possibly because of social protection in some host countries and worse conditions at home. 

\section{International Migration: Aggregate Patterns\label{Sec:AggregatePatterns}}

The main properties of the dataset that we employ here have been extensively studied in a number of recent papers \cite{Ozden_etal_data_2011,Parsons_2012,PRB_2008,wmr_2010}. Despite the share of international migrants in the world's population has remained remarkably stable at around 3 percent over the past 50 years, the total number of foreign-born residents steadily grew over time, reaching about 190 million people in 2010 \cite{wmr_2010}. Between 1960 and 2000, the global migrant stock increased from 93 to 167 million. A large part of such migrant growth is explained by flows to Western Europe and the United States (50\%) and the break-up of the Soviet Union (25\%). The remaining 25\% is accounted for by the Gulf States, within-Africa migration, and flows to Australia, New Zealand and Canada \cite{Ozden_etal_data_2011}. 

This hints to a historical pattern where the distribution of migration stocks in 1960 mainly reflected a post-colonial landscape, with most European- and South-Asia-born migrants. Two big shocks (partitioning of India and break-up of Soviet Union) and the emergence of Gulf States as major migrant destinations are the main forces shaping international migration in the last century. As a result, in 2000 most migrants originated from Latin America, Eastern Europe and Central Asia \cite{Ozden_etal_data_2011}. 

According to recent estimates, the number of world migrants is projected to achieve 405 million in 2050, despite prospective migrants increasingly face government-imposed barriers to movement and a number of nation states that has almost quadrupled in the last century (creating more borders to cross) \cite{wmr_2010}. This positive trend is due to the increasing demand for migrant (especially skilled) labour, triggered by population aging in the developed world; young and still-rising populations in developing countries; cheaper communications and transport; and environmental changes ---especially in less developed countries. 

International migration patterns are remarkably heterogeneous across regions and individual countries \cite{Ozden_etal_data_2011}. Quite counter-intuitively, in 2000 most of total world's migration (60\%) occurred between developing countries (south-south migration) or between developed countries (north-north migration) \footnote{The north-south divide is a generic way of dividing from a socio-economic and political perspective countries between wealthy and developed ones, and poorer and developing, based on the Human Development Index (HDI). Therefore it has not necessarily to do with geographic north and south, even if the majority of wealthy and developed countries are located in the geographic north of the planet.}. Only 70 million people (37\%) moved from a developing to a developed country. Most of south-south migration is driven by the quest for better working opportunities, and the need to escape conflicts or natural disasters. Geography strongly limits south-north migration, due to travel costs and policy-based restrictions on crossing international borders, but constrains mobility in general: 50\% of all migrations occur within regional scopes and 40\% regard neighboring countries. Cultural distance (e.g. common religion and common language) also explains a good deal of migration decisions. Despite south-north migration does not account for a high share of total migration, 75\% of migrants moved to a country with a higher Human Development Index \footnote{See http://hdr.undp.org/en/statistics/hdi/}, i.e. (developing or developed) countries with higher living standards and/or more job opportunities. This suggests that gains in human capital and labour productivity drive most of migration. Over the years, however, north-north, north-south and south-south migration share are slowly declining, at the expenses of south-north migration.

At a country level, population size obviously matters \cite{Ozden_etal_data_2011}. A small bunch of large countries in North America, Europe, and the former Soviet Union concentrates most of the world’s 191 million international migrants. Top destination and origin countries often coincide (e.g. Germany, India, the Russian Federation, Ukraine and the United Kingdom). The United States are still the principal destination, hosting around 20\% of World's migrant population. When one considers migration as a share of country population, however, the picture changes. Only a few large countries (e.g., Australia, Canada and Saudi Arabia) exhibit high shares of migrants, while in some very small countries ---typically Pacific islands or countries plagued by political upheaval and military conflict--- migrants account for more than 50\% of the population. In general, the more remote a small country is, the more people decide to leave.

\section{The International-Migration Network\label{Sec:IMN}}
We employ the 5 origin-destination 226$\times$226 matrices provided by \cite{Ozden_etal_data_2011} to build a time-sequence of weighted-directed networks describing bilateral migration stocks among $N=226$ countries. More precisely, the weighted International-Migration Network (IMN) at time $t=1960,\dots, 2000$ is fully characterized by the $N\times N$ weight matrix $M^t$, whose generic element $m_{i,j}^t$ represents the stock of migrants originated in country $i$ and present at time $t$ in country $j$. Accordingly, we define the binary projection of the IMN through the sequence of adjacency matrices $A^t=\{a_{i,j}^t\}$, where $a_{i,j}^t=1$ iff $m_{i,j}^t=1$, and zero otherwise. Note that since in general $m_{i,j}^t\neq m_{j,i}^t$, and $m_{i,j}^t>0$ does not necessarily imply $m_{j,i}^t>0$, both weighted and binary IMNs are directed (non-symmetric) networks.

\begin{sidewaysfigure}
\includegraphics[width=25cm]{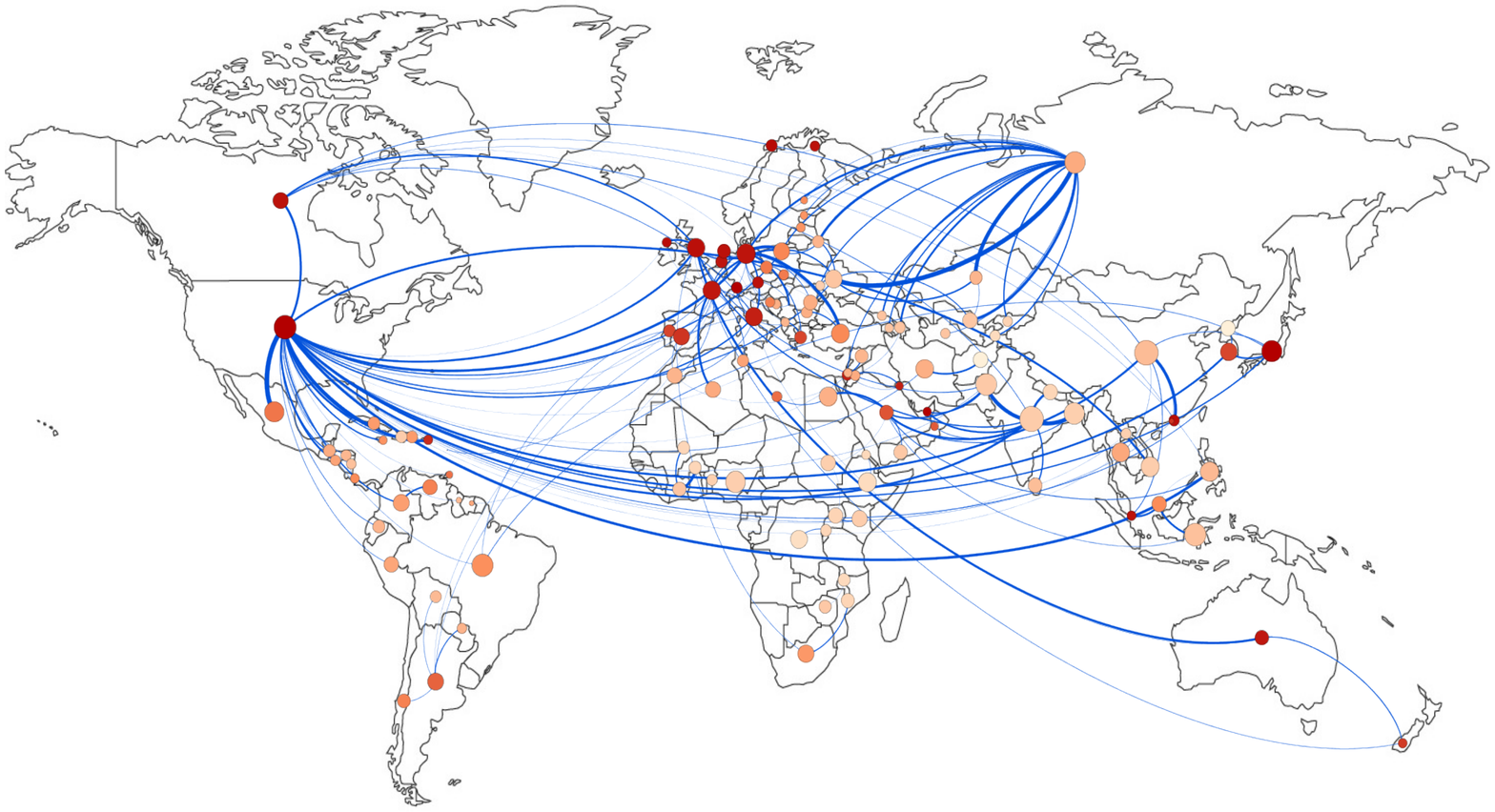}
\caption{The International-Migration Network (IMN) in year 2000. The figure plots the undirected weighted version of the IMN where only bilateral link weights (total number of bilateral migrants) larger than 200000 are reported. Tickness of links in the plot is proportional to the logs of link weights. The size of the nodes is proportional to the log of country population. Node color represents country income, measured by country per-capita Gross Domestic Product (pcGDP). Red=Higher Country pcGDP.}
 \label{fig:map_2000_undir}
\end{sidewaysfigure}

Figure \ref{fig:map_2000_undir} plots the undirected weighted version of the IMN in year $t=2000$. Link directions are suppressed to attain a better visualization of the graph and link thickness is proportional to the logs of total bilateral migrants ($m_{i,j}^t+m_{j,i}^t$). To get a feel of migration determinants, node size is proportional to the log of country population, while node color (from beige to red) represents country income, measured by country per-capita Gross Domestic Product (pcGDP). Only links associated to a number of total bilateral migrants larger than 200000 are shown. The map allows one to appreciate the central role of the US and Russia, the importance of Gulf and Asian countries, the widespread presence of low-income countries, and the local nature of many migration corridors.     

In Table \ref{tab:descriptives} we summarize instead the main properties of the IMN. As discussed in the previous Section, the volume of the network (total number of migrants) almost doubled from 1960 to 2000. The number of directed links (i.e. migration corridors, see \cite{Ozden_etal_data_2011}) and consequently network density also increased by about 45\% in the same period, in line with the density detected in the international-trade network (ITN), see Ref. \cite{Fagiolo2009pre}. Therefore, in the trade jargon, IMN growth was both intensive and extensive. The net increase in terms of average link weight was also positive, from more than 5 thousand to 7 thousand. The median link weight increased, too. However, it remained two orders of magnitude smaller than the mean, hinting to a highly skewed, broad link-weight distribution (see below). This can be seen also from the huge values attained by maximum link weight, which is more than 8 million in 1960 and 1970 (i.e., migration from Pakistan to India), and decreases to 5 million in 1980 and 1990 (i.e. migration from Russian Federation to Ukraine). In 2000, instead, the maximum stock of migrants (more than 9 million) was attained by the Mexico-USA corridor.

An important feature of directed networks is their degree of symmetry \cite{Fagiolo2006EcoBull}. 
From a binary perspective, the number of bilateral links in IMN has steadily increased, leading to a bilateral density (e.g. ratio of bilateral to existing links) of around two thirds, well below the 93\% figure of the ITN. As expected, the IMN is much less symmetric than the ITN: whereas aggregate-trade bilateral relationships can more easily emerge (due to e.g. preferential trade agreements), the directionality of migration corridors cannot be that easily reversed, especially if the drivers for migration reside in human capital and labour productivity differentials. From a weighted perspective, on the contrary, the asymmetry of the IMN has been increasing through time \footnote{We have computed the asymmetry index developed in Ref. \citep{Fagiolo2006EcoBull}. The index ranges between 0 (maximum weighted symmetry) to 1 (maximum weighted asymmetry). A similar result is obtained by computing the Pearsons correlation index between the upper-diagonal and the lower-diagonal of the weight matrix \cite{myreciprocity}.}, meaning that despite a fraction of unilateral migration corridors have been reciprocated across the years, the actual stock of migrants who moved over such bilateral corridors have become more and more asymmetric. Overall, this implies that a weighted directed analysis should be preferred and network symmetrization avoided.

Finally, the IMN appears to be almost fully strongly connected. The largest (strong) component features all countries but a few isolated ones. For example, in 1960 and 1970, only Norfolk Island and Taiwan were not chosen from any other country as destinations (even though they were the origin of some migration corridor). The same happened in 1980 for Taiwan and in 1990 for Belize. The giant component of the IMN displays a small constant diameter and a decreasing average path length (computed on directed paths). This suggests that the increase in density and bilateral links over time, due to historical trends discussed above, has resulted in shortening the overall distances between countries in the IMN, either by creating directed corridors between countries previously not connected, or by forming short-cuts able to bypass longer paths (more on that below).

\medskip

\section{Binary and Weighted Topology\label{Sec:Topology}}
This Section discusses the statistical properties of the binary and weighted architecture of the IMN. We are interested in characterizing the time-evolution of node/link network statistics, and their correlation structure \citep[we refer the reader to Refs.][for a formal discussion of all statistics employed in our analyses]{Fagiolo2009pre,Squartini_etal_2011a_pre,Squartini_etal_2011b_pre}. 

\medskip

\subsection{Link-Weight, Degree and Strength Distributions}
\begin{figure}
\includegraphics[width=9cm]{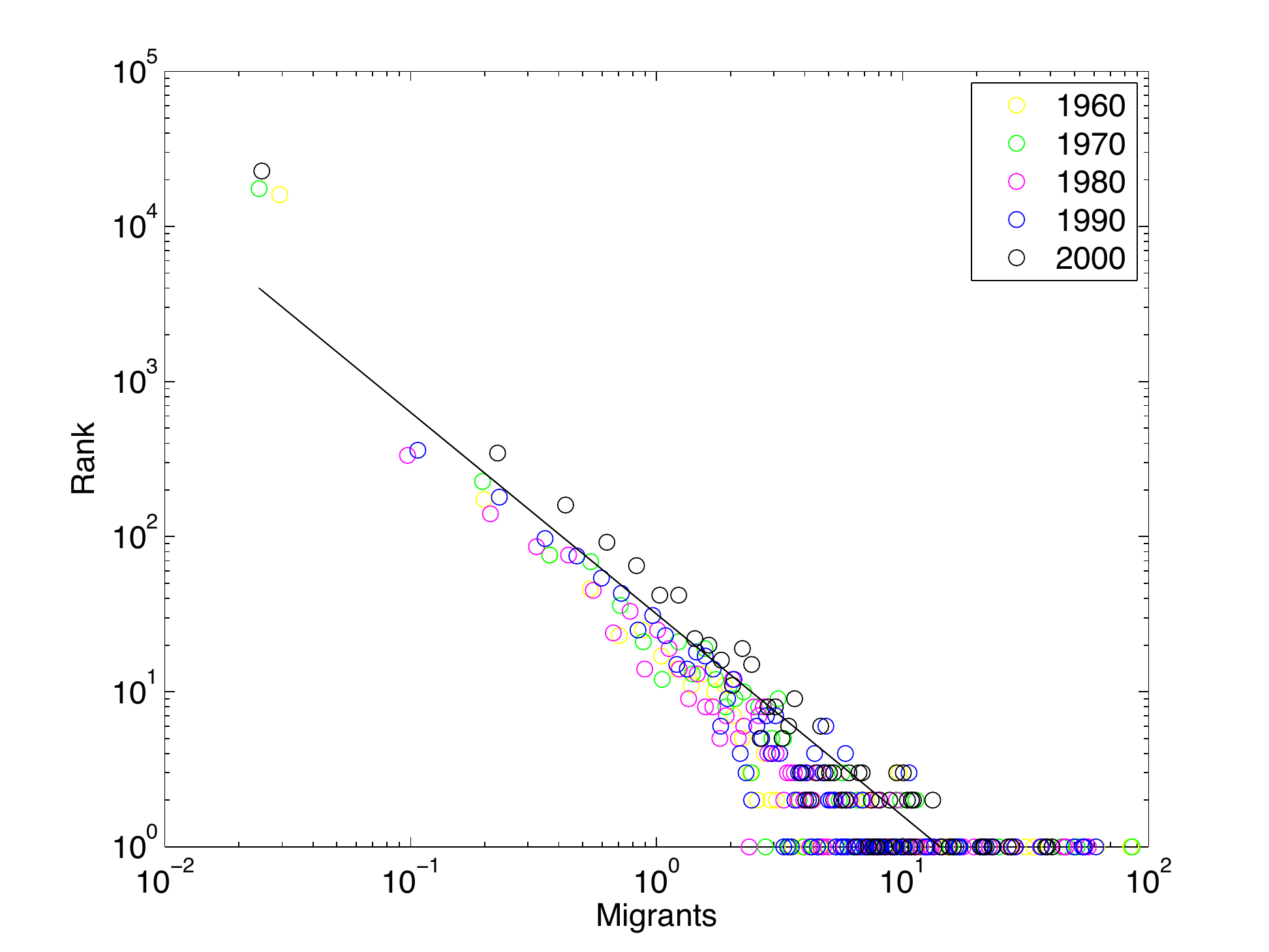}
\caption{Log-log plot of the link-weight distribution (number of migrants) in different years. Link-weights are rescaled by the volume of the IMN in order to wash away the trend. Power-law exponent: 1.36.}
 \label{fig:linkweight_distr}
\end{figure}

A first striking result concerns the extent of heterogeneity across link weight and node statistics in the IMN. As Figure \ref{fig:linkweight_distr} shows, link weights are power-law distributed across all the years with a stable power law exponent of about 1.3. This strongly differs from log-normal weights that one typically finds in the ITN and other macroeconomic networks \citep{Fagiolo2009pre,SRF_2010_QF} and suggests a rich-get-richer process for the accumulation of migration stocks. This is well in tune with the push-pull literature on migration determinants \citep{Lee_1966,GastonNelson_2011}, according to which migration occurs if the reason to emigrate (the push) is balanced by a corresponding pull at destination. Push factors incentivizing emigration include the lack of economic opportunities, religious or political persecution, hazardous environmental conditions, and so on. Pull factors at destination instead comprise availability of jobs, religious or political freedom, the perception of a relatively benign environment. Existence of institutions at the origin and well-established communities of foreign-born people at destination also fosters migration. If pull and push factors induce persistence gaps between origin and destination along their development phases, they are likely to trigger a self-enforcing process where strong migration corridors increasingly attract people.

\begin{figure}[h]
        \begin{minipage}[h]{8.5cm}
        \begin{scriptsize}
        \centering {\includegraphics[height=7cm,width=8cm]{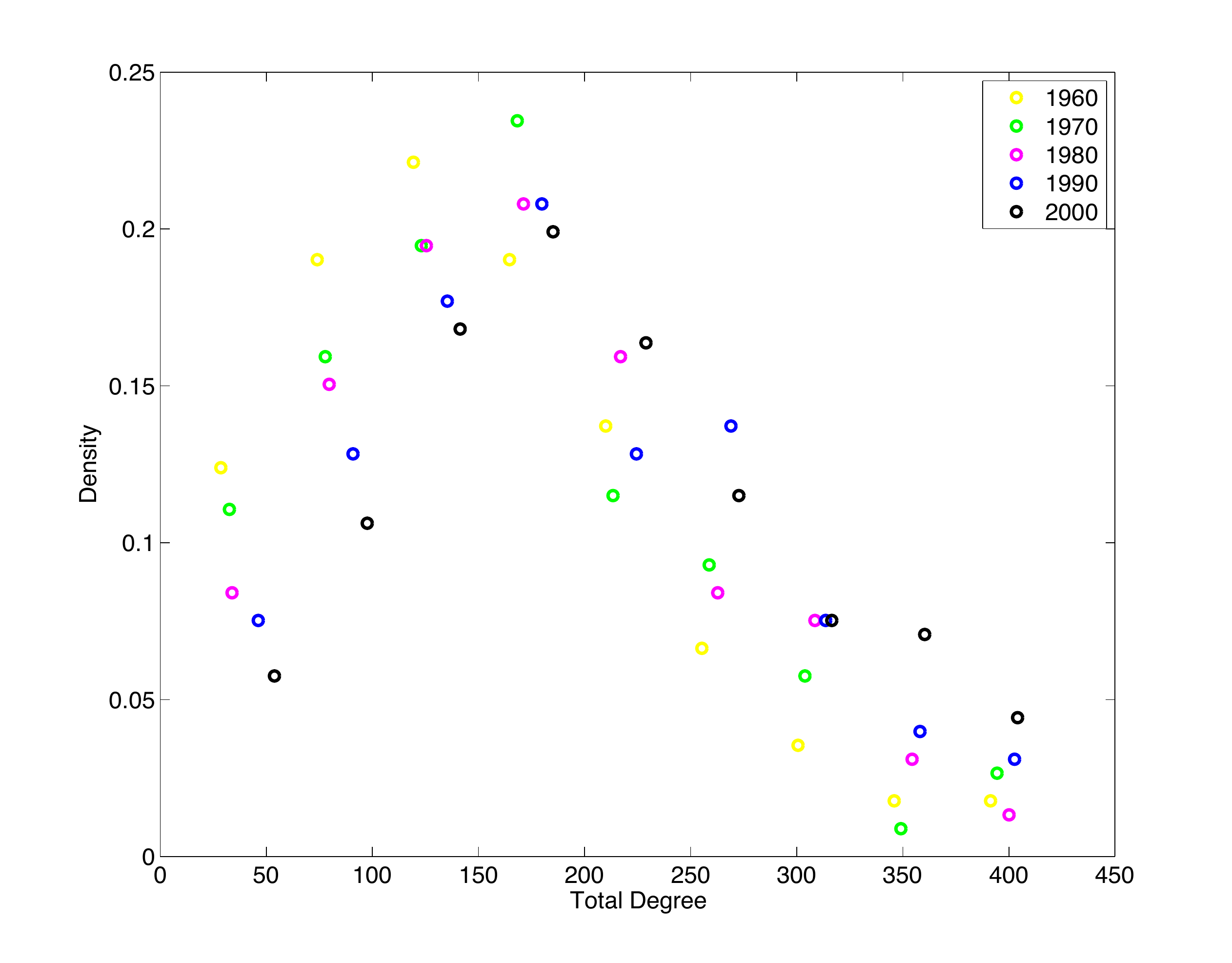}}
        \caption{Total degree distribution of the IMN in different years.}\label{fig:tot_deg_dens_years}
        \end{scriptsize}
        \end{minipage}\hfill
        \begin{minipage}[h]{7.5cm}
        \begin{scriptsize}
        \centering {\includegraphics[height=7cm,width=7.5cm]{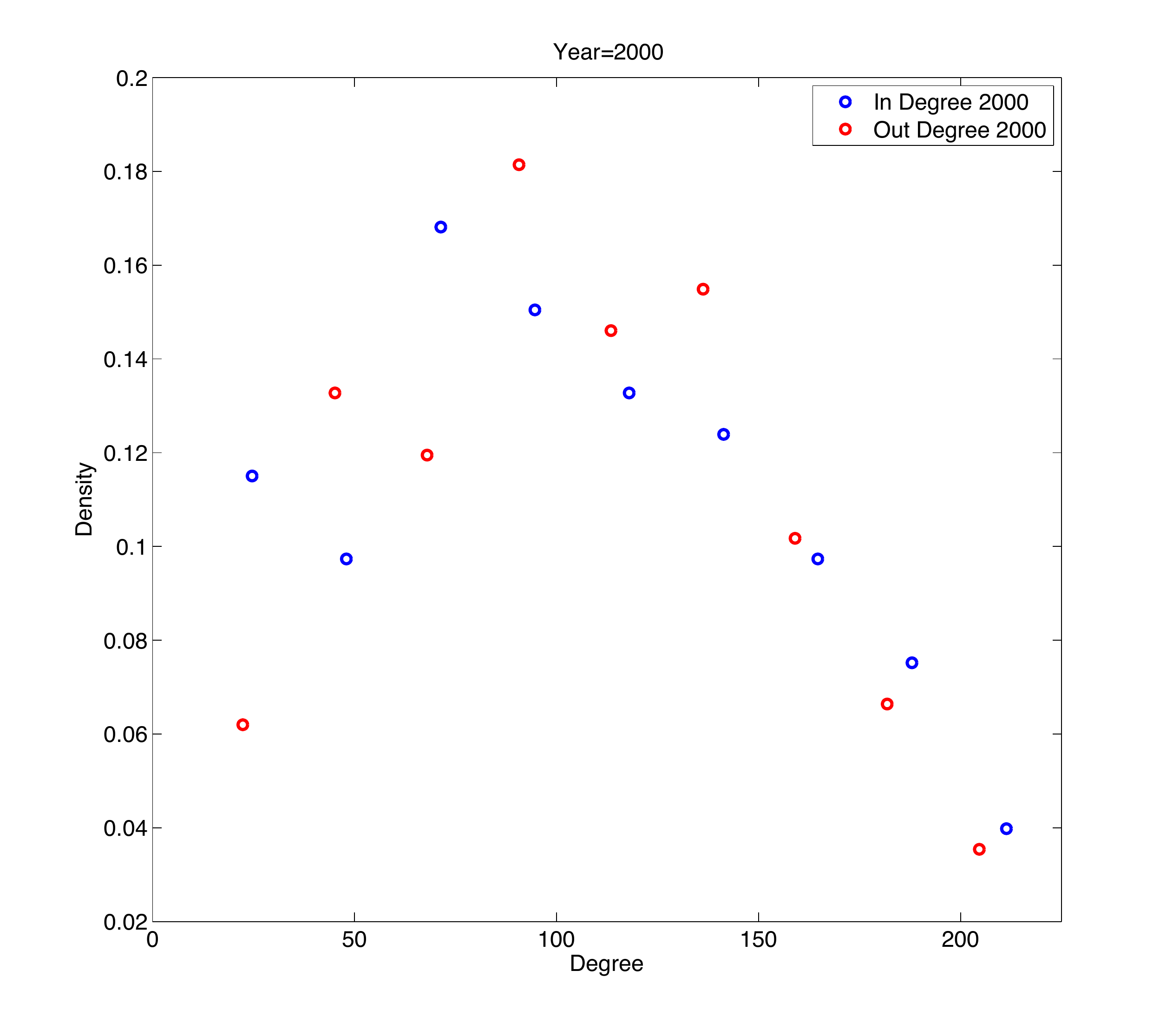}}
        \caption{In-degree and out-degree distribution of the IMN in year 2000.}\label{fig:inout_degree_dens_2000}
        \end{scriptsize}
        \end{minipage}
        \end{figure}

On the contrary, the distributions of in-, out- and total node degree, respectively defined for a given year (superscripts omitted for simplicity) as: 
\begin{eqnarray}
ND^{out}_i=\sum_j{a_{ij}},\\ \nonumber
ND^{in}_i=\sum_j{a_{ji}},\\
ND_i=ND^{out}_i+ND^{in}_i, \nonumber
\end{eqnarray}
exhibit a characteristic scale, see Figures \ref{fig:tot_deg_dens_years}-\ref{fig:inout_degree_dens_2000}. Hence, most countries share an intermediate number of migration corridors (around 170 in total), with a right tail decreasing almost exponentially. 

Degree distributions are remarkably stable across the years. This is because the underlying link formation and deletion process is quite persistent. Over two consecutive decades, the probability of link formation grew from 7.8\% to 12.8\%, whereas the likelihood that an already established corridor is severed went from 6.5\% to 8\%. This means that between any two consecutive decades the IMN binary architecture has remained quite stable: on average 91.4\% of country pairs did not change their status (linked or not linked). These figures closely match those of the ITN \citep{Fagiolo2009pre}.     

We also find that the number of immigration and emigration corridors is identically distributed (this is confirmed also by two-sample Kolmogorov-Smirnov statistical tests) and unimodal across the years, which implies that, despite countries do not reciprocate all the directed bilateral links, they keep on average an identical number of in-ward and out-ward migration links.

\begin{figure}[h]
        \begin{minipage}[h]{7.5cm}
        \begin{scriptsize}
        \centering {\includegraphics[height=7cm,width=7.5cm]{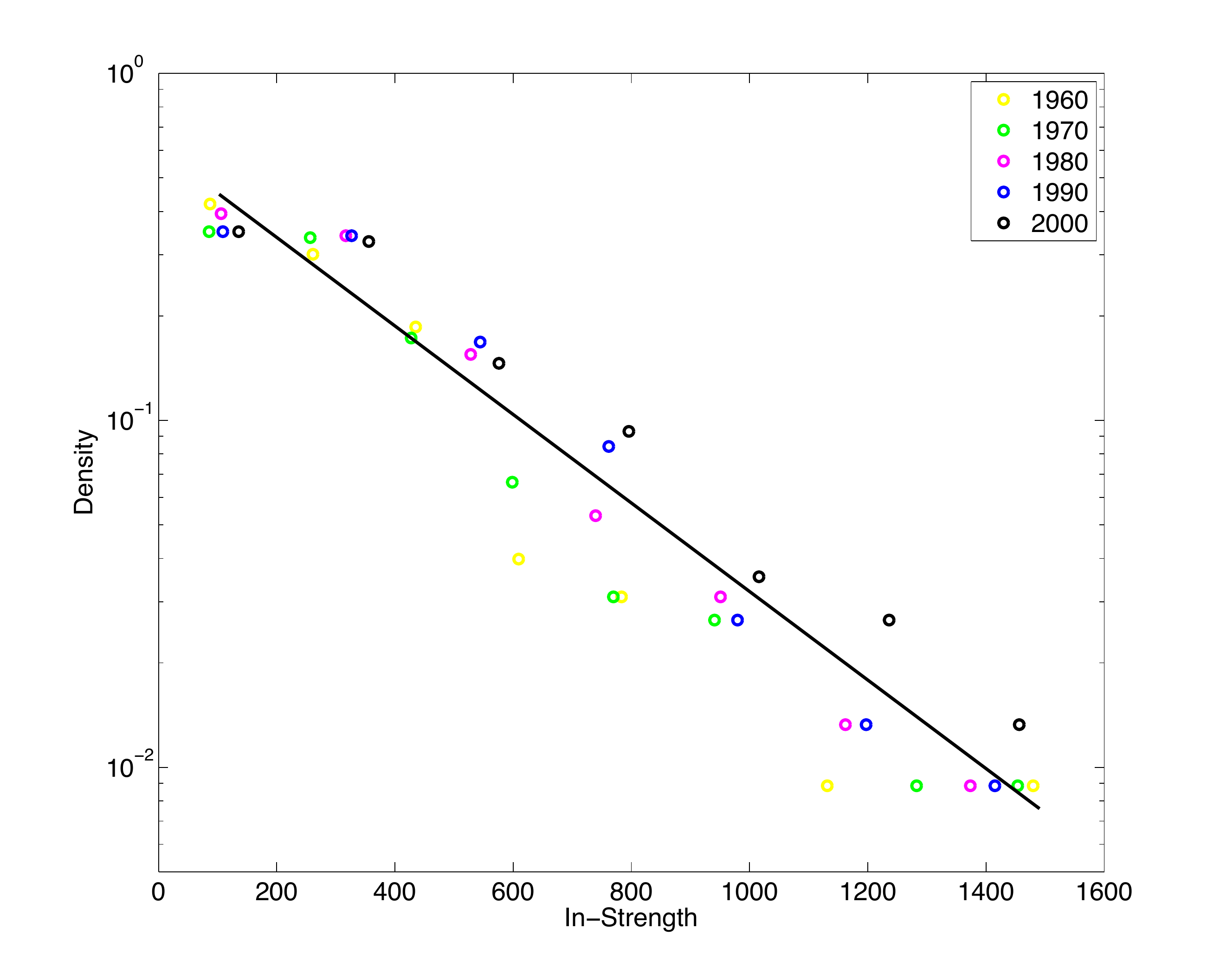}}
        \caption{In-strength distribution of the IMN in year 2000. Strength computed using log of migrants link weights $\tilde{m}_{i,j}^t$.}\label{fig:instrength}
        \end{scriptsize}
        \end{minipage}\hfill
        \begin{minipage}[h]{7.5cm}
        \begin{scriptsize}
        \centering {\includegraphics[height=7cm,width=7.5cm]{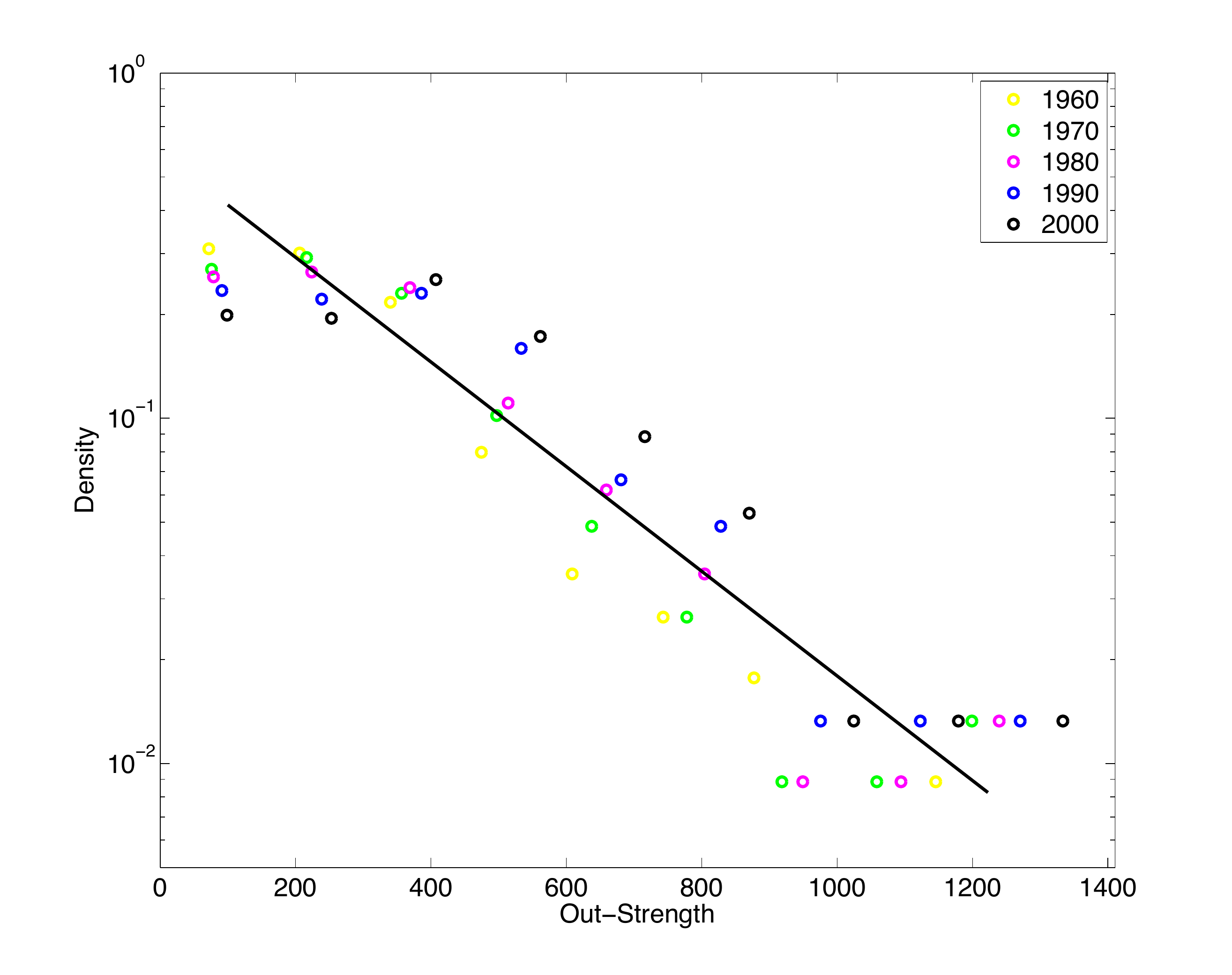}}
        \caption{Out-strength distribution of the IMN in year 2000. Strength computed using log of migrants link weights $\tilde{m}_{i,j}^t$.}\label{fig:outstrength}
        \end{scriptsize}
        \end{minipage}
        \end{figure}

\begin{figure}[h]
        \begin{minipage}[h]{7.5cm}
        \begin{scriptsize}
        \centering {\includegraphics[height=7cm,width=7.5cm]{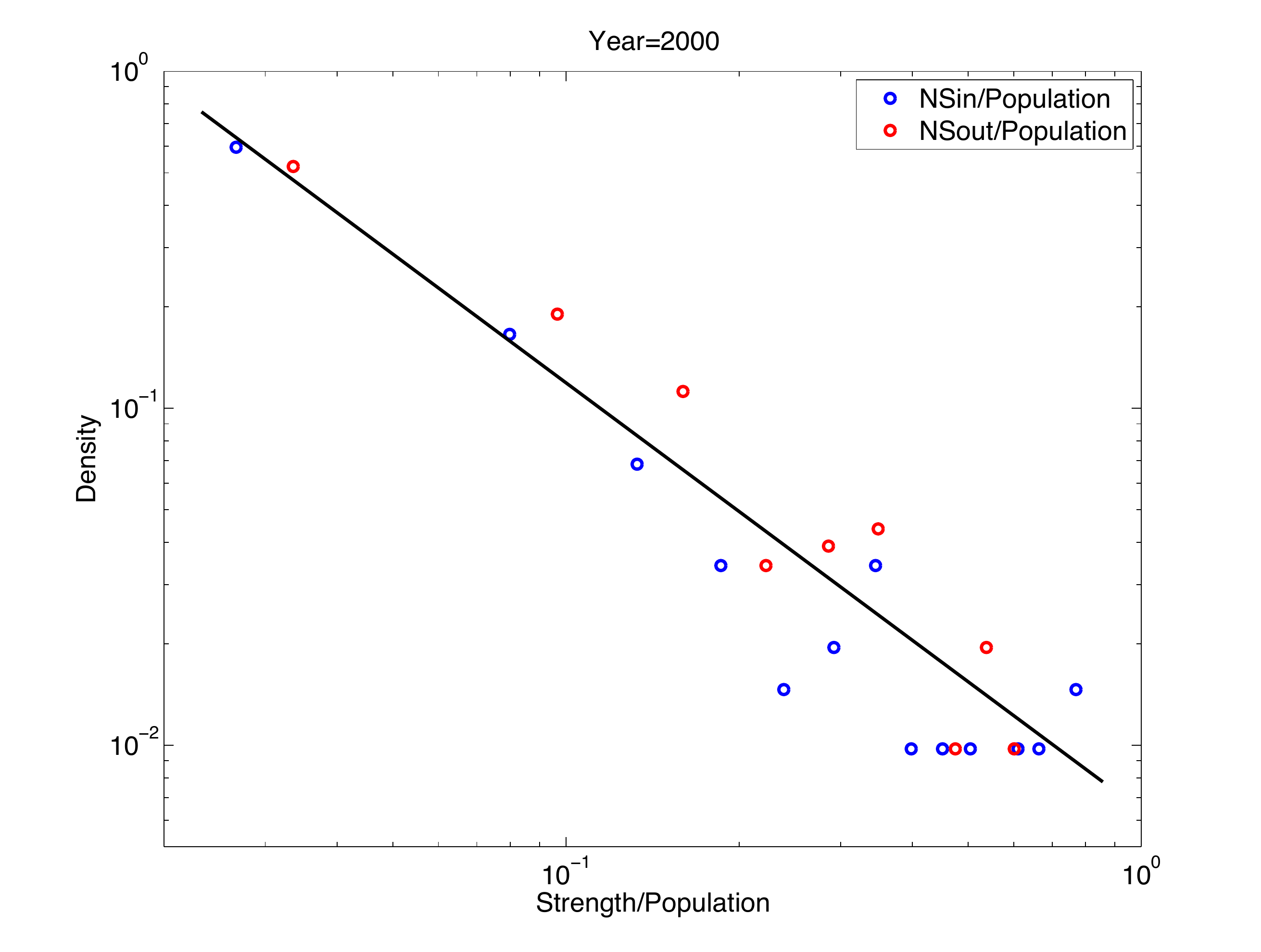}}
        \caption{Log-log plot of in- and out-strength rescaled by country population in year 2000. Rescaling does not alter the power-law behavior of total country immigrant and emigrant stocks. Here we compute in- and out-strength using the original number of bilateral migrant stocks $m_{i,j}^t$.}\label{fig:str_pop}
        \end{scriptsize}
        \end{minipage}\hfill
        \begin{minipage}[h]{7.5cm}
        \begin{scriptsize}
		\vskip -1.5cm
        \centering {\includegraphics[height=7cm,width=8cm]{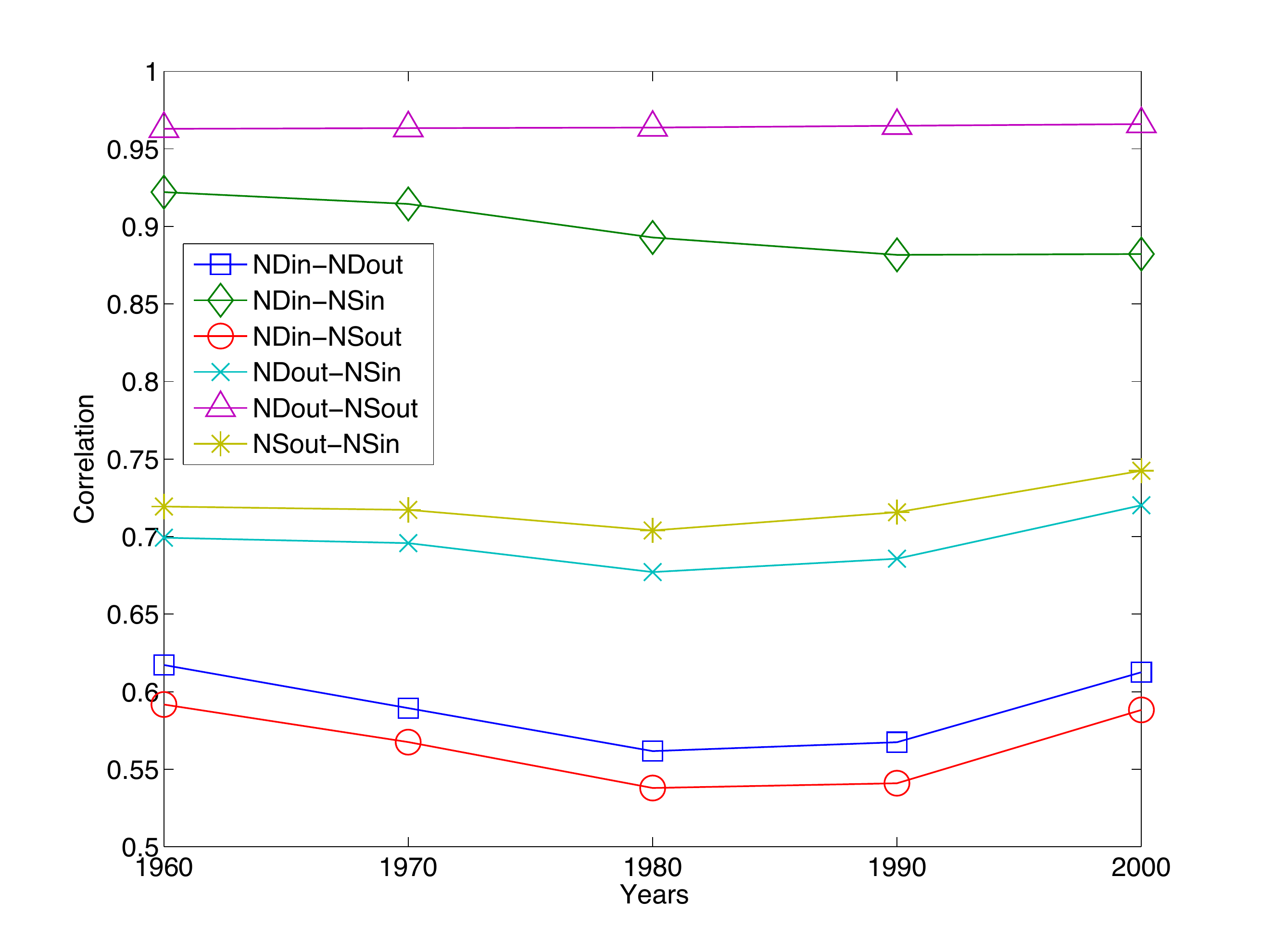}}
        \caption{Correlation coefficients between in/out degree and in/out strength vs. years.}\label{fig:deg_str_corr}
        \end{scriptsize}
        \end{minipage}
        \end{figure}

The high heterogeneity in link weights maps into power-law distributed in, out and total node strength \citep{Barrat2004pnas}.
Note that, in order to mitigate the effect of power-law distributed link weights, in the remainder of the paper we will log-transform the weights, i.e. we shall weight each link by $\tilde{m}_{i,j}^t=\log(m_{i,j}^t)$. In, out and total node strength read:
\begin{eqnarray}
NS^{out}_i=\sum_j{\tilde{m}_{ij}},\\ \nonumber
NS^{in}_i=\sum_j{\tilde{m}_{ji}},\\ \nonumber
NS_i=NS^{out}_i+NS^{in}_i,
\end{eqnarray}
As a result, one typically finds a linear-log relation in both in- and out-strength, which in this case is a transformation of total number of immigrants and emigrants of a given country, see Figures \ref{fig:instrength}-\ref{fig:outstrength}. Therefore, country immigrant and emigrant stocks are broadly distributed, with a small bunch of countries that host very large stocks of migrants. Interestingly, this power-law behavior is not the effect of heterogeneous country size. Indeed, as Figure \ref{fig:str_pop} shows, a power law distribution for both in- and out-strength is obtained also when one rescales the latter by country size: fat-tails emerge also in the distribution of the shares of total immigrant and emigrant to country population.

Degree and strength profiles are positively correlated in the IMN. Figure \ref{fig:deg_str_corr} plots Pearson's correlation coefficients between in/out degree and strength. As expected, the more inward (outward) migration corridors a country has, the more immigrants (emigrants) it channels (correlation close to one). However, countries that are chosen as destination by many other countries are not necessarily the origin of many emigration corridors (positive but milder correlation between in and out degree). Correlation levels increase when one considers inward and outward migrant stocks, because of the effect of large and high-income countries that both receive a lot of immigrants and send to many world destinations a lot of emigrants. Finally, countries with many inward (outward) corridors are not necessarily associated to more emigrants (immigrants).

\begin{figure}[h]
        \centering {\includegraphics[width=10cm]{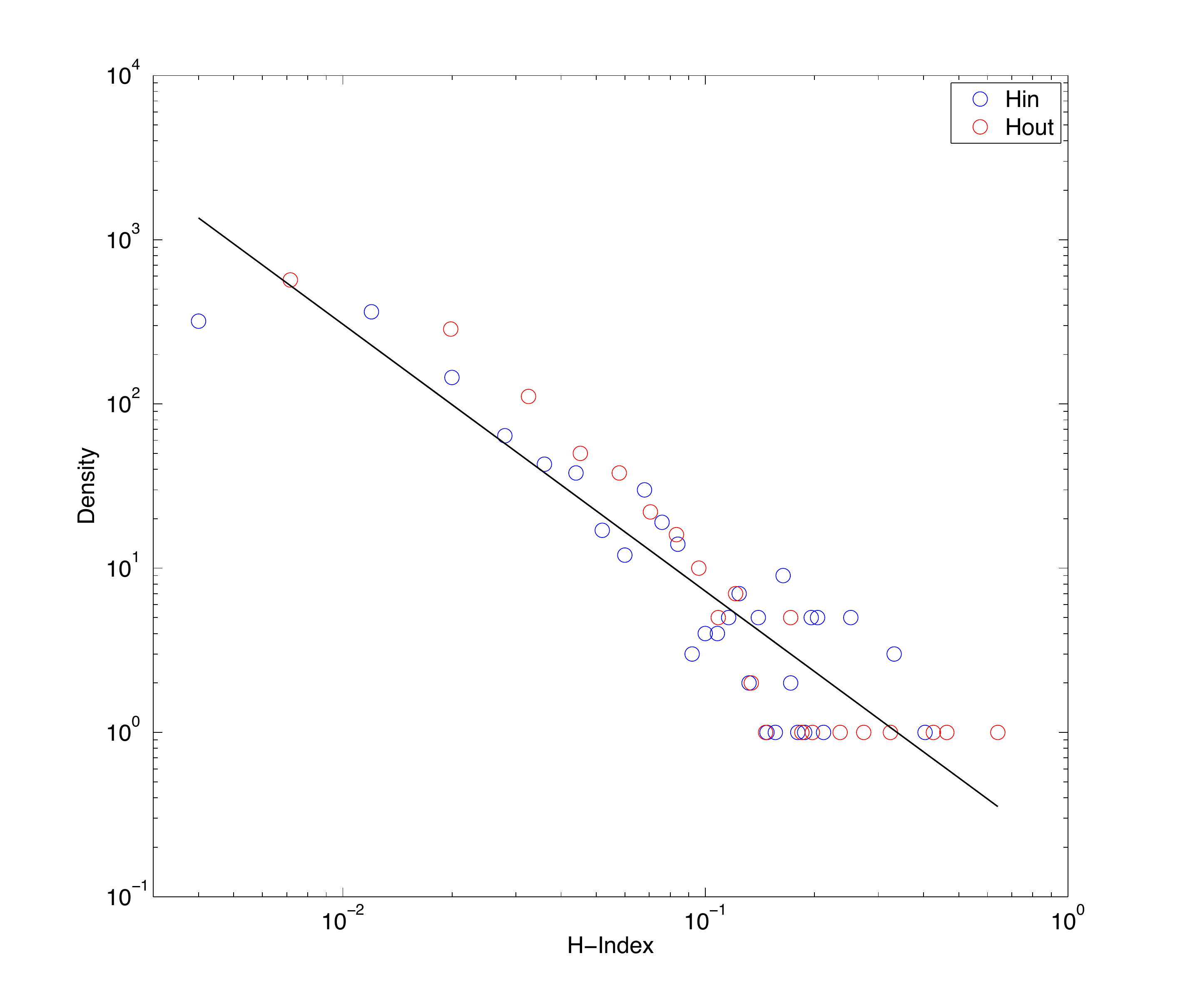}}
        \caption{The distribution of link-weight concentration indices. For each given country, $H^{in}$  measures the concentration of its incoming link-weight (immigration) portfolio, whereas $H^{out}$ measures the concentration of its outgoing link-weight (emigration) portfolio. Valued are pooled across all the 5 waves.}\label{fig:h_index_powerlaw}
        \end{figure}

\begin{figure}[h]
        \begin{minipage}[h]{7.5cm}
        \begin{scriptsize}
        \centering {\includegraphics[height=7cm,width=8cm]{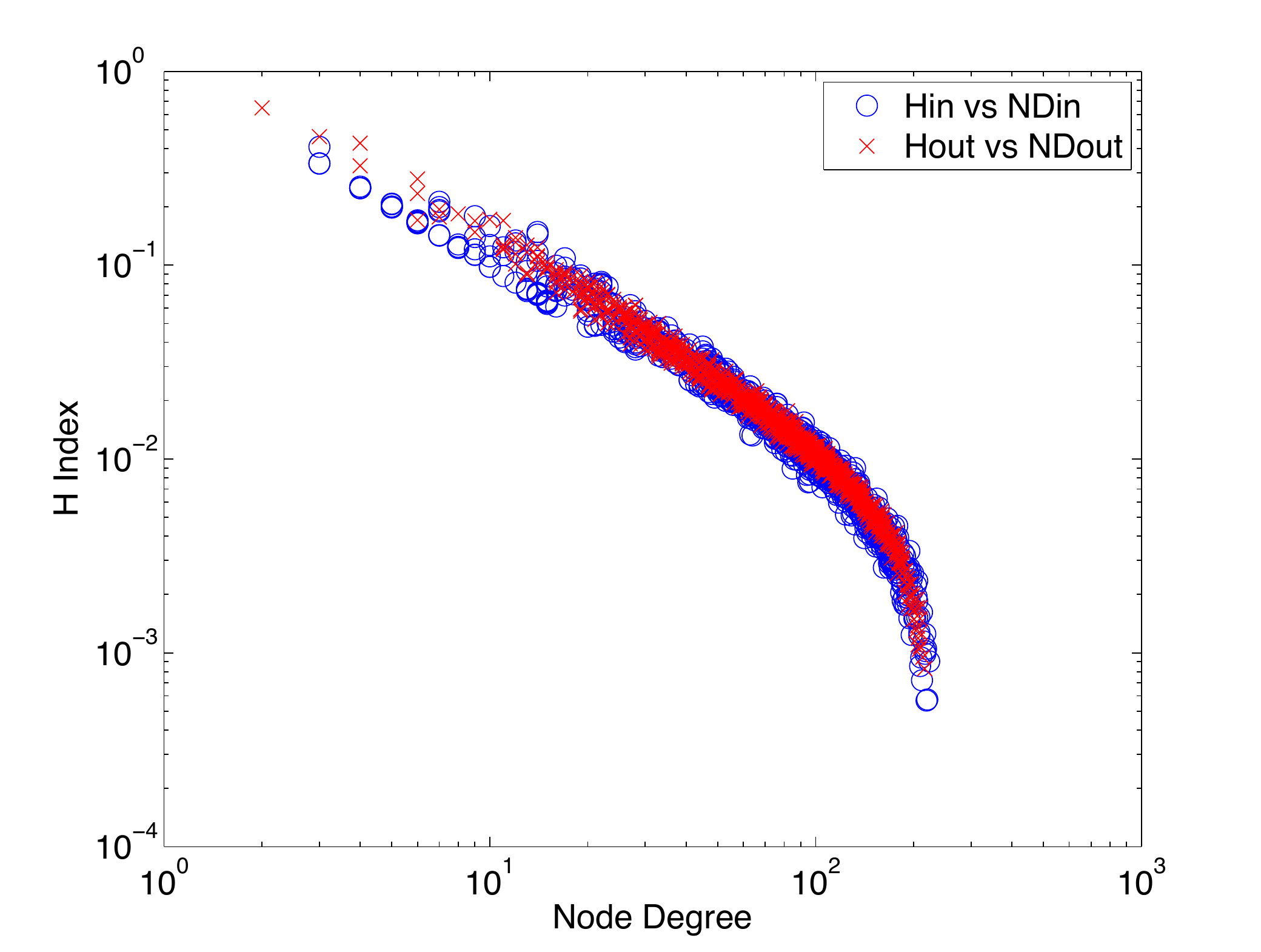}}
        \caption{Correlation between link-weight concentration indices and node in/out degrees. For each given country, $H^{in}$  measures the concentration of its incoming link-weight (immigration) portfolio, whereas $H^{out}$ measures the concentration of its outgoing link-weight (emigration) portfolio. Valued are pooled across all the 5 waves.}\label{fig:nd_h}
        \end{scriptsize}
        \end{minipage}\hfill
        \begin{minipage}[h]{7.5cm}
        \begin{scriptsize}
        \centering {\includegraphics[height=7cm,width=8cm]{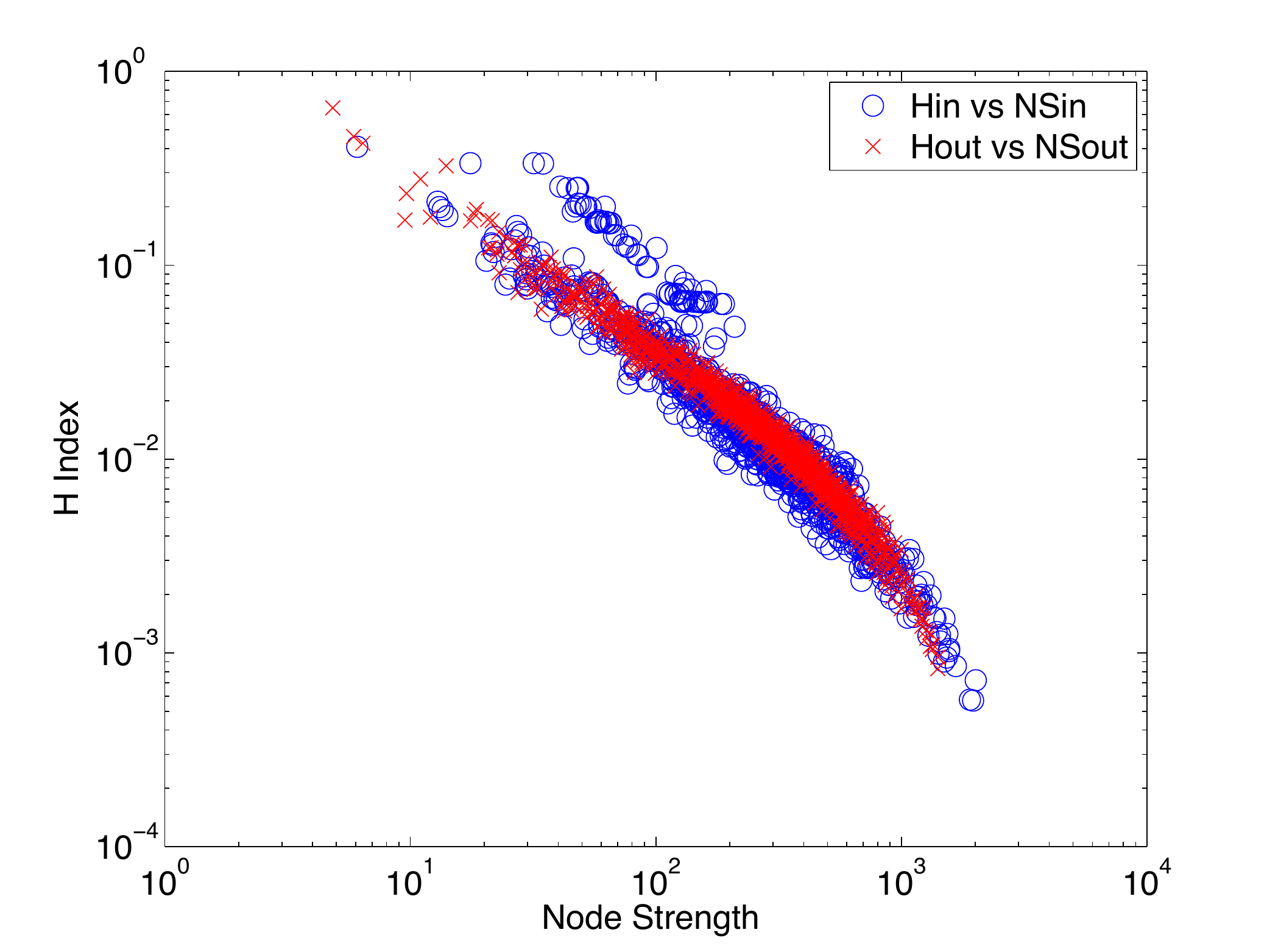}}
        \caption{Correlation between link-weight concentration indices and node in/out strengths. For each given country, $H^{in}$  measures the concentration of its incoming link-weight (immigration) portfolio, whereas $H^{out}$ measures the concentration of its outgoing link-weight (emigration) portfolio. Valued are pooled across all the 5 waves.}\label{fig:ns_h}
        \end{scriptsize}
        \end{minipage}
        \end{figure}

We now investigate the extent to which countries in the IMN hold concentrated or diversified portfolios of migrant stocks. For each country $i$, we compute the concentration of its incoming link-weight (immigration) portfolio as:
\begin{equation}
H_i^{in}=\frac{(N-1)\sum_j{\left(\frac{\tilde{m}_{ji}}{NS_i^{in}}\right)^2}-1}{N-2},
\end{equation}
and of its outgoing link-weight (emigration) portfolio as: 
\begin{equation}
H_i^{out}=\frac{(N-1)\sum_j{\left(\frac{\tilde{m}_{ij}}{NS_i^{out}}\right)^2}-1}{N-2}.
\end{equation}
In other words, we compute Herfindahl concentration indices \cite{herf_1964} on in- and out- link weight country portfolios. H indices range from 0 (homogeneous distribution) to 1 (fully concentrated distribution). We find that countries heavily differ in the concentration profiles robustly over time. Indeed, both $H^{in}$ and $H^{out}$ are power-law distributed, see Figure \ref{fig:h_index_powerlaw}. This suggests that countries with a fairly uniform distribution of immigrant or emigrant stocks among migration corridors co-exist with countries that possess a strongly concentrated portfolio of migrant stocks. Furthermore, there exists a negative power-law relation between $(H^{in},H^{out})$ and node connectivity (Figures \ref{fig:nd_h} and \ref{fig:ns_h}). Countries that hold more inward (outward) migration corridors or larger stocks of immigrants (emigrants) are typically less concentrated, i.e. they hold more even portfolios of immigrants (emigrants). While high concentration of migrant portfolios is somewhat expected for small players in the IMN, the power-law decrease in concentration as connectivity increases strongly supports the view that big players in the IMN tend to diversify a lot in their migration patterns, fully exploiting all migration corridors.

\subsection{Assortativity, Clustering, Path Length}
An important feature of complex networks is their assortative or disassortative pattern \citep{newman02assortative,MEJN-mixing03}. Here we study whether countries characterized by more inward/outward corridors or large emigrant/immigrant stocks are connected with countries that in turn are more connected either in terms or degrees or strength. We capture this by computing Pearson's correlation coefficient between node total degree or strength (ND or NS), and average nearest-neighbor degree or strength (ANND or ANNS), defined as the average degree (strength) of the neighbors of a given country.

\begin{figure}[h]
        \begin{minipage}[h]{7.5cm}
        \begin{scriptsize}
        \centering {\includegraphics[height=7cm,width=7.5cm]{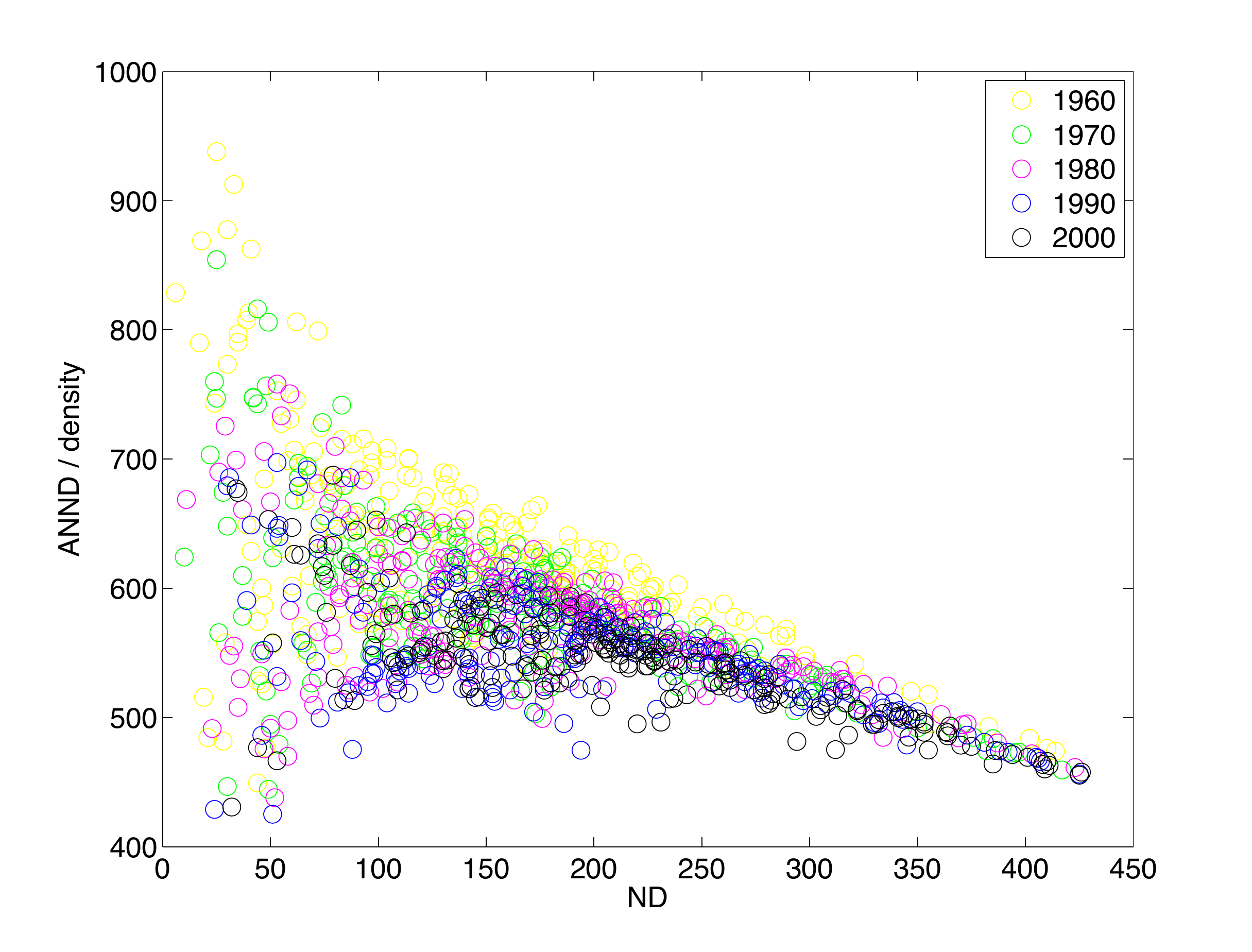}}
        \caption{Disassortative behavior in the binary IMN. Average neraest-neighbor degree (ANND) rescaled by network density vs total node degree in different years.}\label{fig:annd_nd}
        \end{scriptsize}
        \end{minipage}\hfill
        \begin{minipage}[h]{7.5cm}
        \begin{scriptsize}
        \centering {\includegraphics[height=7cm,width=8cm]{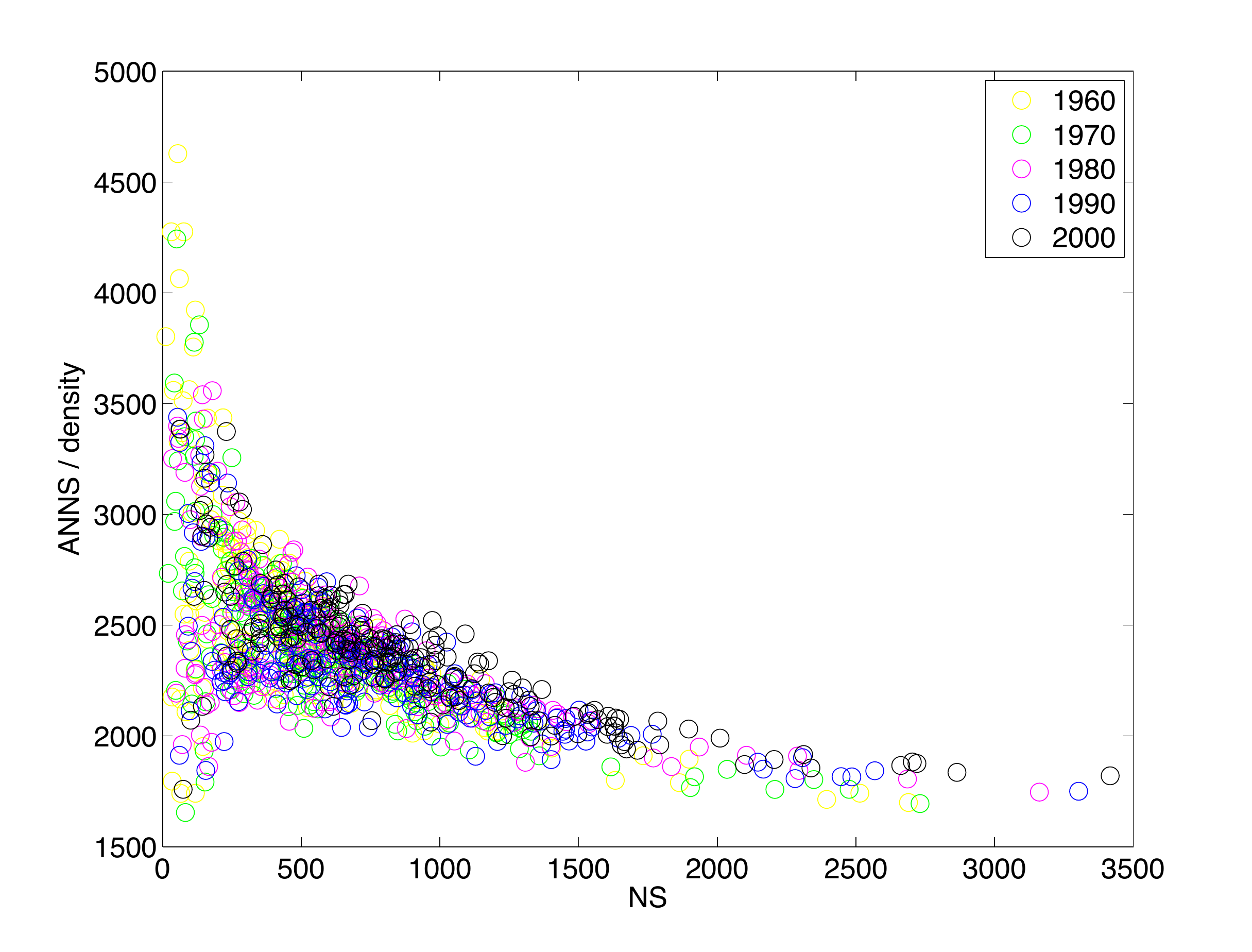}}
        \caption{Disassortative behavior in the weighted IMN. Average neraest-neighbor strength (ANNS) rescaled by network density vs total node strength in different years.}\label{fig:anns_ns}
        \end{scriptsize}
        \end{minipage}
        \end{figure}

As Figures \ref{fig:annd_nd}-\ref{fig:anns_ns} show, both the binary and the weighted versions of the IMN are (weakly) disassortative. Large players in the IMN are typically connected to countries that hold on average a small number of migration corridors or channel a small stock of migrants. Notice that instead small players are very heterogeneous. Mostly due to geo-political constraints, some small countries, typically Pacific islands, are locally connected to few other geographically-close countries that are themselves poorly connected. Other small countries instead hold privileged corridors to big players (e.g. European and North-American countries) and therefore display a disassortative mixing. This evidence is very robust when one disaggregates ANND and ANNS in order to take into account directionality.

In many networks, disassortative patterns come together small level of binary or weighted clustering \cite{Fagiolo2008physa}. To check if this is the case for the IMN, we compute binary total node-clustering coefficient (BCC) as:
\begin{equation}
BCC_i=\frac{(A+A^T)_{ii}^3}{2[ND_i(ND_i-1)-2d_i^{\leftrightarrow}]},
\label{Eq:CC_BDN}
\end{equation}
where $A$ is the adjacency matrix, $A^T$ is its transpose and $d_i^{\leftrightarrow}$ is the number of reciprocated edges between $i$ and its neighbors. Similarly, weighted total node-clustering (WCC) reads:
\begin{equation}
WCC_i=\frac{(\hat{M}+\hat{M}^T)_{ii}^3}{2[ND_i(ND_i-1)-2d_i^{\leftrightarrow}]},
\label{Eq:CC_BDN}
\end{equation}
where $\hat{M}=M^{[1/3]}$. 
We find that the IMN displays a very large level of average clustering. Average BCC increases from 0.64 to 0.69 in our sample year. Correspondingly, average WCC goes from 0.13 to 0.15 (a relatively high value, comparable to that in the ITN when logs of export flows are employed to weigh the links).

\begin{figure}[h]
        \centering {\includegraphics[height=7cm,width=8cm]{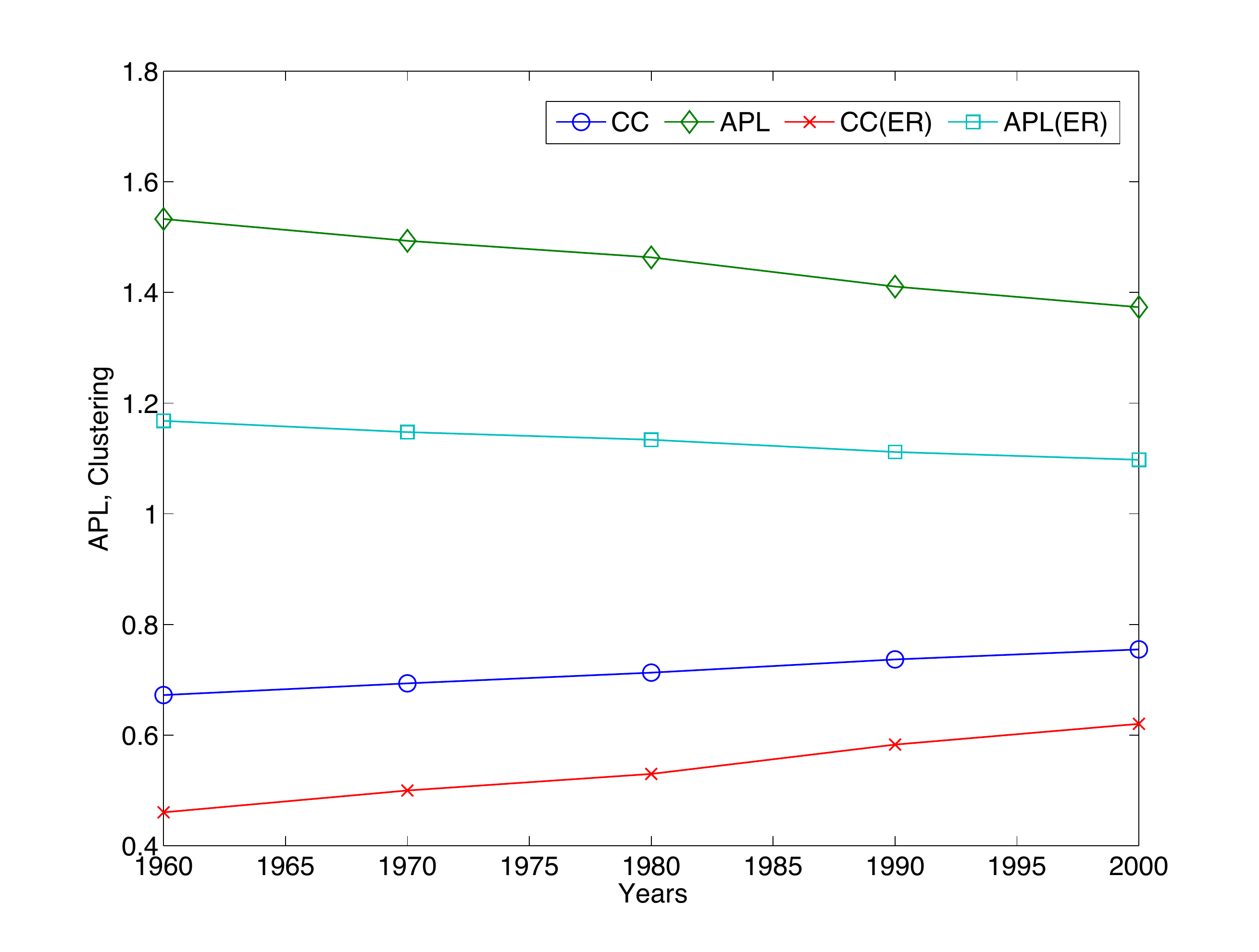}}
        \caption{Small-world behavior of the IMN. We plot network-wide binary clustering coefficient \cite{WattsStrogatz1998} and average-path length in the binary undirected IMN together with their values in random Erdos-Renyi (ER) graphs preserving observed IMN density in each year.}\label{fig:cc_apl}
        \end{figure}

We already know from Table \ref{tab:descriptives} that directed average-path length (APL) is very low and has been decreasing over time. Putting these two pieces of evidence together, we conclude that the binary IMN exhibits small-world features \cite{WattsStrogatz1998}. This is exemplified in Figure \ref{fig:cc_apl}, where we plot the global binary clustering coefficient (CC) \citep{Luce_Perry_1949} and (undirected) average-path length against their expected values in corresponding Erdos-Renyi (ER) random graphs where observed density is preserved. Note that APL is larger than its expected value and decreases over time, whereas the clustering coefficient is 30\% larger than its expected value in ER graphs and keeps increasing over time from 0.67 to 0.76. This means that, as we approach year 2000, the formation of new links has gradually filled previously uncompleted triplets, thus contributing to the decrease of geodesic distances and to the increase in clustering. A larger number of previously unconnected countries have opened migration corridors between them. This has increased the probability that any two countries that shared a migration corridor with a target country became themselves migration partners.

\begin{figure}[h]
        \begin{minipage}[h]{7.5cm}
        \begin{scriptsize}
        \centering {\includegraphics[height=7cm,width=8cm]{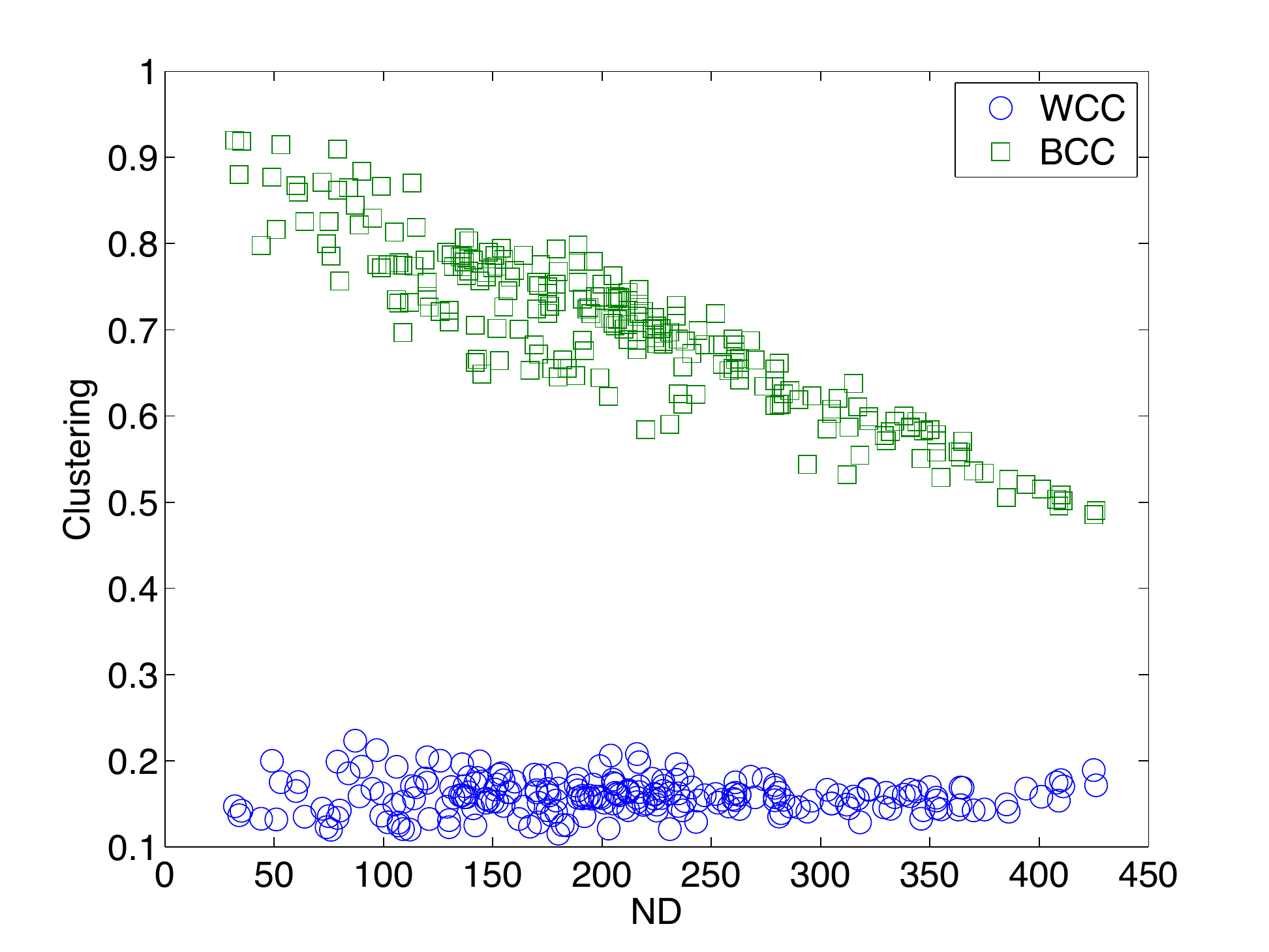}}
        \caption{Node clustering coefficients vs. node degree (ND) in year 2000. WCC: Weighted clustering coefficient. BCC: binary clustering coefficient.}\label{fig:cc_nd}
        \end{scriptsize}
        \end{minipage}\hfill
        \begin{minipage}[h]{7.5cm}
        \begin{scriptsize}
        \centering {\includegraphics[height=7cm,width=8cm]{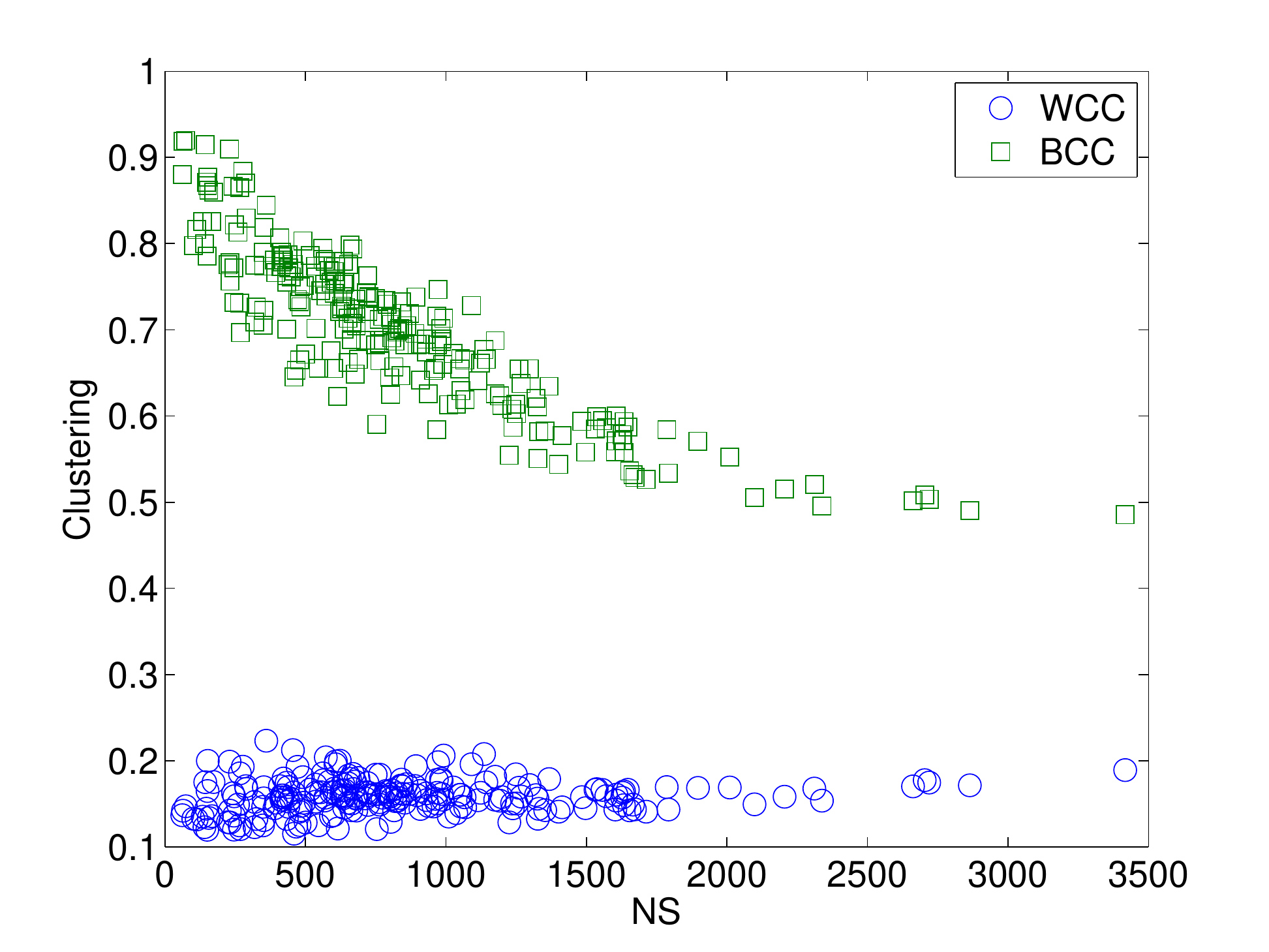}}
        \caption{Node clustering coefficients vs. node strenght (NS) in year 2000. WCC: Weighted clustering coefficient. BCC: binary clustering coefficient.}\label{fig:cc_ns}
        \end{scriptsize}
        \end{minipage}
        \end{figure}

We now explore whether country local clustering is associated to country connectivity, i.e. whether country binary and weighted clustering coefficients (BCC and WCC) are correlated with degrees and strengths. Two interesting facts emerge (see Figures \ref{fig:cc_nd}-\ref{fig:cc_ns} for an example). First, node BCC is negatively correlated with both ND and NS. This hints to a hierarchical structure where very connected nodes, in terms of both ND and NS, typically form few triangles with their neighbors --- a phenomenon possibly driven by the few large hubs present in the IMN. Second, we find that nodes that either hold a small number of migration corridors, or a small number of migrants, or both, are typically connected with pairs of countries that are also connected. However, the intensity of these triangles is the same of those created by nodes holding a lot of intense connections --- i.e. most of the existing triplets are really of a weak intensity. It is important to notice that the absence of a positive significant correlation between weighted clustering and node strength is partly due to the log-transformation of link weights. Indeed, the WCC-NS correlation computed using migrant levels (and not their logs) to weigh the links become significantly positive. This is because nodes that are very connected (very high NS) typically hold many strong pairs of links. If only a few of these pairs form a triangle, their weights amplify that of the triangle itself, thus leading to very high WCC. This is not true if one log-linearizes link weights, as the link-weight distribution displays a characteristic scale, with tails decaying almost exponentially. The same mismatch happens in the ITN: when one weights the links using export levels, more intensively-trading countries for more heavily clustered neighborhoods \cite{Fagiolo2008physa}, while a negative correlation is found using logs of exports as link weights. 

\subsection{Rich Club, Core-Periphery and Community Structure}

\begin{figure}[h]
        \begin{minipage}[h]{7.5cm}
		\vskip -1.8cm
        \begin{scriptsize}
        \centering {\includegraphics[height=7cm,width=8cm]{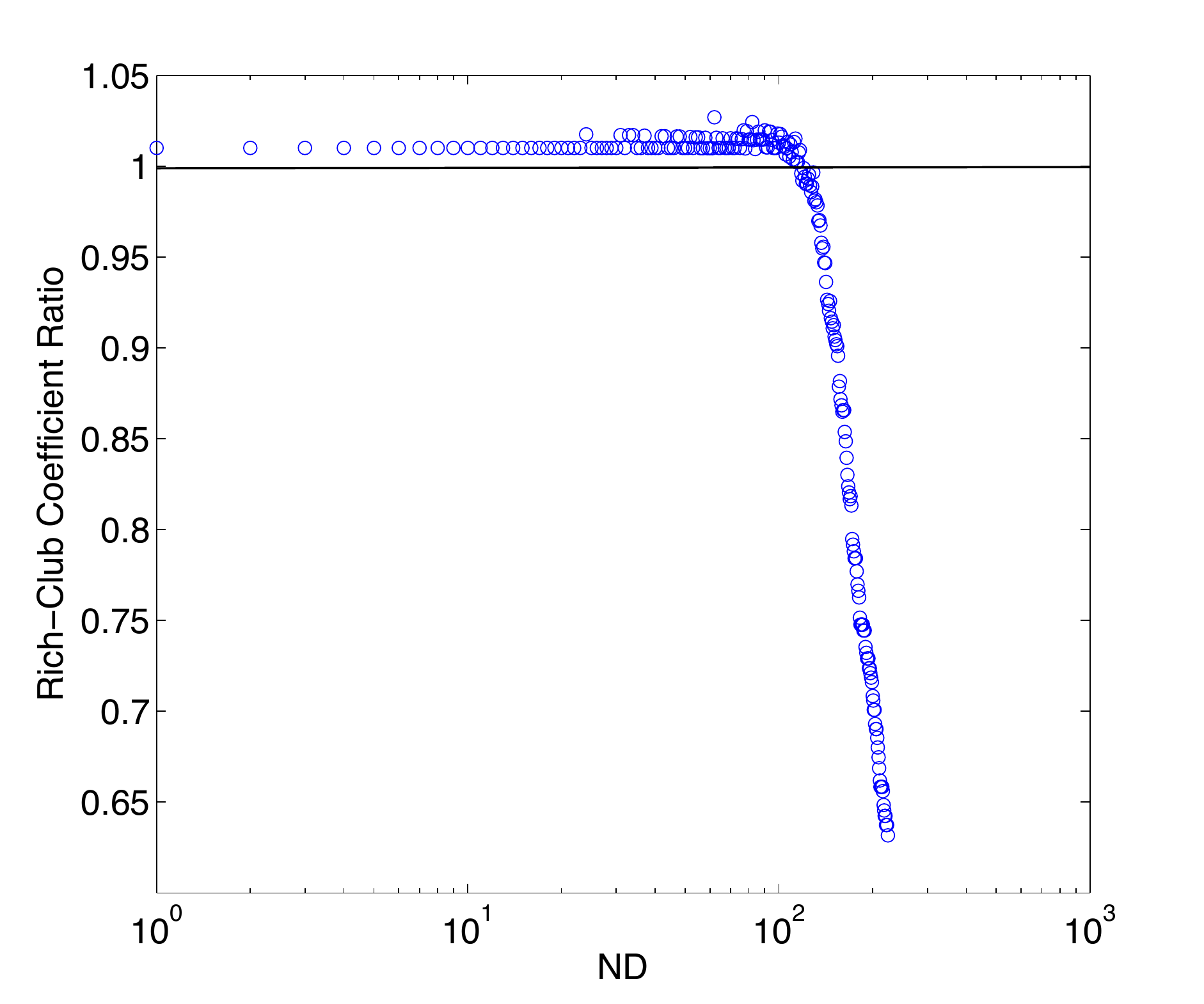}}
        \caption{Binary rich-club coefficient as a function of node degree for the undirected version of the binary IMN (year 2000). The coefficient is defined as the percentage of edges in place among the nodes having degree higher than a given node degree \cite{RichClub1}, divided by its value in random uncorrelated networks \cite{RichClub2}.}\label{fig:bin_rich_club_2000}
        \end{scriptsize}
        \end{minipage}\hfill
        \begin{minipage}[h]{7.5cm}
        \begin{scriptsize}
        \centering {\includegraphics[height=7cm,width=8cm]{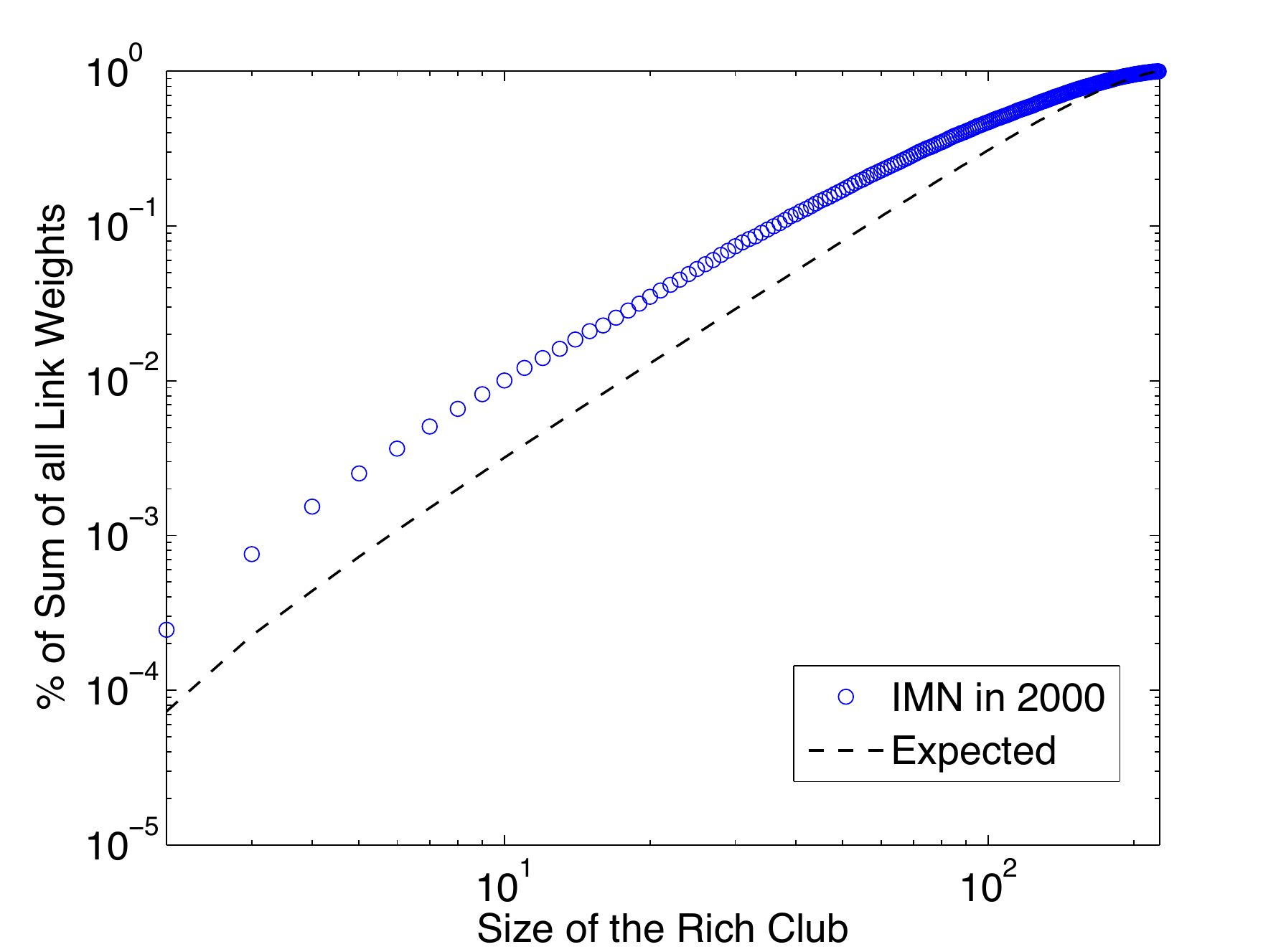}}
        \caption{Weighted rich-club coefficient \cite{Opsahl_etal_2008} as a function of node strength for the undirected version of the IMN (year 2000). We plot for year 2000 the relation between the size $M$ of the would-be rich club and the weighted rich-club ratio (WRCR), defined as the percentage of total migrants carried by the links between these $M$ countries, where countries have been sorted in a descending order according to their total strength. Expected: value of the WRCR in random networks where link weights are reshuffled over the existing binary architecture.}\label{fig:wei_rich_club_2000}
        \end{scriptsize}
        \end{minipage}
        \end{figure}

The existence of a negative correlation between connectivity and binary clustering, and the absence of such association for weighted clustering, coupled with binary and weighted disassortativity, points to an overall topological structure for the IMN where concentration of strong migration corridors between a small number of rich-club countries is quite unlikely. To explore this evidence, we have computed the binary rich-club coefficient $R^t(k)$ for the undirected version of the IMN, defined, for each time period $t$ and node degree $ND=k$, as the percentage of edges in place among the nodes having degree higher than $k$ 
. Since a monotonic relation between $k$ and $R^t(k)$ is to be expected in many networks, due to the intrinsic tendency of hubs to exhibit a larger probability of being more interconnected than low-degree nodes, $R^t(k)$ must be corrected for its version in random uncorrelated networks (see Ref. \cite{RichClub2} for details). If the resulting (corrected) rich-club index $\tilde{R}^t(k)>1$, especially for large values of $k$, then the corresponding graph will exhibit statistically-significant evidence for rich-club behavior. The binary IMN does not seem to show any clear rich-club ordering, as Figure \ref{fig:bin_rich_club_2000} shows for year 2000. This is in line with what one typically observes for the ITN \cite{Fagiolo2009pre}. Interestingly, unlike the ITN, the IMN does not exhibit any weighted rich-club structure either \cite{Opsahl_etal_2008}. To get a feel, Figure \ref{fig:wei_rich_club_2000} plots for year 2000 the relation between the size $M$ of the club and the weighted rich-club ratio (WRCR), defined as the percentage of total migrants carried by the links between these $M$ countries, where countries have been sorted in a descending order according to their total strength (our measure of ``richness'') \footnote{More precisely, let $\{i_1^t,i_2^t,...,i_N^t\}$ the labels of the $N$ nodes sorted in a descending order according to their year-$t$ NS. The WRCR for a given rich-club size $M>1$ is computed as the ratio between $\sum_{j=1}^{M}\sum_{h=1}^{M}{\tilde{m}_{i_j^t i_k^t}}$ and the sum of all entries of the weight matrix.}. We have also added the expected value of the WRCR in comparable networks where the binary structure is taken as given and weights are reshuffled uniformly at random. The plot shows that in year 2000 the WRCR curve increases very slowly with $M$, although is always larger than expected in comparable random networks (where link weights are reshuffled over the existing binary architecture). For example, to make up for 40\% of total migrants one would need a rich-club of about 90 countries (contrast this with the ITN, where 40\% of total trade is accounted for by only 10 countries).     

Interestingly, the IMN does not exhibit a core-periphery structure (CPS), either at the binary or at the weighted level. To check for CPS, we employed the node coreness concepts \cite{core_peri}
, finding weak evidence for any clear CPS, with the purported core containing a very large number of countries (about 90) and the detected periphery displaying signs of modular structure.

All this suggests that the IMN is a relatively poorly concentrated network, with a small-world structure where, especially at the weighted level, a few hubs and a lot of local structures coexist, generating high clustering and disassortativity. Motivated by this observation, we explore the community structure of the weighted IMN, using the standard community-detection Newman-Girvan modularity algorithm \cite{newman_girvan2004}
. The value of modularity $Q$ in a weighted-directed network is computed as the sum of link weights $\tilde{m}_{i,j}$ of the nodes within modules, net of their expected value $\tilde{m}^e_{i,j}$ according to a chosen null-random model. More formally:

\begin{equation}
Q = \frac{1}{V} \sum_{ij}\left[ \tilde{m}_{i,j} - \tilde{m}^e_{i,j}\right]\delta_{c_i,c_j}
\label{eq:modularity}\end{equation} 
where $V=\sum_h\sum_k{\tilde{m}_{h,k}}$ is the volume of the network and $\delta_{c_i,c_j}$ is 1 if nodes $i$ and $j$ are in the same community and 0 otherwise. In our study, we have employed the null model \cite{Arenas_etal_2007}: 
\begin{equation}
\tilde{m}^e_{i,j}=\frac{NS_i NS_j}{2V}.
\end{equation}   
This model approximately preserves node strengths, while placing links and weights at random. Modularity optimization was performed using a ``Tabu Search'' algorithm \cite{glover1998}. 

In Table \ref{tab:commstruct} we summarize our results. The IMN displays relatively high values of optimized modularity ($Q$) and features as expected quite a rich structure of clusters. The number of communities decreases across time, suggesting that globalization has made the architecture of the IMN less fragmented and modules more strongly interconnected between them. The distribution of community sizes has also shifted to the right because clusters within communities have correspondingly enlarged. 

We have also tested the stability of community partitions over time, computing the normalized mutual information (NMI) \cite{nmi} index. Given two community partitions $\mathcal{P}_A$ and $\mathcal{P}_B$, the NMI reads
\begin{equation*}
\label{eq:nmi}
\mbox{NMI}(\mathcal{P}_A,\mathcal{P}_B) =
\frac{-2\displaystyle\sum_{i=1}^{C_A}\sum_{j=1}^{C_B}N_{ij}log\left(\frac{N_{ij}N}{N_{i.}N_{.j}}\right)}{\displaystyle\sum_{i=1}^{C_A}N_{i.}log\left(\frac{N_{i.}}{N}\right)+\displaystyle\sum_{j=1}^{C_B}N_{.j}log\left(\frac{N_{.j}}{N}\right)}
\end{equation*}
where $C_A$ and $C_B$ are respectively the number of communities in
$\mathcal{P}_A$ and $\mathcal{P}_B$; $N_{ij}$ is the number of nodes in the community $i$ of the partition $\mathcal{P}_A$ that appear in the community $j$ of the partition $\mathcal{P}_B$; and finally $N_{i.} = \sum_j N_{ij}$,
$N_{.j} = \sum_i N_{ij}$ and $N = \sum_i\sum_j N_{ij}$. The NMI is equal to 1 if $\mathcal{P}_A$ and $\mathcal{P}_B$ are identical and assumes a value of 0 if the two partitions are independent. As the last column of Table \ref{tab:commstruct} shows, the community structure of the IMN has remained relatively stable until 1990, and then underwent some reorganization in the last decade of our sample, as the smaller NMI value indicates.

\begin{figure}[h]
        \begin{minipage}[h]{12cm}
        \begin{scriptsize}
        \centering {\large{{(a) Year=1960}} \vskip 0.5cm}{\includegraphics[width=12cm]{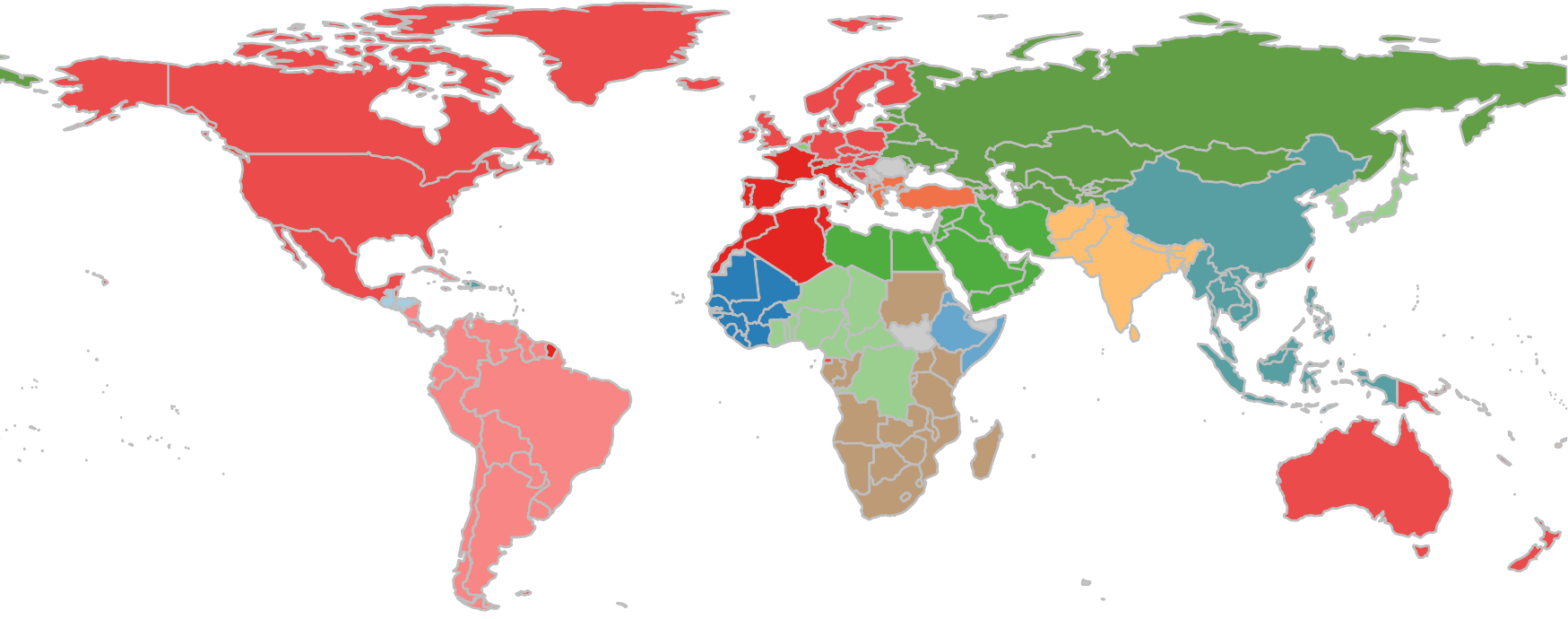}}
        \end{scriptsize}
        \end{minipage}\hfill
		\vskip 0.5cm
        \begin{minipage}[h]{12cm}
        \begin{scriptsize}
        \centering {\large{{(b) Year=2000}} \vskip 0.5cm} {\includegraphics[width=12cm]{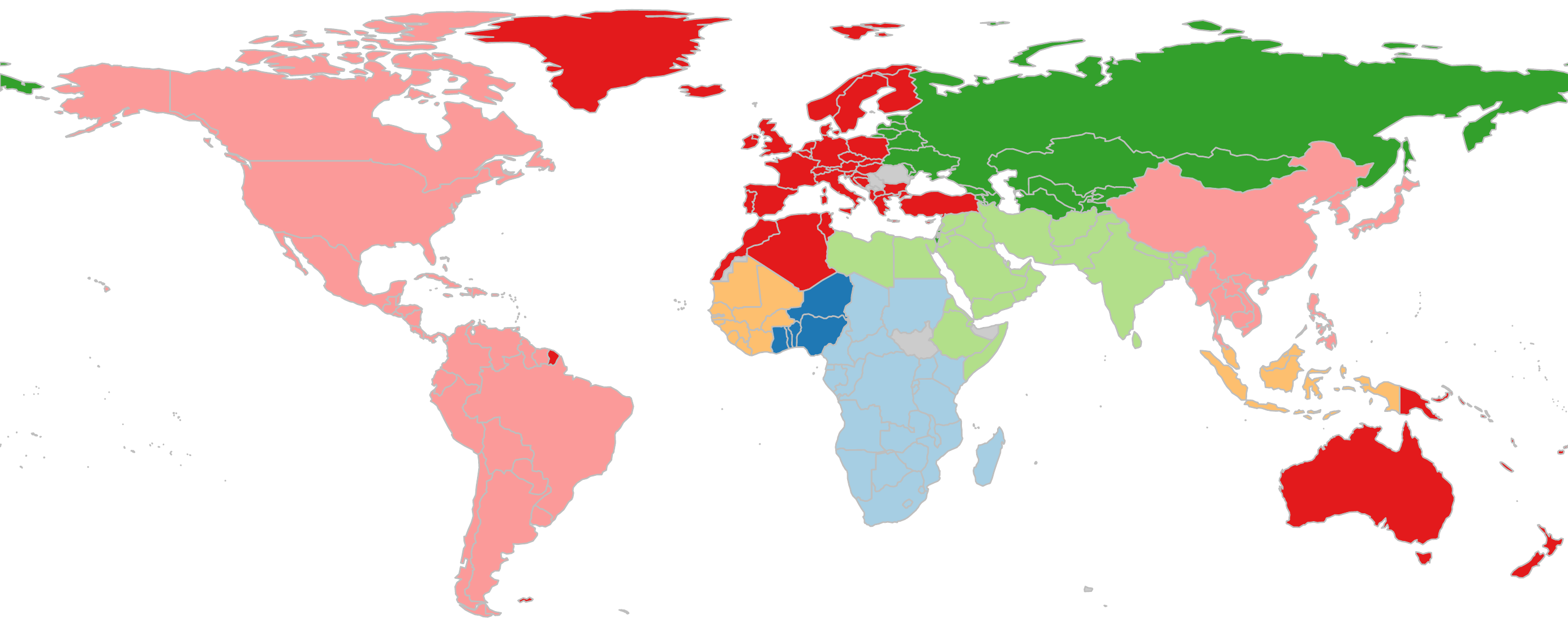}}
        \caption{Community detection in the weighted IMN. World maps show in 1960 (top) and 2000 (bottom) communities detected using Newman-Girvan modularity algorithm \cite{newman_girvan2004} using Tabu Search \cite{glover1998}. Countries belonging to same community are plotted in the same color. Grey: Not classified countries.}\label{fig:world_map_commstruct}
        \end{scriptsize}
        \end{minipage}
        \end{figure}

To better appreciate the community structure of the IMN, Figure \ref{fig:world_map_commstruct} compares world maps in 1960 and 2000 where countries are colored according to the community they belong to. In 1960, IMN community structure was much more geographically fragmented. South and north american countries belonged to different communities. Europe was also separated in two clusters (Italy, France, Spain and Portugal vs. central and eastern countries). Turkey, Greece and Rumania were also forming a separate group, as it happened for Eastern asian countries and the indian block. In 2000, there existed a unique big American cluster. Another big cluster ranging from India to Lybia and involving Gulf states was formed. European countries became integrated (including Turkey), and so did central and southern African states. Interestingly, a cluster of countries gravitating around Indonesia got formed, probably due to the increasing economic role of that country in the Asian scenario. 

\begin{figure}[h]
        \centering {\includegraphics[height=6.5cm,width=8cm]{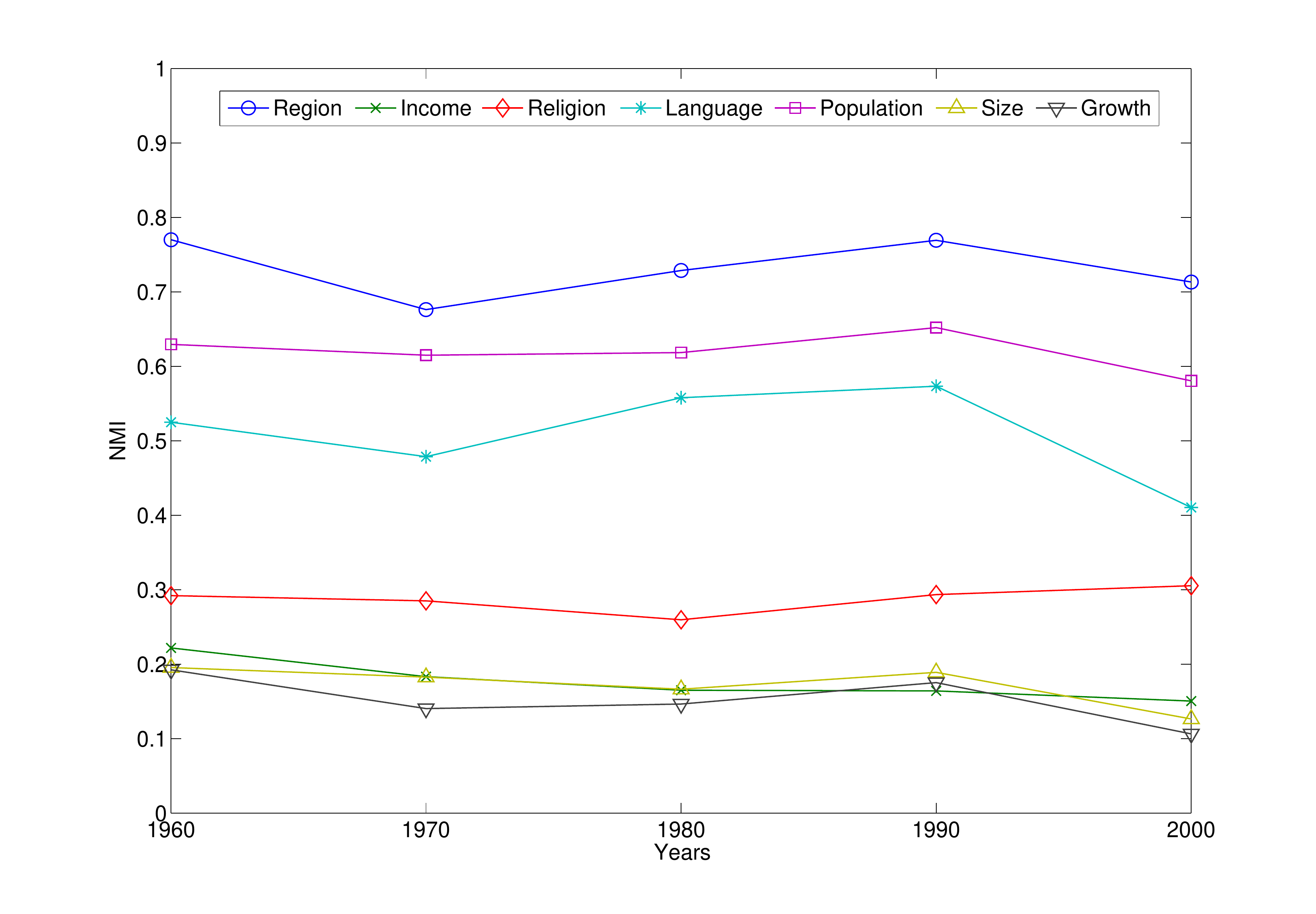}}
        \caption{Explaining IMN weighted community structure. We plot the normalized mutual information (NMI) \cite{nmi} index to compare community-structure partitions and country partitions induced by a number of explanatory variables \cite{Barigozzi_etal_2010physa}. Values close to 1 indicate that the partition induced by explanatory variable is similar to that obtained applying Newman-Girvan modularity algorithm \cite{newman_girvan2004} on IMN data. Macro Area: UN subregions according to M.49 classification (see http://unstats.un.org/unsd/default.htm). Income: World Bank Atlas Method based on  per-capita gross national income (GNI). The groups are: low income, \$1,025 or less; lower middle income, \$1,026 - \$4,035; upper middle income, \$4,036 - \$12,475; and high income, \$12,476 or more (see http://data.worldbank.org/about/country-classifications). Religion: country official religion (Cia Factbook, www.cia.go). Language: country official language (Cia Factbook, www.cia.go). Population: Country population, World Bank (http://data.worldbank.org/country). Size: Country Real Gross Domestic Product (GDP), World Bank (http://data.worldbank.org/country). Growth: Yearly GDP growth (http://data.worldbank.org/country).}\label{fig:expl_commstruct_vars}
        \end{figure}

Overall, the role of geographical constraints emerges very clearly, as expected. IMN clusters are quite geographically concentrated. Theoretical work, however, suggests that other socio-economic covariates may affect migration \cite{GastonNelson_2011}, including language, religion, income, population, economics size and growth. Following Ref. \cite{Barigozzi_etal_2010physa}, Figure \ref{fig:expl_commstruct_vars} explores this link by computing in each year the NMI index between the partition obtained via community detection in the weighted IMN, and the partitions of the set of $N=226$ countries according to geographical and socio-economic variables \footnote{See Figure \ref{fig:expl_commstruct_vars} caption for details on data sources. Country partitioning for categorical variables (region, GNI class, language) has been perfomed by assigning a class to each value of the variable. Continuous-valued variables (population, real GDP, GDP growth) have been disaggregated in decile classes.}. A larger value of NMI for a given variable hints to a better explanatory power of that variable in describing the observed partition of the set of $N$ countries in the database \cite{Barigozzi_etal_2010physa} induced by the IMN architecture. It is easy to see that geography (captured by sub-region breakdown) explains most of the variation in migration community structure. Sheer country size, captured by total population, and to a lesser extent official country language, are also able to provide a good characterization of observed IMN modules. Conversely, religion, economic size (real GDP), income (per capita gross national income, GNI) and GDP growth have instead a poor explanatory power. The association between explanatory variables and IMN community structure is also confirmed by Pearson's Chi-Squared tests of independence, which show that the partition induced by any variable considered in this study is not statistically independent on the IMN-induced partition. 

These results strengthen most of the findings obtained by standard literature on migration patterns \cite{wmr_2010,Ozden_etal_data_2011}, which mostly analyzed bilateral migration stocks as they were independent of each other. Here we show that from a more global perspective some of the most commonly discussed determinants of migration may help in accounting for the creation of country clusters.

Notice also that our results bring support to the theory of international migration systems \citep{zlotnik_empirical_1992}. Such a theory suggests that migration patterns attain some stability and structure over space and time, allowing for the identification of stable groups of countries characterized by relatively large migrant relationships between member countries, as compared to flows from outside the system. Community-structure detection can therefore help to operationalize this concept, which remained mostly qualitative since it was firstly proposed \citep{Fawcett_1989}.

Finally, the community structure of the IMN appears to be strongly different from the one found in the ITN \citep{Barigozzi_etal_2010physa}. For example, the ITN exhibits a much less fragmented and modular structure. Most of Asian and Sub-Saharian African countries form a single cluster. The same happens for countries in the former Soviet Union with east and west European countries. The only similarity concerns the large American cluster. These differences and similarities can be explained by two related factors. First, the ITN features a prominent rich-club structure, which tends to keep together many countries, especially the largest and richest ones. Second, geography plays a relevant role also in the ITN \citep{abbate_etal_2012}, which explains why European and American clusters are separated in both networks.

\subsection{Centrality}
The analysis on country connectivity has highlighted a very asymmetric role played by different nodes in the weighted IMN, with a few big players coexisting with many small ones. This is true both when considering migrant stocks or their share to country population. One can interpret node degree and strength as local measures of country centrality, which however do not consider the global embeddedness of each country in the IMN. 

We first ask whether in the IMN global centrality may convey different information as compared to local centrality. To explore this issue, we have computed on both the binary and the weighted directed IMN a battery of centrality indicators belonging to two main families: (i) path-length-based centrality measures (in- and out-closeness centrality \cite{closeness,Opsahl2010b} and betweenness centrality \cite{betweenness}); (ii) eigenvector-based centrality statistics  (Bonacich \cite{bonacich1,bonacich2}, Katz \cite{katz}, in/out PageRank \cite{pagerank}, and hub/authority centrality \cite{Kleinberg_1999,Perra_Fortunato_2008}).

The correlation analysis in Tables \ref{table:centrality_corr_2000} and \ref{table:centrality_corr_ND_NS_2000} for year 2000 shows that in the IMN path-length- and eigenvector-based  centrality indicators are highly and positively correlated. Furthermore, local centrality indicators control for a large portion of the variation of global centrality indicators across the nodes. Results are robust over the years. The reason for the overlapping between local and global centrality measures lies in the high IMN density coupled with the observed degree distribution. Indeed, if one controls for the observed degree distribution (and thus for density) fitting a Maslov-Sneppen null model to the data \cite{maslov,msz}, all observed correlation structure between global and local centrality measures are almost perfectly reproduced. Furthermore, at the binary level, some of the observed high and positive correlation (i.e. correlation between binary H/A centrality and in/out degrees) can be replicated only by fixing the density of the graph and fitting an Erdos-Renyi model to the data \cite{ErdosRenyi1960}. 

As a result, all centrality measures almost univocally indicate the most central countries in the network, either as origins of emigration flows (out-centrality measures, including hubness centrality) or as destinations of immigration flows (in-centrality measures, including authority centrality). For example, in year 2000 top destinations included the U.S., France, Canada, Germany, Australia, the Netherlands, whereas top origins featured the U.K.; the U.S., France and Germany (three big players in both rankings), India and China.         

\section{Determinants of IMN structure: Null vs. Econometric Models\label{sec:Determinants}}

In this Section we investigate whether the topological structure of the IMN, as explored in Section \ref{Sec:Topology}, can be explained by simple statistical or econometric models. We begin by fitting to the IMN null network models. Then we try to assess the extent to which a set of reasonable determinants of migrations (geographical distance, country income, language, religion, etc.), which are typically identified to be the most relevant drivers of migration stocks \cite{Ozden_etal_data_2011,wmr_2010,GastonNelson_2011}, can account for the observed structure of the IMN. As target variables to be explained, we focus in particular on node connectivity, ANND/ANNS and clustering levels, and on assortative mixing and cluster-degree/strength correlation (leaving for a subsequent study the exploration of additional topological properties, e.g. motifs \cite{motifs}).

\subsection{Null Models}\label{SubSec:NullModels}
In line with the recent literature on trade networks \cite{Squartini_etal_2011a_pre,Squartini_etal_2011b_pre}, we ask whether the observed IMN binary and weighted topology can be replicated by fixing its local structure only (degree and strength sequences) and leave everything else at random. More precisely, we employ the analytical method developed in Ref. \cite{Squartini_Garlaschelli_2011} to compute expected statistics (and confidence intervals) for higher-order network statistics (assortativity, clustering) when one constrains both the in-degree and out-degree (or in- and out-strength) to be equal to the observed ones. Expected values are then computed over the grand-canonical ensemble of all possible random graphs obeying on average the constraints. 
The method is equivalent to Maslov-Sneppen rewiring algorithm \cite{maslov,msz} in the binary case, but it does not require simulations and works for both sparse and dense networks (as in the IMN case). In the weighted case, it allows to overcome some limitations of alternative methods \cite{Fronczak_Fronczak_2012,Bargigli_Gallegati_2011,Ansmann_Lehnertz_2011}, see Ref. \cite{Squartini_etal_jeic} for a discussion. 

\begin{figure}[h]
        \begin{minipage}[h]{8.5cm}
        \begin{scriptsize}
        \centering {\includegraphics[height=5cm]{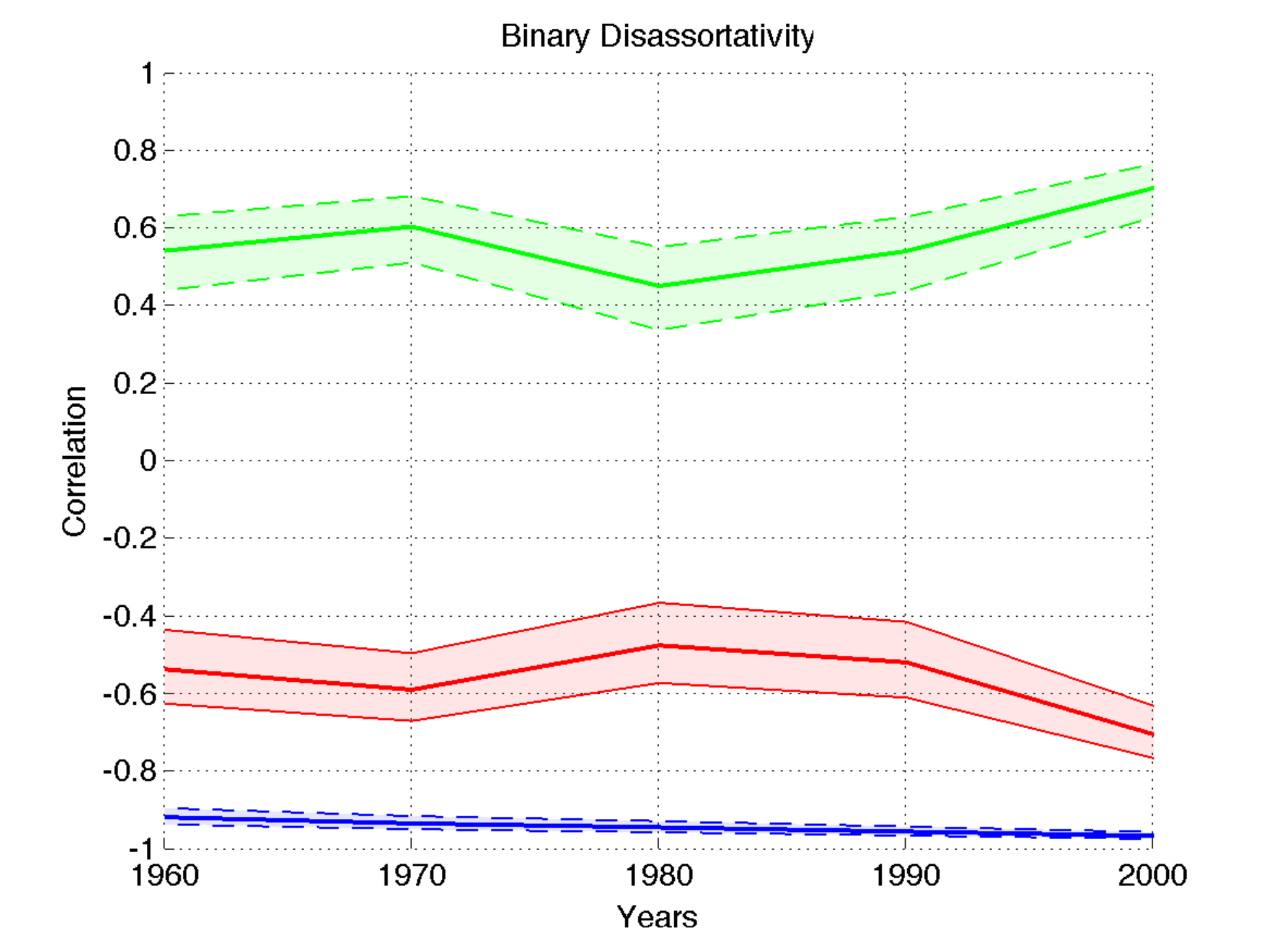}}
        \caption{Null-model analysis of the binary directed IMN. Pearsons correlation coefficients between: observed ANND vs. observed ND (red), expected ANND vs. observed ND (blue), observed vs. expected ANND (green). Expected network statistics computed fixing NDin and NDout sequences and applying the method in \cite{Squartini_Garlaschelli_2011}. 95\% confidence bands shown as shaded areas.}\label{fig:null_models_ass_bin}
        \end{scriptsize}
        \end{minipage}\hfill
        \begin{minipage}[h]{8.5cm}
        \begin{scriptsize}
        \centering {\includegraphics[height=5cm]{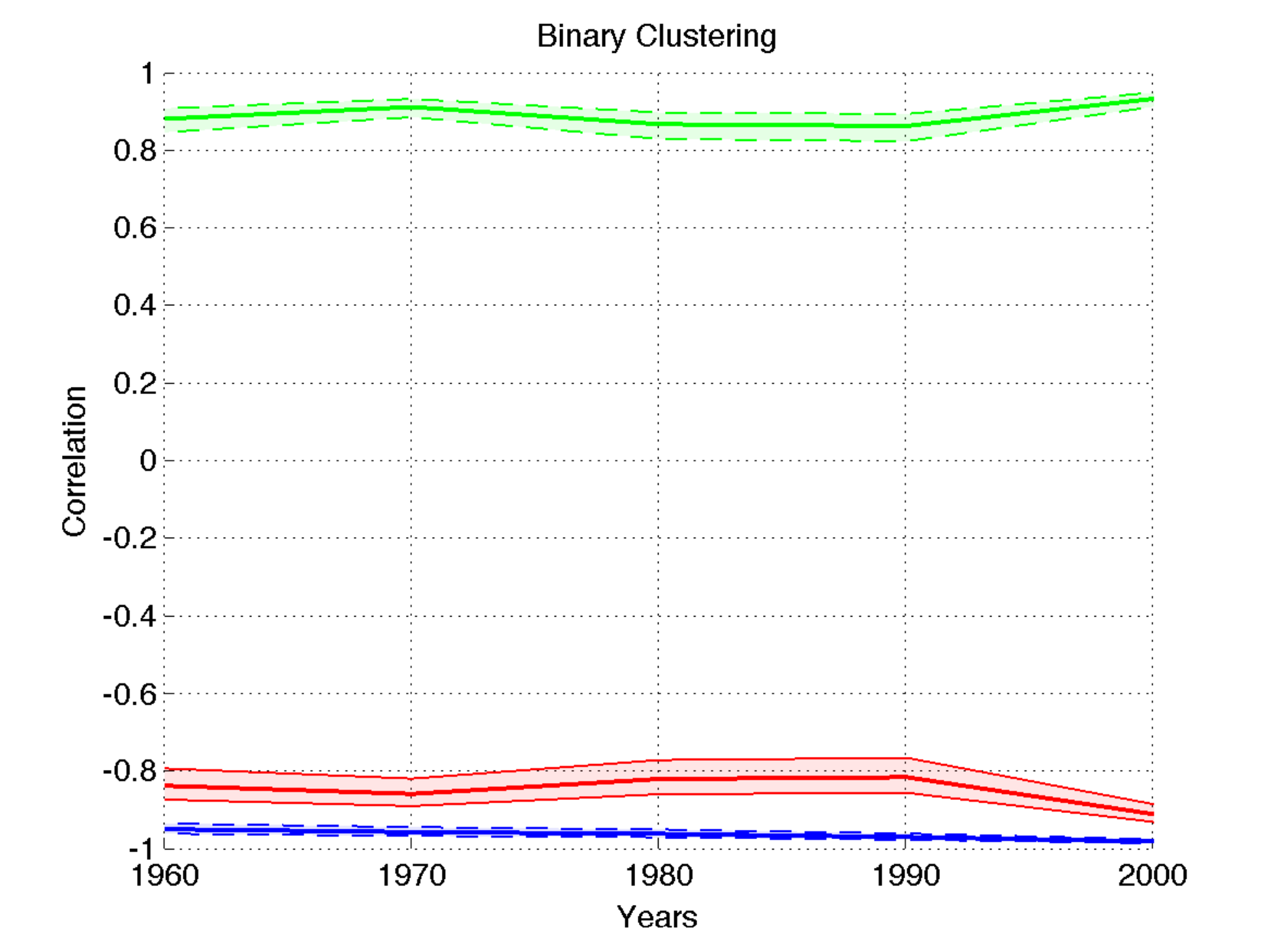}}
        \caption{Null-model analysis of the binary directed IMN. Pearson correlation coefficients between: observed BCC vs. observed ND (red), expected BCC vs. observed ND (blue), expected vs. observed BCC (green). Expected network statistics computed fixing NDin and NDout sequences and applying the method in \cite{Squartini_Garlaschelli_2011}. 95\% confidence bands shown as shaded areas.}\label{fig:null_models_clu_bin}
        \end{scriptsize}
        \end{minipage}
		\vskip 1cm
        \begin{minipage}[h]{8.5cm}
        \begin{scriptsize}
        \centering {\includegraphics[height=5cm]{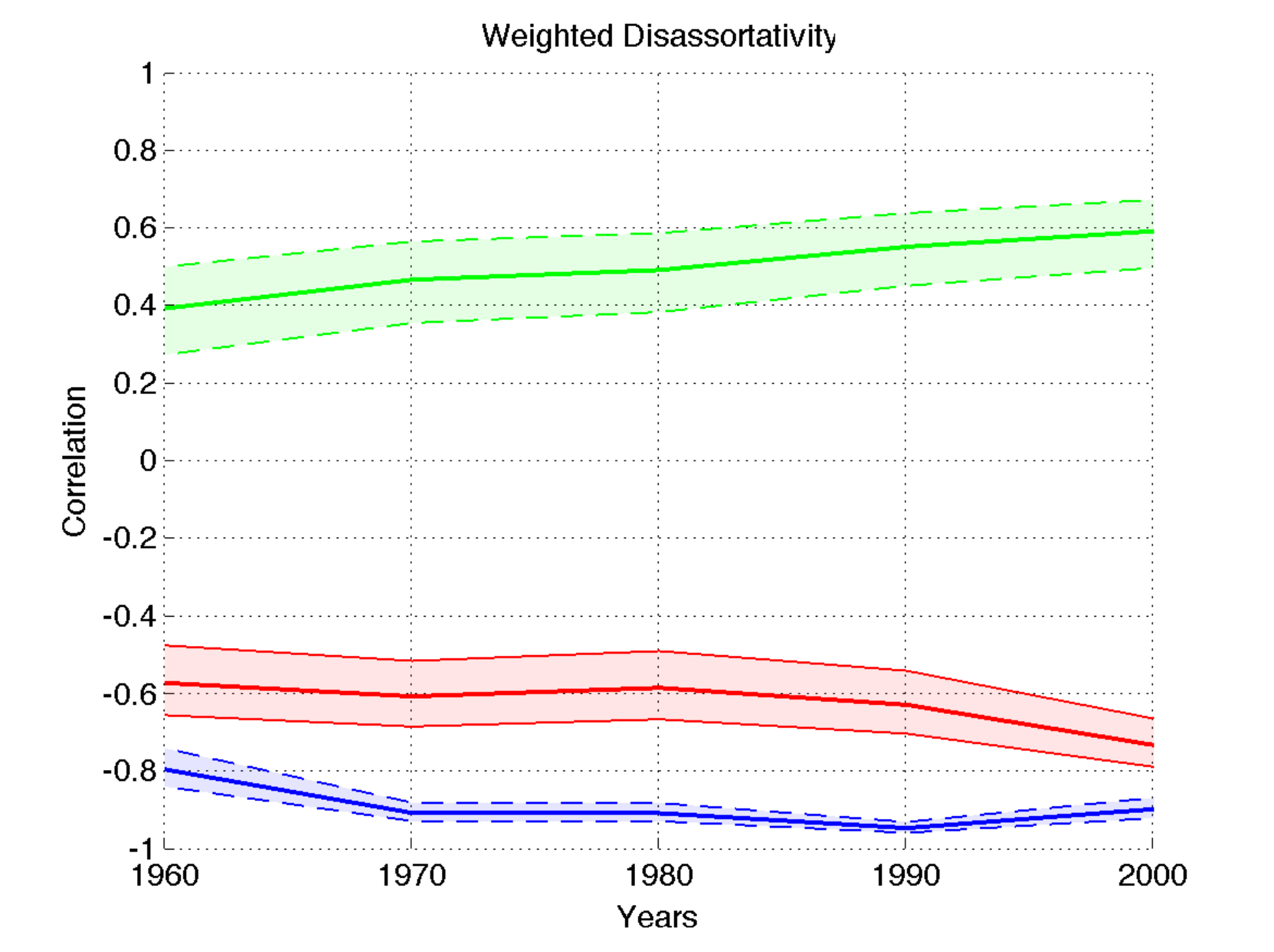}}
        \caption{Null-model analysis of the weighted directed IMN. Pearsons correlation coefficients between: observed ANNS vs. observed NS (red), expected ANNS vs. observed NS (blue), observed vs. expected ANNS (green). Expected network statistics computed fixing NSin and NSout sequences and applying the method in \cite{Squartini_Garlaschelli_2011}. 95\% confidence bands shown as shaded areas.}\label{fig:null_models_ass_wei}
        \end{scriptsize}
        \end{minipage}\hfill
        \begin{minipage}[h]{8.5cm}
        \begin{scriptsize}
        \centering {\includegraphics[height=5cm]{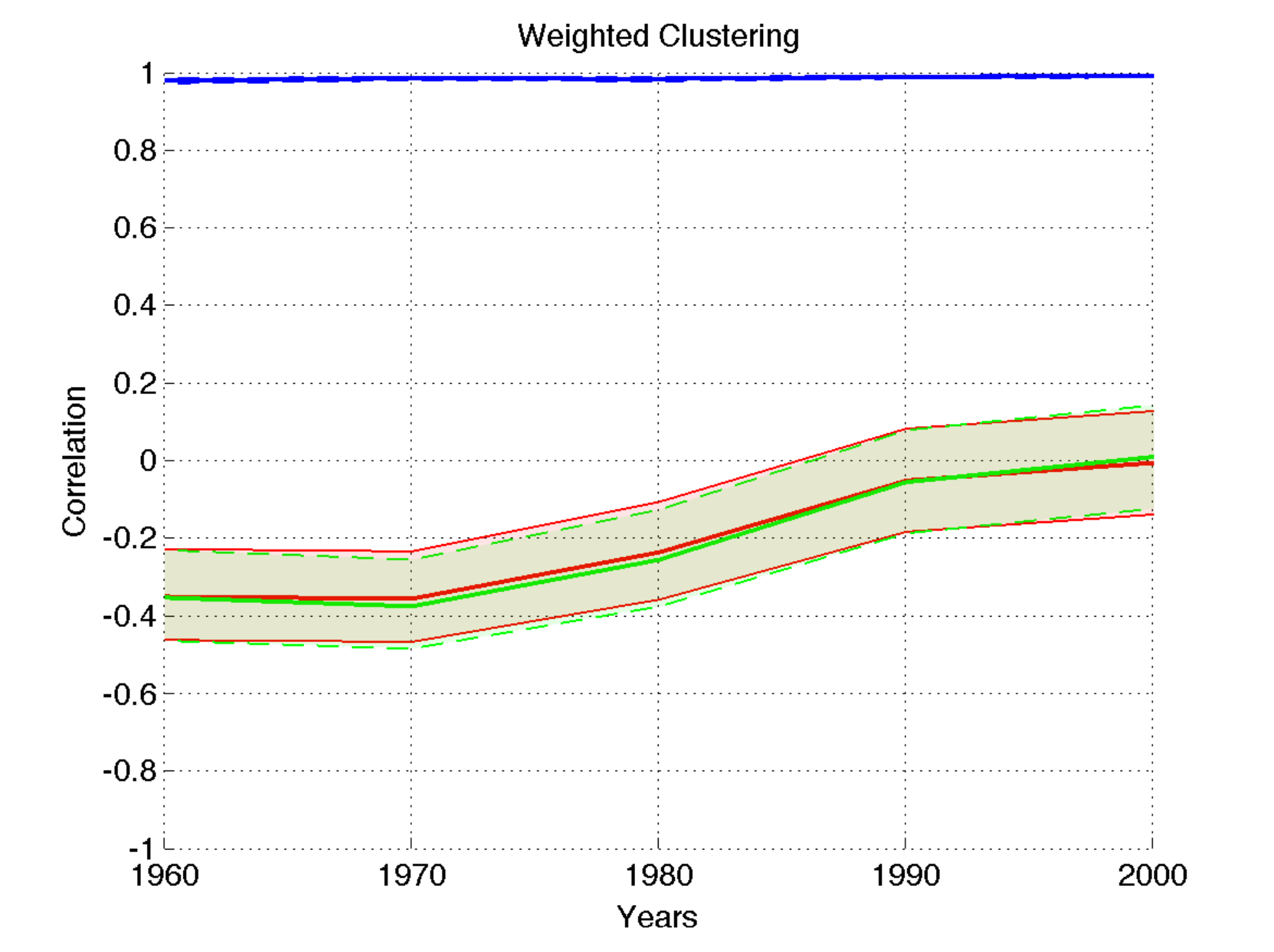}}
        \caption{Null-model analysis of the weighted directed IMN. Pearsons correlation coefficients between: observed WCC vs. observed NStot (red), expected WCC vs. observed NS (blue), expected vs. observed WCC (green). Expected network statistics computed fixing NSin and NSout sequences and applying the method in \cite{Squartini_Garlaschelli_2011}. 95\% confidence bands shown as shaded areas.}\label{fig:null_models_clu_wei}
        \end{scriptsize}
        \end{minipage}
        \end{figure}

Our main results are reported in Figures \ref{fig:null_models_ass_bin}-\ref{fig:null_models_clu_wei}, where we plot the evolution over time of correlation coefficients between expected and observed quantities in both the binary and the weighted (directed) IMN. Contrary to what happens in the ITN \cite{Squartini_etal_2011a_pre,Squartini_etal_2011b_pre,Squartini_etal_jeic}, local binary statistics (degrees) are not sufficient to control for higher-order statistics such as binary disassortativity and binary clustering. Furthermore, fixing in- and out-strength sequences is not enough to account for weighted disassortativity and clustering-strength correlation (which, as observed above, is almost always weak and not significant). The only topological feature that can be somewhat reproduced by the null model is the strong binary clustering-degree correlation, although from a statistical point of view expected correlation is significantly different from the observed one in all the years. We interpret these results in the light of the underlying complexity of IMN structure. Despite some of the global structure like node centrality can be explained in terms of local node information, local network structure is not enough to explain most of the correlation structure between higher-order statistics and topological features. This seems to suggest that other determinants may be important to explain the topological structure of the IMN. This is the point we shall explore in the next sub sections.
 
\subsection{Geographical Distance}
\begin{figure}[h]
        \centering {\includegraphics[height=9cm]{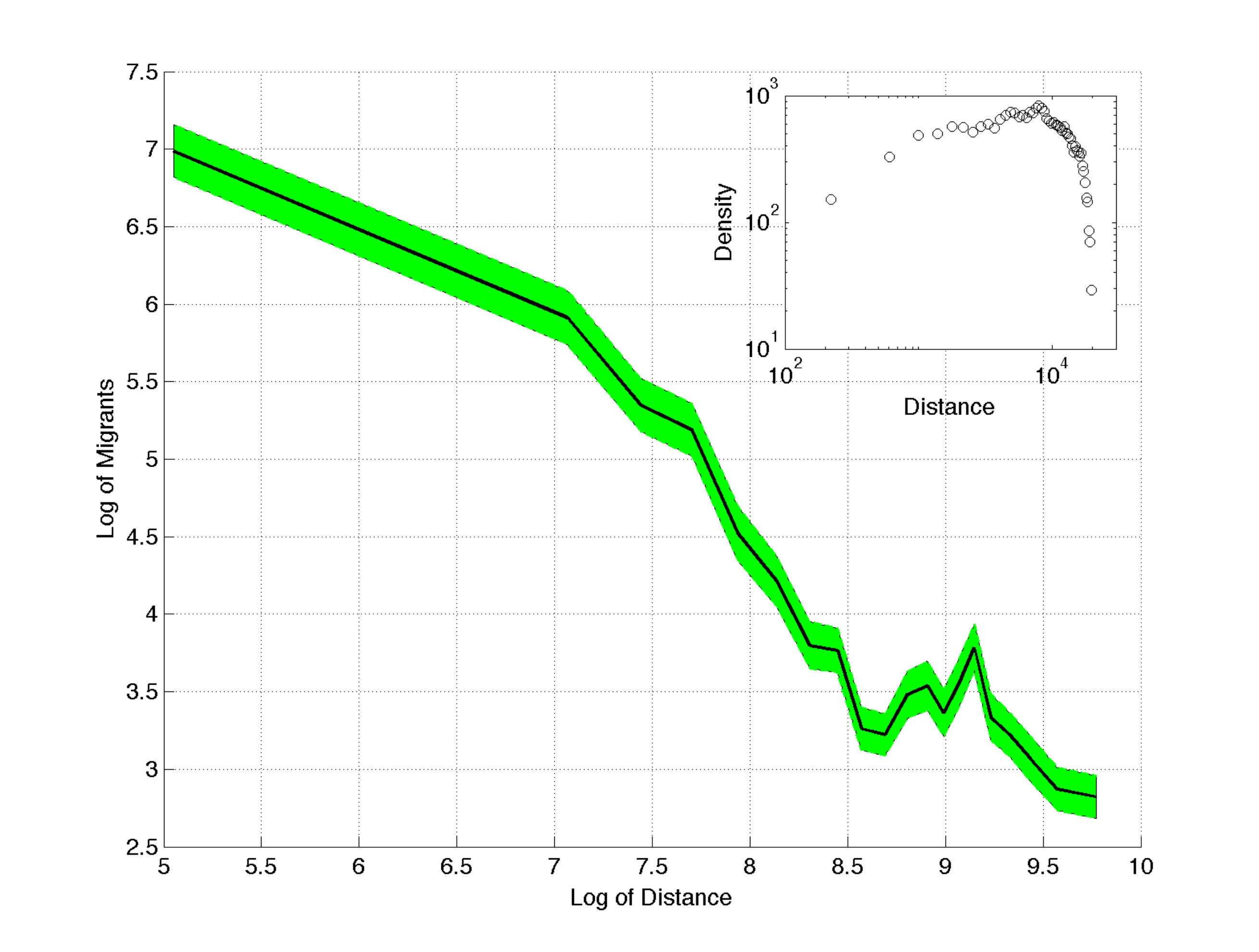}}
        \caption{Link weights in the IMN (year 2000) vs. log of geographical distance (geodist). Solid black line: conditional mean binned over the quantiles of geodist distribution. Shaded area: 95\% confidence band for the conditional mean. Inset: geodist distribution (loglog scale).}\label{fig:dist_migr_2000_inset}
        \end{figure}

We begin by asking whether geographical distance between countries ($d_{ij}$) correlates with the topological properties of the IMN. We build a $N\times N$ geo-distance matrix $D=\{d_{ij}\}$ using data from CEPII \footnote{See http://www.cepii.fr/anglaisgraph/bdd/distances.htm} complemented by our own great-circle distance calculations using geographical country coordinates \footnote{See https://www.cia.gov/library/publications/the-world-factbook/fields/2011.html}. Note that, in order to focus on geography only, we do not use time-dependent distance notions using e.g.\ data on population of major cities to compute country barycenters. We first ask whether link weights in the IMN are related to geographical distance. Figure \ref{fig:dist_migr_2000_inset} shows for year 2000 that the average number of migrants is sharply decreasing with the logs of geographical distance. This confirms a well-known stylized fact: the longest the trip, the less likely migration. This evidence is extremely robust across the years. Furthermore, the figure inset indicates that the distribution of geographical distance is rather unimodal, with tails rapidly decaying towards the bounds dictated by geographical constraints. Most of country pairs are located at intermediate distances, and very close or very far countries are less and less likely.

To understand how geographical distance affects IMN structure, we perform two simple exercises. In the first one, we calculate a measure of country remoteness as:
$$R_i=\frac{1}{N-1}\sum_{i\neq j}{\log(d_{ij})},$$  
and we correlate this measure with country-specific network binary and weighted statistics, see Table \ref{table:corr_distance_nodevars}. We see that local connectivity measures are negatively correlated with remoteness, both in the binary and weighted case, consistently across the years. This means that the more remote a country is, the less migration corridors it holds and the less migrant stocks it channels. Average neighbor connectivity is instead uncorrelated with distance. This is because, on the one hand, more remote countries tend to be less connected and possibly form links with nearby countries, which are mostly weakly connected themselves. On the other hand, this local pattern is counterbalanced by the need to find better opportunities far away, which mostly happens with very connected countries. Finally, more remote countries typically form less intensive triangles, as it is very unlikely that any two of their neighbors are strongly connected between themselves.      

\begin{figure}[h]
        \begin{minipage}[h]{8.5cm}
        \begin{scriptsize}
        \centering {\includegraphics[height=5cm]{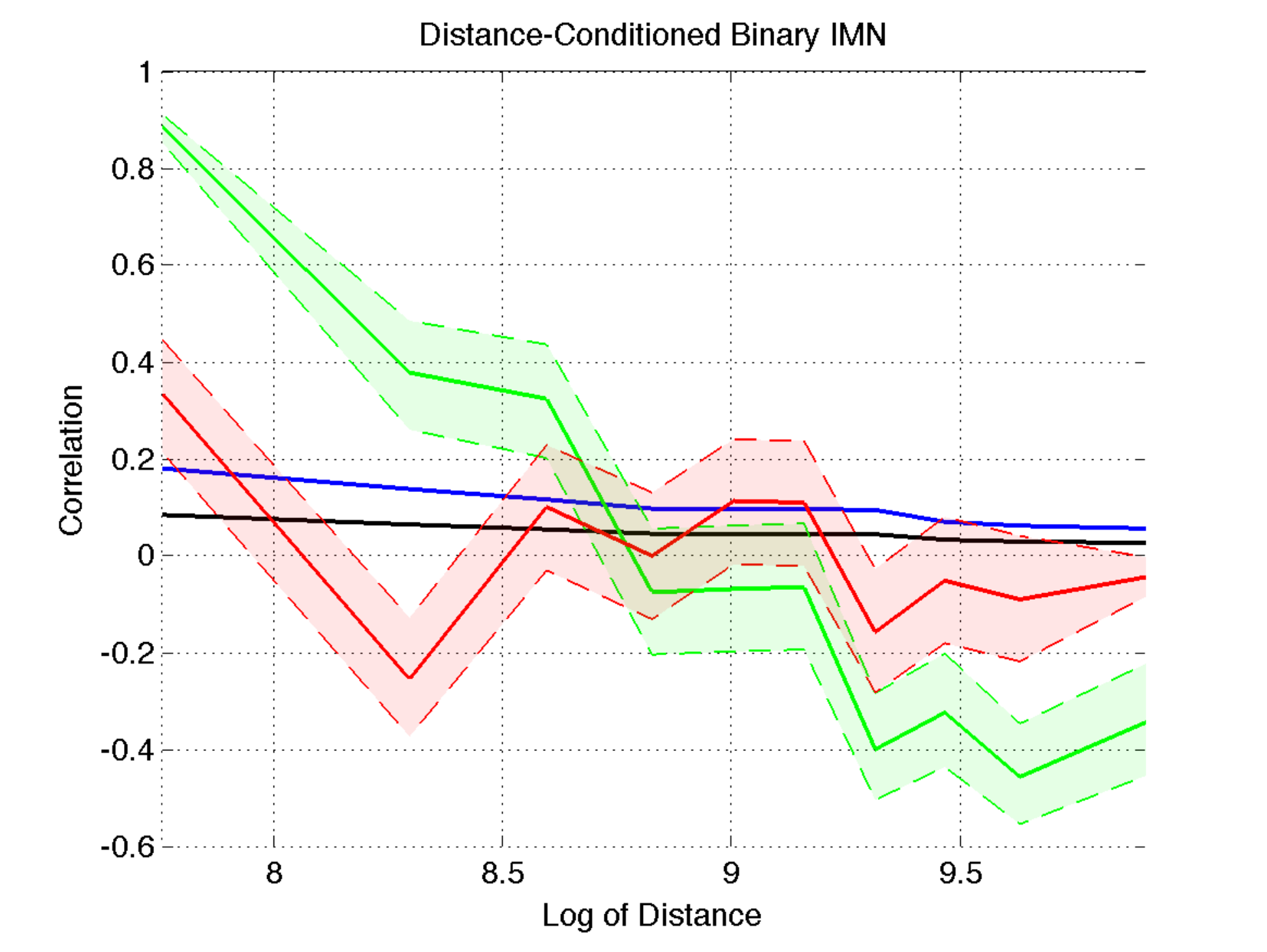}}
        \end{scriptsize}
        \end{minipage}\hfill
        \begin{minipage}[h]{8.5cm}
        \begin{scriptsize}
        \centering {\includegraphics[height=5cm]{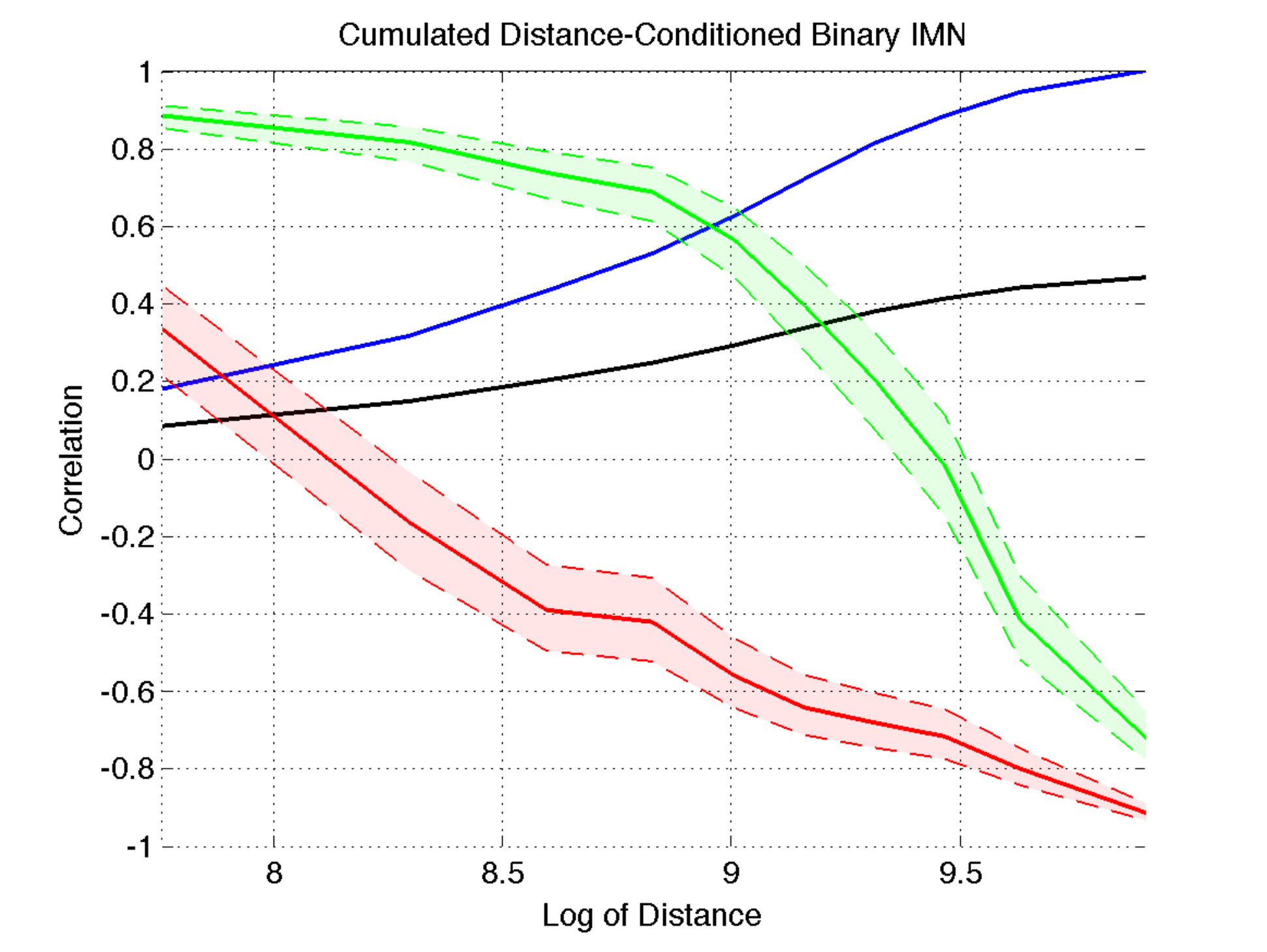}}
        \end{scriptsize}
        \end{minipage}
		\vskip 1cm
        \begin{minipage}[h]{8.5cm}
        \begin{scriptsize}
        \centering {\includegraphics[height=5cm]{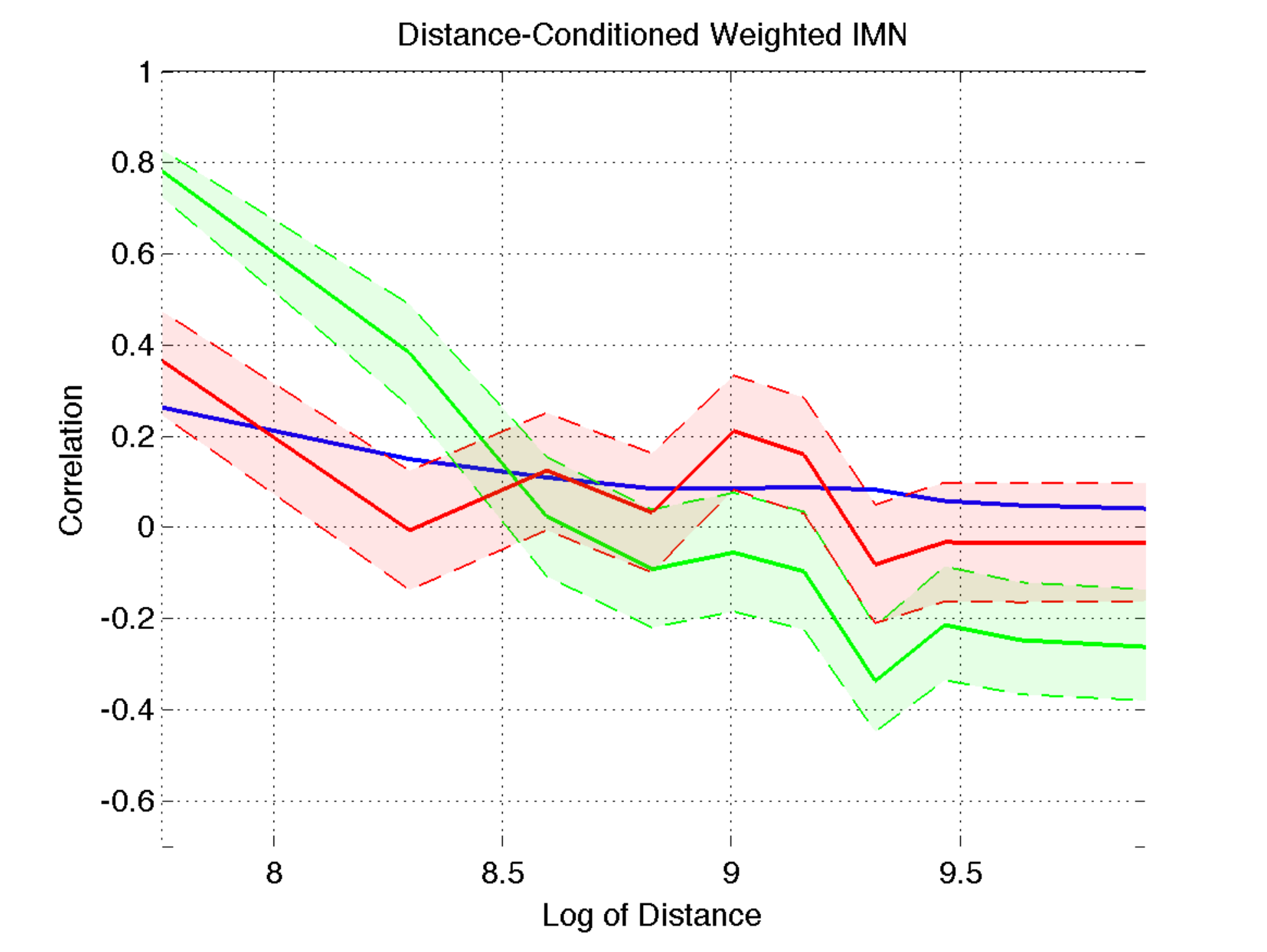}}
        \end{scriptsize}
        \end{minipage}\hfill
        \begin{minipage}[h]{8.5cm}
        \begin{scriptsize}
        \centering {\includegraphics[height=5cm]{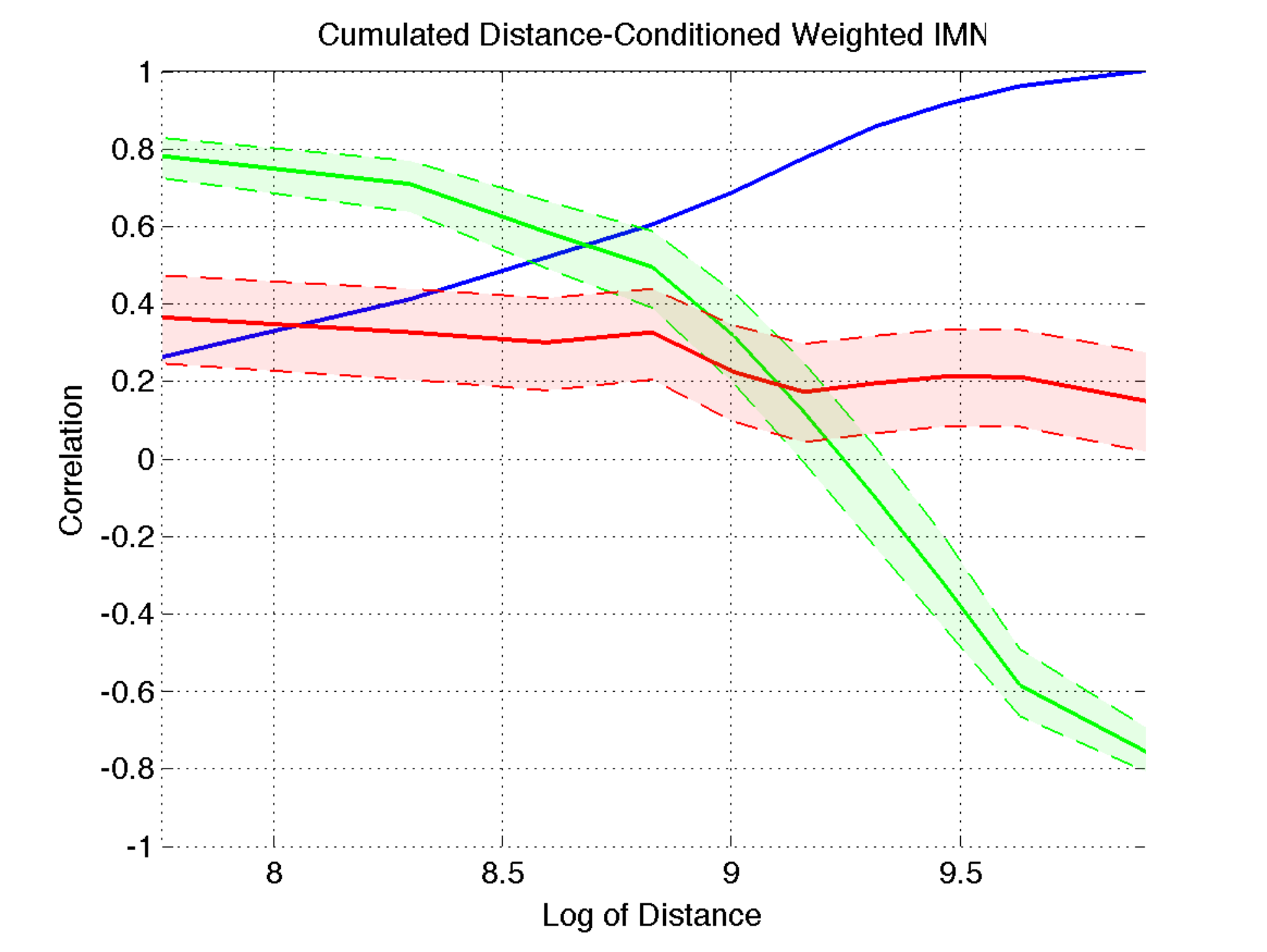}}
        \end{scriptsize}
        \end{minipage}
        \caption{Patterns of assortativity and clutering-connectivity in distance-conditioned binary and weighted IMNs (year: 2000). Left panels: distance-conditioned networks. Right panels: cumulated distance-conditioned networks. Top panels: binary networks. Bottom panels: weighted networks. Black solid line: network density. Blue solid line: in binary IMNs represents the fraction of all existing links in the IMN present in the distance-conditioned network; in weighted IMNs is the fraction of the overall network volume explained by distance-conditioned networks. Green line: correlation between ANND (ANNS) and ND (NS). Red line: correlation between BCC (WCC) and ND (NS). Green and red shaded patch: 95\% confidence intervals.}\label{fig:distcond_stats}
        \end{figure}

The second exercise aims at understanding whether assortativity and clustering-connectivity patterns depend on geographical distance \cite{abbate_etal_2012}. In each year, we partition the distribution of geographical distance into $K$ quantile classes ${\Delta_1,\dots,\Delta_{K}}$, where $\Delta_k=[\delta_{k-1},\delta_{k})$,  $k=1,\dots,K$ are quantile values and $(\delta_{0},\delta_{K})$ are the minimum and the maximum of the distance distribution. We then define, for each quantile class $\Delta_k$, the \textit{weighted} distance-conditioned IMN as the network with weight matrix $W^k$, where only the links $ij$ such that $d_{ij}\in \Delta_k$ are present, with their corresponding weight $\tilde{m}_{ij}$ from the original IMN weight matrix. Therefore, in the distance-conditioned weight matrix $W^k$, a zero represents either a missing link in the full IMN or a link connecting any two countries whose geographical distance is either $d_{ij}<\delta_{k-1}$ or $d_{ij}\geq\delta_{k}$. In other words, $W^k$ represents the IMN in place if one only considers links between countries whose geographical distance lies within the $k$-th quantile class. Given $W^k$, one can define the \textit{binary} distance-conditioned IMN as the network whose adjacency matrix $B^k$ is the binary projection of $W^k$. Finally, from binary and weighted distance-conditioned IMNs, one can define a \textit{cumulated} distance-conditioned network as the network whose weight matrix $W_c^k$ contains links between countries $ij$ whose geographical distance is $d_{ij}\leq\delta_{k}$, with $W_c^K$ and $B_c^K$ being the full weighted and binary IMN. We investigate if the magnitude of the correlations between ANND(S) and ND(S), and (W)BCC and ND(S) are invariant across space. Figure \ref{fig:distcond_stats} shows our main results for year 2000 (again, results are stable over time). To begin with, note that both network density (black line) and the fraction of links or total volume explained, slowly decrease as distance increases in distance-conditioned networks (and steadily and weakly increase in cumulated ones, see right panels). This is expected, as we already know that a larger distance implies less connectivity on average. What is instead remarkable is how assortativity and clustering-connectivity correlations change with distance. At short distances, indeed, both the binary and the weighted IMN are assortative: countries holding more corridors or larger stocks of migrants typically interact with more connected countries. This may be explained by recalling that the IMN is almost-entirely organized across tightly interconnected modules that are geographically constrained. Within these local modules, countries are assortatively mixed. Conversely, long-distance corridors are responsible of most of disassortative mixing: less-connected countries or modules far apart in space interact with more-connected countries or modules. The overall disassortative patterns is therefore attained by adding less and less assortative-mixing patterns as distance increases. A similar pattern emerges also for clustering-connectivity correlation (red curve). At the binary level, the high and negative correlation between BCC and ND mostly depend on the aggregation of very heterogeneous distance-conditioned classes. Again, at short distances, more connected countries are typically more clustered, whereas at higher distances the sign is reversed. This is not so in the weighted network, where the initially positive correlation ends up in the aggregate in a not-statistically significant relation between connectivity and clustering. 

\subsection{Country-Specific Variables and Node Network Statistics}

In this sub section we investigate whether social, macroeconomic and demographic country characteristics, which are usually found to have some impact on migration patterns in the literature \cite{Ozden_etal_data_2011}, correlate with the observed level of the most relevant node-specific network statistics. We divide our set of controls in two groups, according to the type of variable. The first group features continuous country variables: population, gross domestic product (GDP), per-capita GDP and GDP growth (yearly country growth rate of GDP). Note that whereas population is a sheer measure of country size, GDP proxies economic size. Per-capita GDP instead controls for country income. Finally, GDP growth expresses the growth potential of a country. The second group comprises country-specific  categorical variables, controlling for country sub-regional breakdown according to the M.49 classification (see http://unstats.un.org/unsd/default.htm), language \footnote{We control for the most officially-spoken languages in a binary fashion: English, Spanish, French, Arabic.}, religion, and the north-south divide \footnote{Cf. \ref{Sec:AggregatePatterns} for a discussion of the role of this divide for migrations. Cf. also the CIA Factbook, https://www.cia.gov/library/publications/the-world-factbook/. See caption of Figure \ref{fig:expl_commstruct_vars} for the sources of the other variables employed.}.

We begin by computing Pearson's correlation coefficients between continuous country variables and network statistics. Table \ref{table:corr_netvars_contvars} presents our results for year 2000 (similar findings are obtained throughout the years). Three findings stand out. First, country growth does not correlate with any network variable. This is true even if one considers long-run growth indicators instead of yearly ones, i.e. average growth between 1960 and 2000. This means that country potential cannot explain country connectivity and clustering in the IMN. Second, the larger (both demographically and economically) a country  is, and the higher its income, the more central it is (both locally and globally) in the network, but the weaker the connectivity of its partners. This is because large and rich countries are typically the target of many immigration corridors from small and poor origin countries. This explains also the negative relation of these variables with binary clustering. Note that instead weighted clustering is not correlated at all with country size and income, as more connected countries are also larger and richer, and we already saw that weighted clustering is not correlated with local connectivity (cf. Figures \ref{fig:cc_nd} and \ref{fig:cc_ns}).       

To study the impact of categorical country-specific variables, we run multivariate analysis of variance (MANOVA) \footnote{MANOVA \cite{montgomery2001design} is a testing procedure that allows one to compare multivariate means across several groups, where these groups can be associated to (categorical) factors attaining a certain number of possible states
. This permits to test the statistical significance of the mean differences across several sub-groups, using the variance-covariance between factors.}. We find that geographical areas and the north-south divide control for most of the variance of node directed connectivity both at the binary and the weighted level. Religion is instead almost uncorrelated with node statistics. Finally, only English accounts for a significant positive difference in country connectivity, especially at the weighted level. Interestingly, none of the above variables consistently affect country clustering and ANND/ANNS. 

We finally bring continuous and categorical country-specific variables together and ask whether, when considered as simultaneous covariates, they are able to explain the level of node-network statistics. We run a battery of simple ordinary-least square (OLS) regressions, where dependent variables are binary and weighted node-specific network statistics and independent regressors include the log of country population, country GDP growth, a north-south dummy controlling for developed/developing variability \footnote{We exclude from the regressions GDP and pcGDP as country size and country income are respectively controlled for by population and north-south divide.} and the already mentioned sets of sub-region, language and religion dummies. Regression results for year 2000 are reported in Table \ref{table:country_vars_regressions}. 

Firstly notice that, according to regression diagnostics, the model is well specified and attains quite a satisfactory (adjusted) $R^2$. The table of estimated coefficients (and related t-test statistic significance, reported using asterisks) confirms the strong role played by country population in explaining both incoming and outgoing migration binary and weighted links: larger countries are more and stronger connected (both as origin and target countries), typically connect with less connected countries and are less clustered at a binary level. As expected, being in the socio-economic north of the planet attracts more migrants and correlates with more inward corridors, but not necessarily entails more emigrants and more  emigration corridors, being connected with very/poorly connected countries, or being more/less clustered. Country economic prospects, as measured by GDP growth, are confirmed not to play an important role in explaining country profiles in the IMN. Similarly, country religion does not seem to affect country connectivity and higher-order network statistics. English-spoken countries seem to be correlated with higher in-ward and out-ward migrant stocks, hinting to the important role of that language as aggregator of different cultures. The crucial role of geographic breakdown is overall confirmed also by OLS regressions. Indeed, omitting sub-region dummies leads to a sharp decrease of adjusted $R^2$ by 25\% on average, compared to an average 5\% decrease due to dropping religion or language dummies. Apart from Australia and New Zealand, north America and north/western Europe, most of sub-regional dummies are more significant in explaining outward statistics than inward ones. This is expected as these regions may most likely control from emigration flows than signaling attractive countries. Finally, weighted clustering seem to be affected by all regional dummies, even if with very weak coefficients. 

\medskip

\subsection{Gravity Models and IMN Architecture}
In the foregoing analysis, we asked whether node-specific variables may explain the observed network structure at the node level. We now explore if one can explain bilateral migrant stock levels by both country-specific variables and bilateral country relationships, and use the prediction of such a model to build a predicted IMN whose topological structure successfully fits the observed one. To do that, we employ a gravity-model (GM) specification, which is traditionally used to explain bilateral migrant flows and stocks \citep{Zipf46,Zipf49,Cohen_etal_2008,Letouze_2009}. The migration-based GM modifies the original version employed to explain international-trade flows \citep{GravityBook} by positing that bilateral flows or stock of migrants from country $i$ to country $j$ in a given year depends in a gravity-like fashion on country $i$ and $j$ sizes, geographical distance, and a series of country-specific variables, dummies and bilateral interaction terms accounting for e.g. relative income, common language and religion, contiguity, etc.. More specifically, following Refs. \cite{Cohen_etal_2008,Letouze_2009}, for any given year $t$ in the sample we fit the following specification to the data (we omit time superscripts for simplicity):

\begin{equation}
m_{i,j}= P_i^{\alpha_1} P_j^{\alpha_2} d_{ij}^{\alpha_3} rY_{ij}^{\alpha_4} \exp\{ \beta_i + \beta_j + \boldsymbol{\gamma} \boldsymbol{Z}_{ij}\} \eta_{ij}, \label{gravity}
\end{equation}        
where: $P_i$ and $P_j$ are country population; $d_{ij}$ is geographical distance between $i$ and $j$; $rY_{ij}$ is relative per-capita GDP between $i$ and $j$; $\eta_{ij}$ is an error term uncorrelated with the independent variables; $\beta_i$ and $\beta_j$ are country dummies; $\boldsymbol{Z}_{ij}$ is a vector of additional explanatory variables including a contiguity dummy (CONT), equal to one if two countries share a border; a common-language dummy (COMLANG), equal to one if two countries have the same official language; a common-religion dummy (COMREL), equal to one if two countries have the same official religion; a colony dummy (COMCOL), equal to one if two countries share a colonial relationship; and three dummies controlling for south-south (SS), south-north (SN) and north-north (NN) migration (e.g. $SS=1$ if countries $i$ and $j$ are both in the south of the World).  The parameters to be estimated are therefore $\boldsymbol{\alpha}=(\alpha_1,\dots,\alpha_4)$, $(\beta_i,\beta_j)$ and $\boldsymbol{\gamma}=(\gamma_1,\dots,\gamma_7)$.   

Estimation of Eq. \eqref{gravity} is not straightforward. Since $m_{i,j}$ is zero when two countries are not linked in the IMN, one cannot simply log-linearize Eq. \eqref{gravity} and apply standard ordinary-least square (OLS) techniques. Doing that would indeed imply to discard zero observations, which from a network perspective means keeping the binary structure as given \citep{Duenas_Fagiolo_2011}. Therefore, we would employ here a two-step procedure known as zero-inflated (ZI) estimation \citep{Burger2009,Linders2006}
. In a nutshell, ZI regressions require to first estimate the probability that a link exists, i.e. $p_{ij}=Prob\{m_{i,j}>0\}=Prob\{a_{i,j}=1\}$. This is done by fitting a logit model where the right-hand side is the same of Eq.\eqref{gravity}. This generates a predicted-probability matrix $\hat{P}=\{\hat{p}_{ij}\}$, where each entry is the predicted probability that a link exists in the IMN. Given $\hat{P}$, one can simulate many instances of the predicted binary IMN $\hat{A}^h=\{\hat{a}_{ij}^h\}$, where each entry $\hat{a}_{ij}^h$ is drawn (independently across all $i$ and $j$) from a Bernoulli r.v. with parameter $\hat{p}_{ij}$, and $h=1,\dots,H$ are Montecarlo replications. In the second step, for each simulated binary matrix $\hat{A}^h$, one estimates the weights to be assigned to the links that have been predicted to exist in matrix $\hat{A}^h$. As customary in this literature, we employ a full-sample Poisson pseudo-maximum likelihood estimation (PPML) \citep{Santos2006} to perform such task (see Table \ref{Table:Gravity}). This allows us to obtain, for each binary-matrix simulation, a predicted migrant-stock matrix $\hat{M}^h=\{\hat{m}_{ij}^h\}$, where each $\hat{m}_{ij}^h$ is equal to the corresponding PPML estimate when $\hat{a}_{ij}^h>0$ and zero otherwise. Notice that by construction the binary projection of $\hat{M}^h$ is equal to $\hat{A}^h$, which however may be different from the observed binary IMN matrix $A$.

Given the sets of predicted matrices $(\hat{A}^h,\hat{M}^h)$, $h=1,\dots,H$, we perform two related exercises. First, we ask whether on average the predicted-binary matrices are similar to the observed one ($A$), in terms of both the percentage of successfully-predicted zeros and ones and the topological structure of the binary IMN. Second, we investigate whether the weighted architecture of the IMN can be successfully predicted by the gravity model. For space reasons, we present results for year 2000 only (similar results hold in all the other years in the sample). To begin with, notice that the ZI Poisson model (ZIP) performs very well in fitting the data (cf. Table \ref{Table:Gravity} for estimated coefficients and diagnostics when one employs all observations --- positive and zero). The model is well-specified according to Wald tests. Poisson estimations (with constant) reach very high pseudo $R^2$ levels (0.6477), in line with migration-based GM exercises in the literature. The signs of estimated coefficients in the second step are in line with expectations. In particular, migrant stocks increase the higher source and receiver country population, and the lower distance and origin-destination relative income. In addition, sharing a common language or religion, a colonial relation, and a common border enhances migration. Also, if the origin country is in the south and the destination is either in the north or in the south is beneficial for migration. Positive but not significant effect is instead found if both countries are in the north. Finally, in line with most of the existing literature, one does not find any asymmetric (origin-destination) population effect.                 

\begin{figure}[h]
        \begin{minipage}[h]{8.5cm}
        \begin{scriptsize}
        \centering {\large{{(a)}}}
        \centering {\includegraphics[height=6cm]{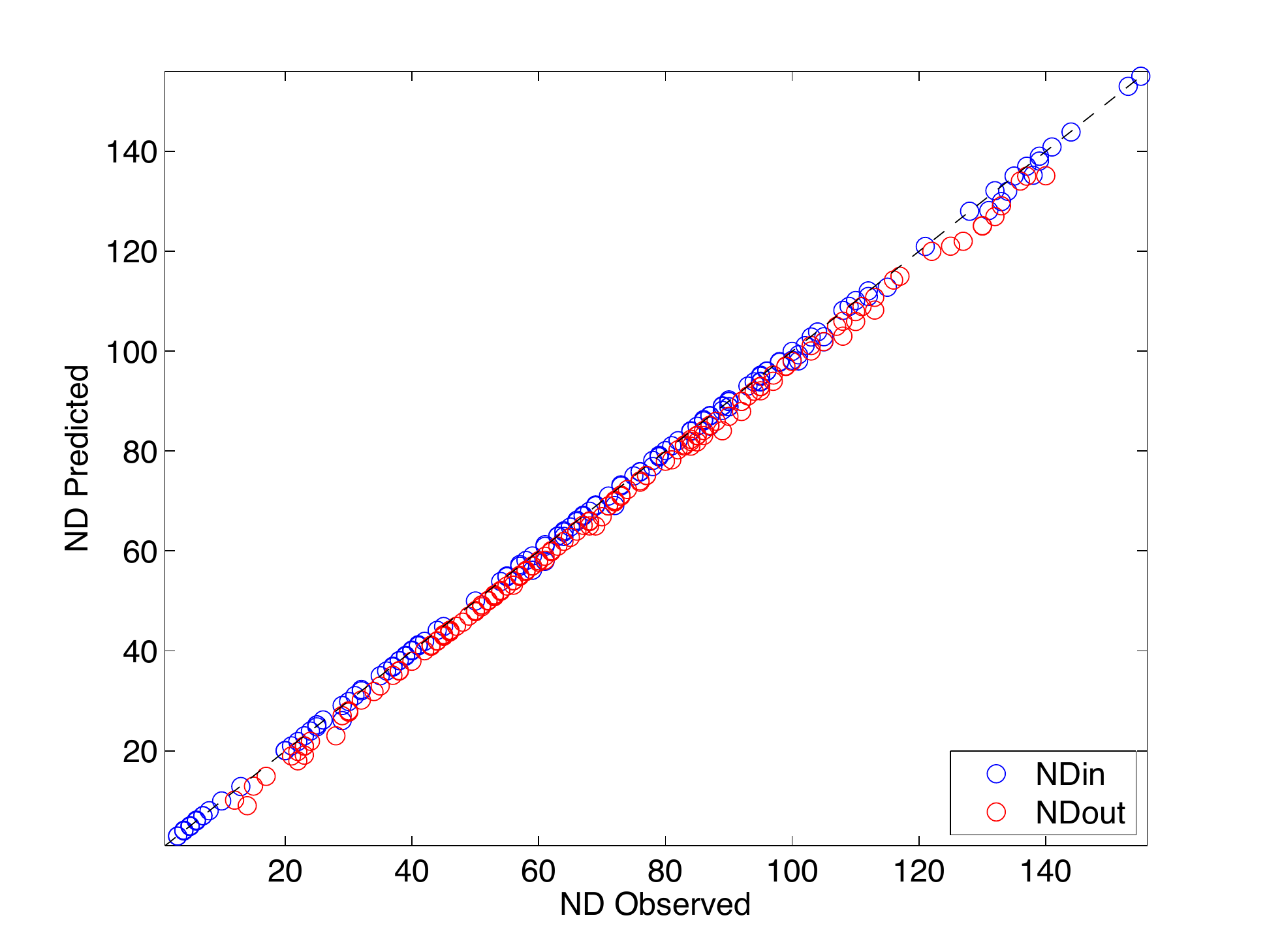}}
        \end{scriptsize}
        \end{minipage}\hfill
        \begin{minipage}[h]{8.5cm}
        \begin{scriptsize}
        \centering {\large{{(b)}}}
        \centering {\includegraphics[height=6cm]{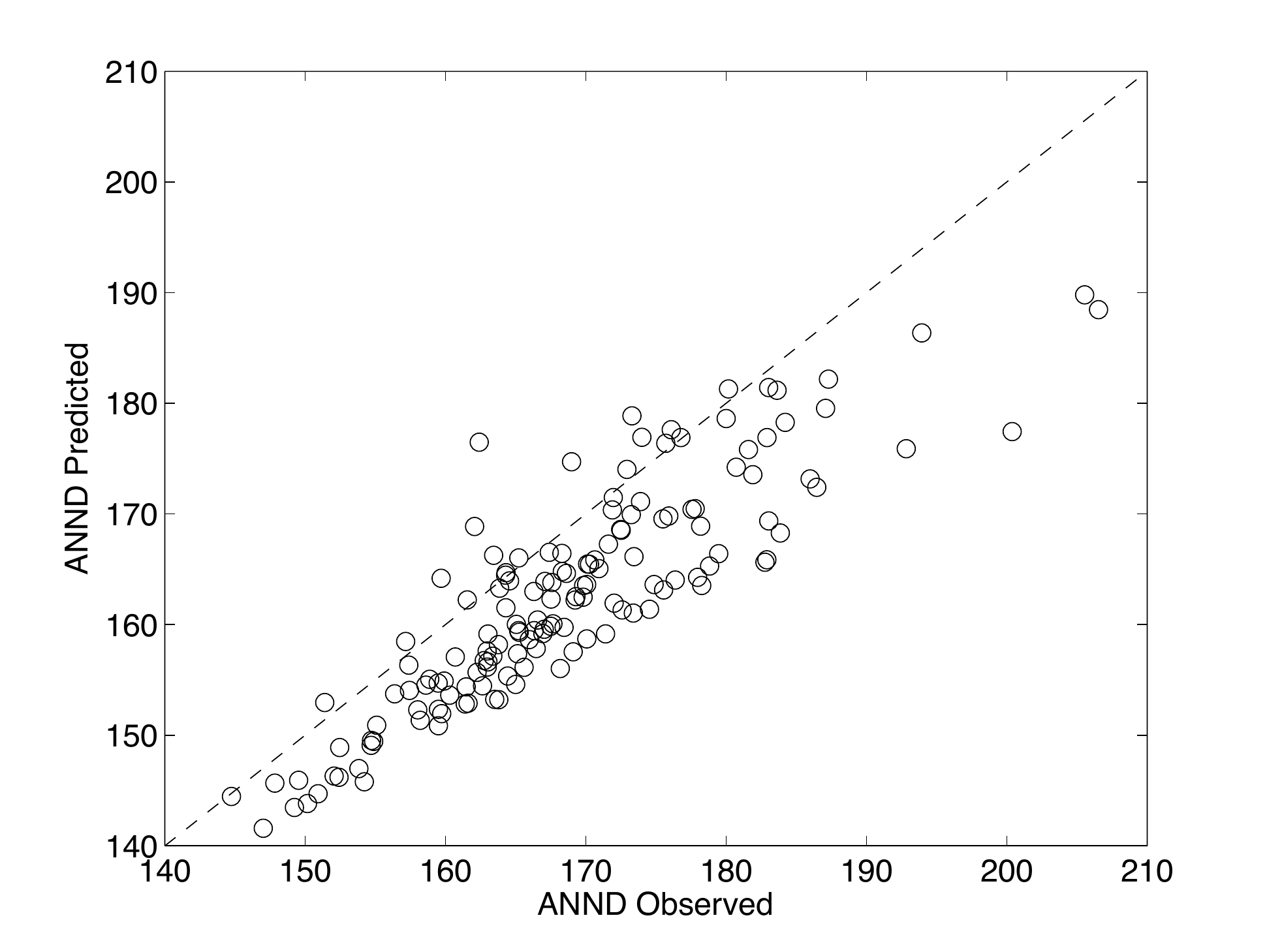}}
        \end{scriptsize}
        \end{minipage}
		\vskip 1cm
        \begin{minipage}[h]{8.5cm}
        \begin{scriptsize}
        \centering {\large{{(c)}}}
        \centering {\includegraphics[height=6cm]{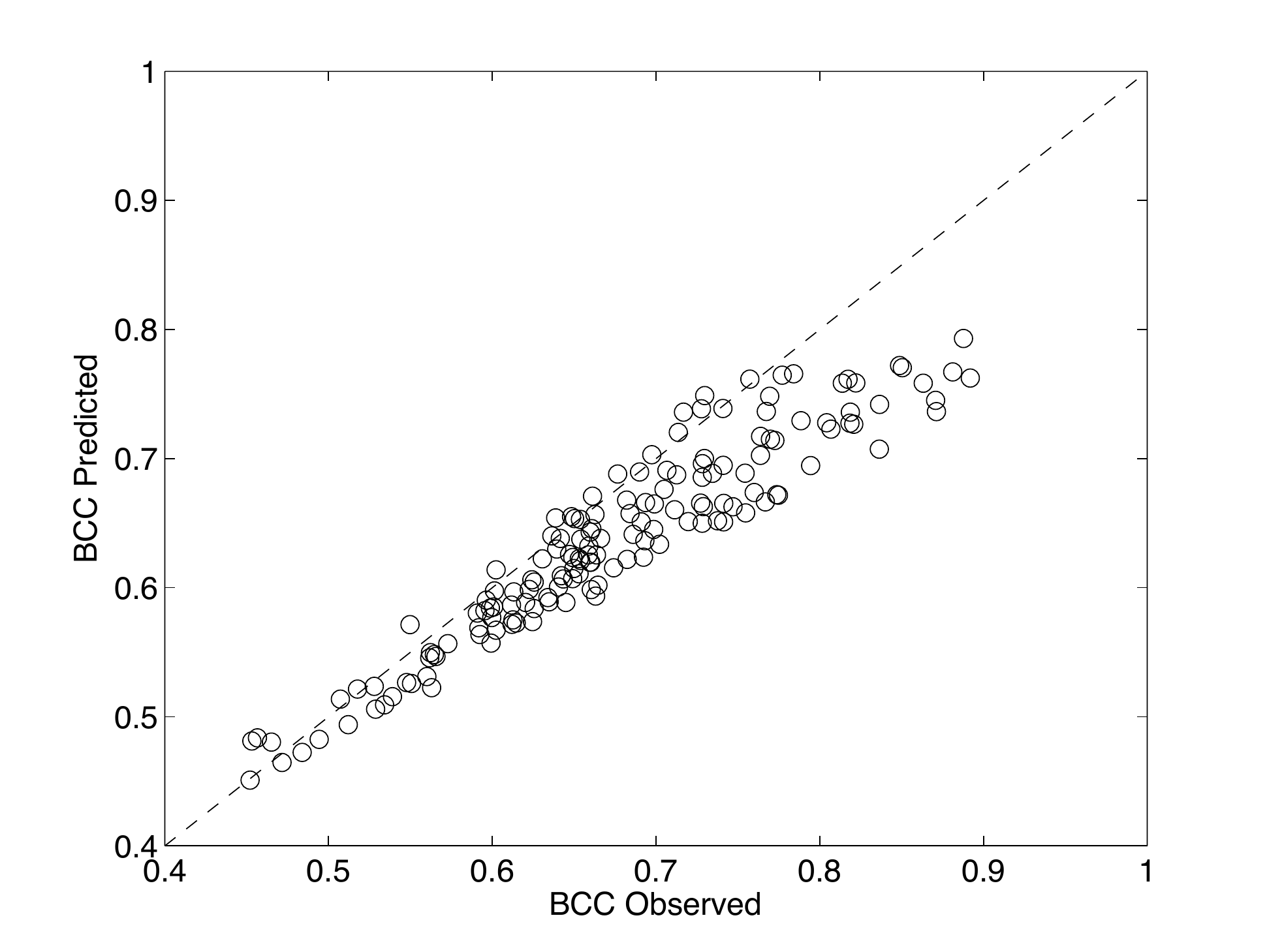}}
        \end{scriptsize}
        \end{minipage}\hfill
        \begin{minipage}[h]{8.5cm}
        \begin{scriptsize}
        \centering {\large{{(d)}}}
        \centering {\includegraphics[height=6cm]{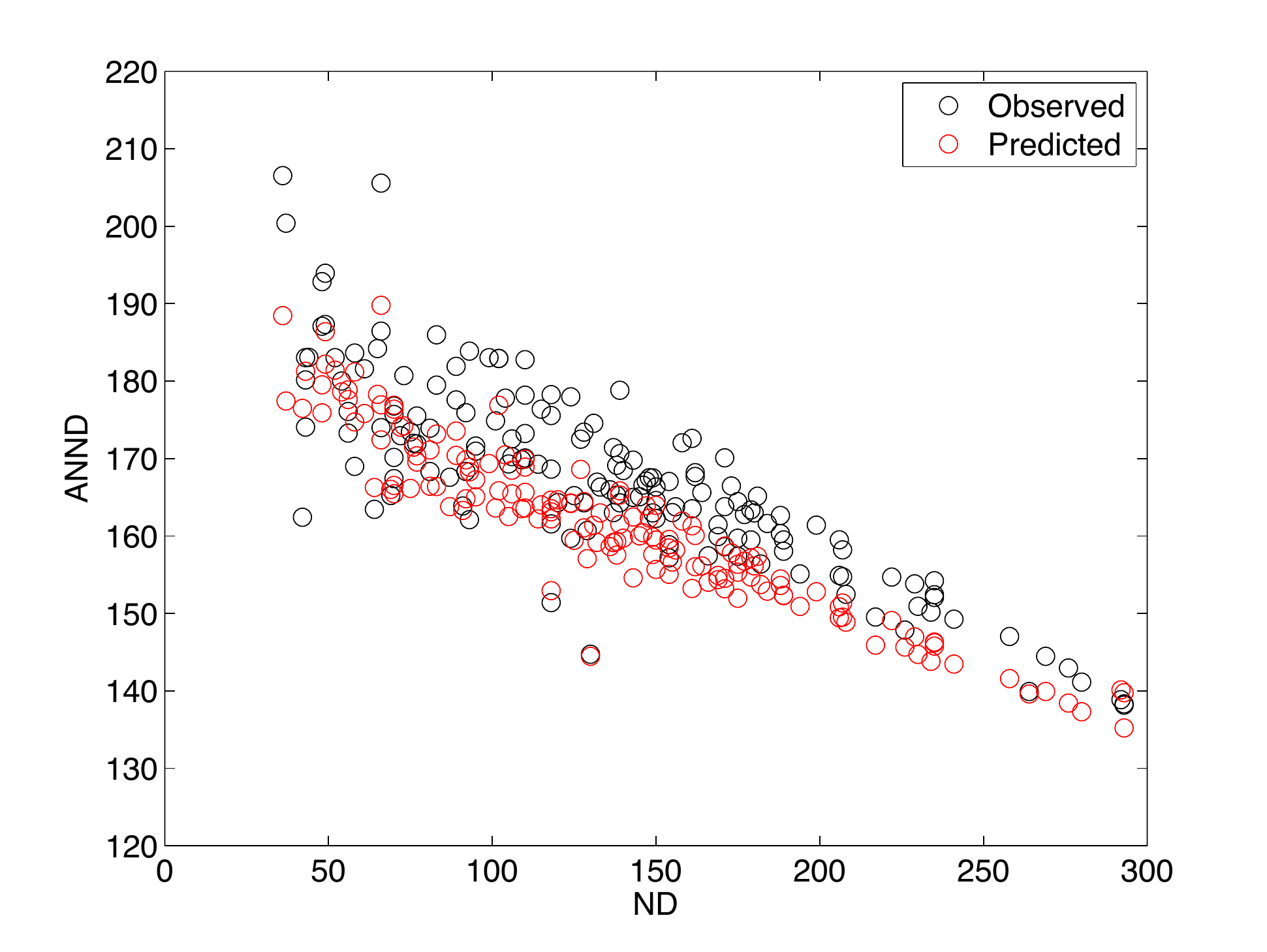}}
        \end{scriptsize}
        \end{minipage}
		\vskip 1cm
        \begin{minipage}[h]{8.5cm}
        \begin{scriptsize}
        \centering {\large{{(e)}}}
        \centering {\includegraphics[height=6cm]{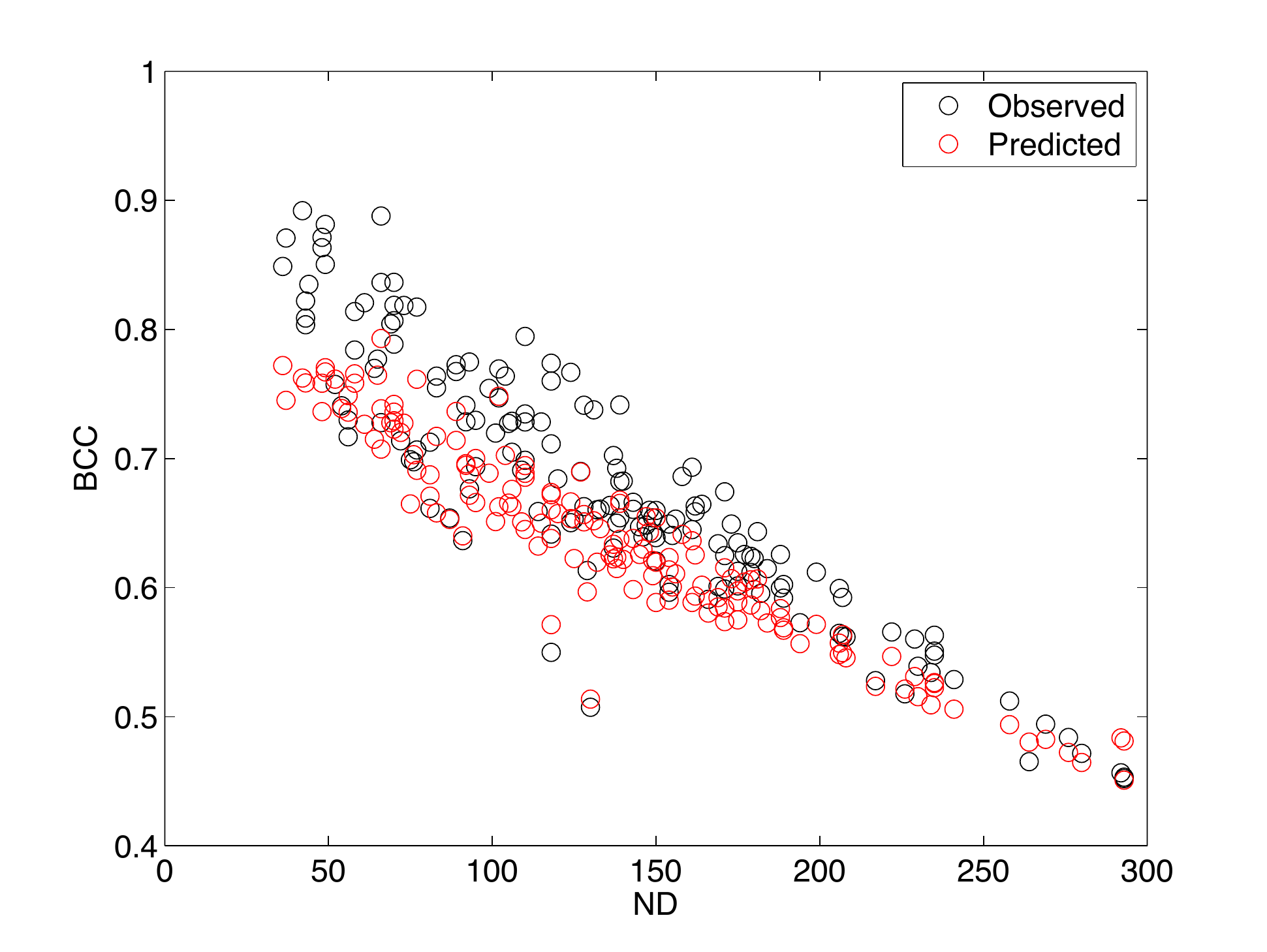}}
        \end{scriptsize}
        \end{minipage}
        \caption{Gravity model estimation of the binary structure of the IMN. (a) Observed vs. predicted node in-degree (NDin) and out-degree (NDout); (b) Observed vs. predicted average nearest-neighbor degree (ANND); (c) Observed vs. predicted binary clusterinc coefficient (BCC); (d) Observed vs. predicted correlation between ANND and total node degree (ND); (e) Observed vs. predicted correlation between binary clustering coefficient (BCC) and total node degree (ND). Each dot corresponds to the average across H=1000 simulated binary structures using gravity-based logit predicted link probabilities.}\label{fig:grav_binary}
        \end{figure}

Our first result is that the gravity model provides a good explanation of the binary structure of the IMN. The logit specification is able on average to correctly predict the presence and the absence of a link in, respectively, 73.4\% and 82.3\% of cases, with an overall success ratio of 78.7\%. This implies that on average the gravity model correctly reproduces both in-degree and out-degree observed IMN sequences, cf Figure \ref{fig:grav_binary}(a). Both observed ANND and BCC are also well reproduced, albeit for large observed ANND and BCC values some underestimation occurs -- cf. panels (c) and (d). Nevertheless, observed disassortativity and BCC-ND correlations are quite-satisfactorily matched by the model, see panels (d) and (e). This result is somewhat surprising if one compares it with similar findings in the ITN \citep{Duenas_Fagiolo_2011}, where instead a gravity specification mostly fails in reproducing the binary structure, while the knowledge of the local binary architecture is sufficient to reproduce high-order topology \citep{Squartini_etal_jeic}. In the IMN, on the contrary, local topology is not enough to replicate higher-order structure (see sub-section \ref{SubSec:NullModels}), whereas country-specific and bilateral country-interaction effects are able to account for most of the topological (binary) structure of migration corridors.

\begin{figure}[h]
        \begin{minipage}[h]{8.5cm}
        \begin{scriptsize}
        \centering {\large{{(a)}}}
        \centering {\includegraphics[height=6cm]{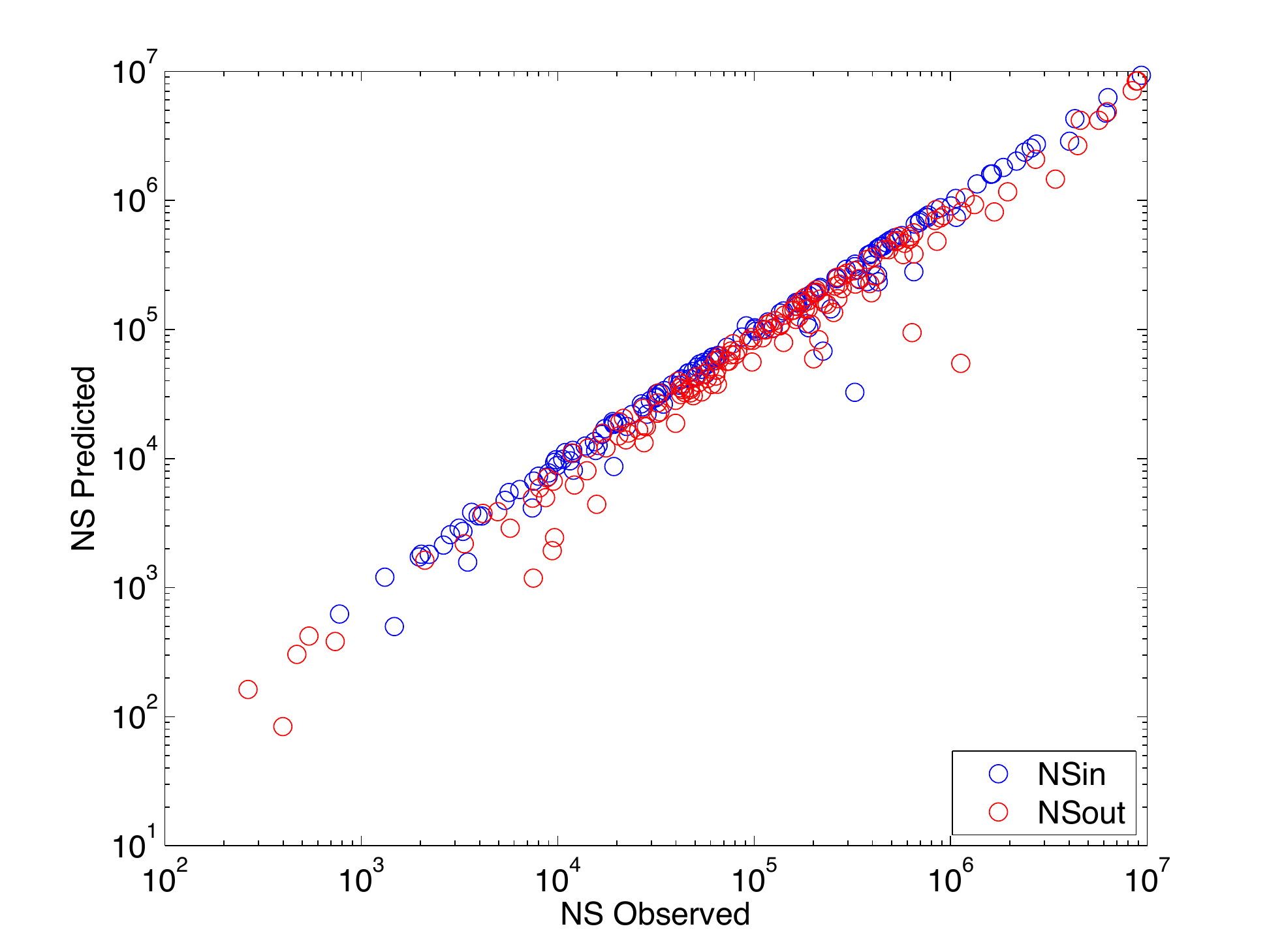}}
        \end{scriptsize}
        \end{minipage}\hfill
        \begin{minipage}[h]{8.5cm}
        \begin{scriptsize}
        \centering {\large{{(b)}}}
        \centering {\includegraphics[height=6cm]{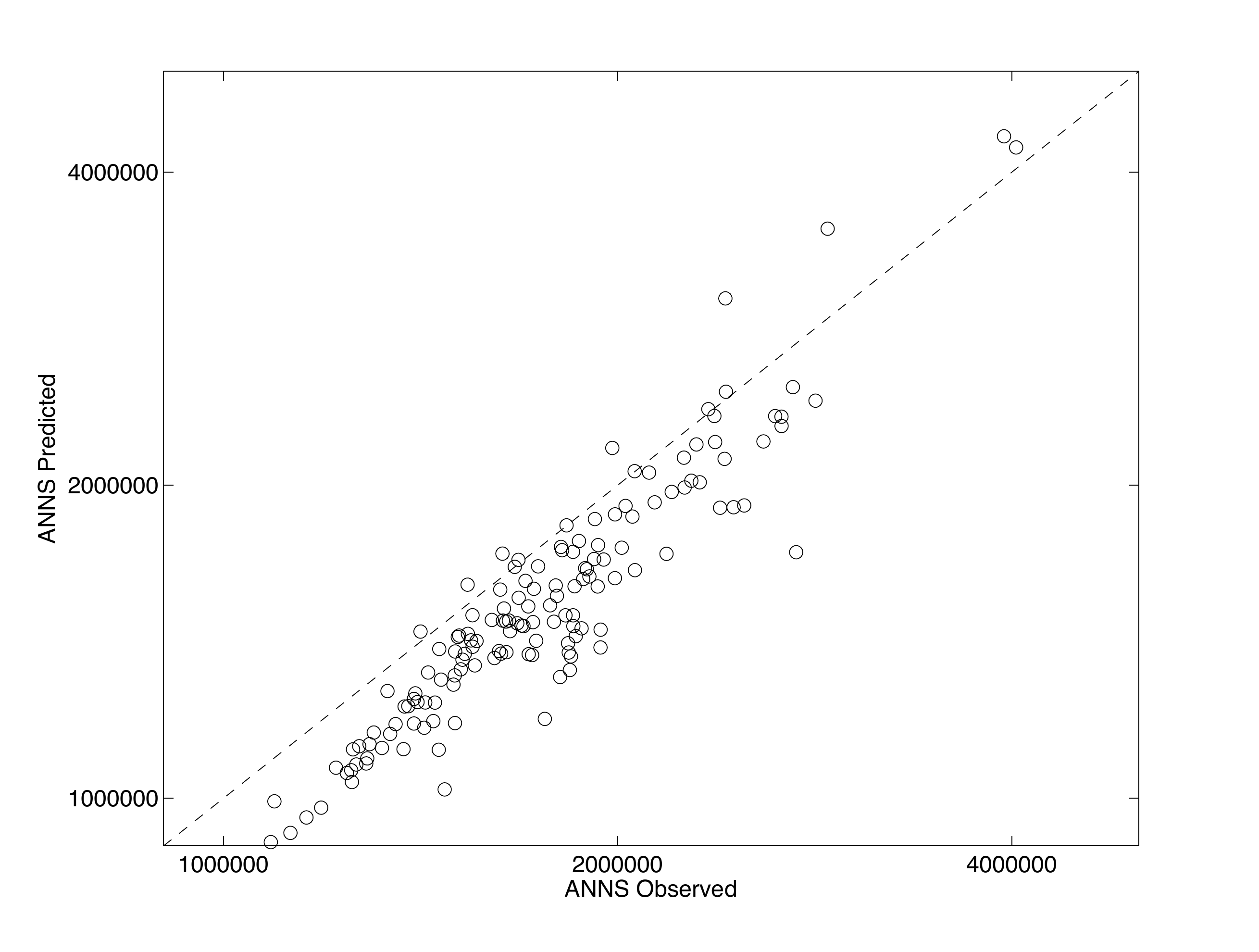}}
        \end{scriptsize}
        \end{minipage}
		\vskip 1cm
        \begin{minipage}[h]{8.5cm}
        \begin{scriptsize}
        \centering {\large{{(c)}}}
        \centering {\includegraphics[height=6cm]{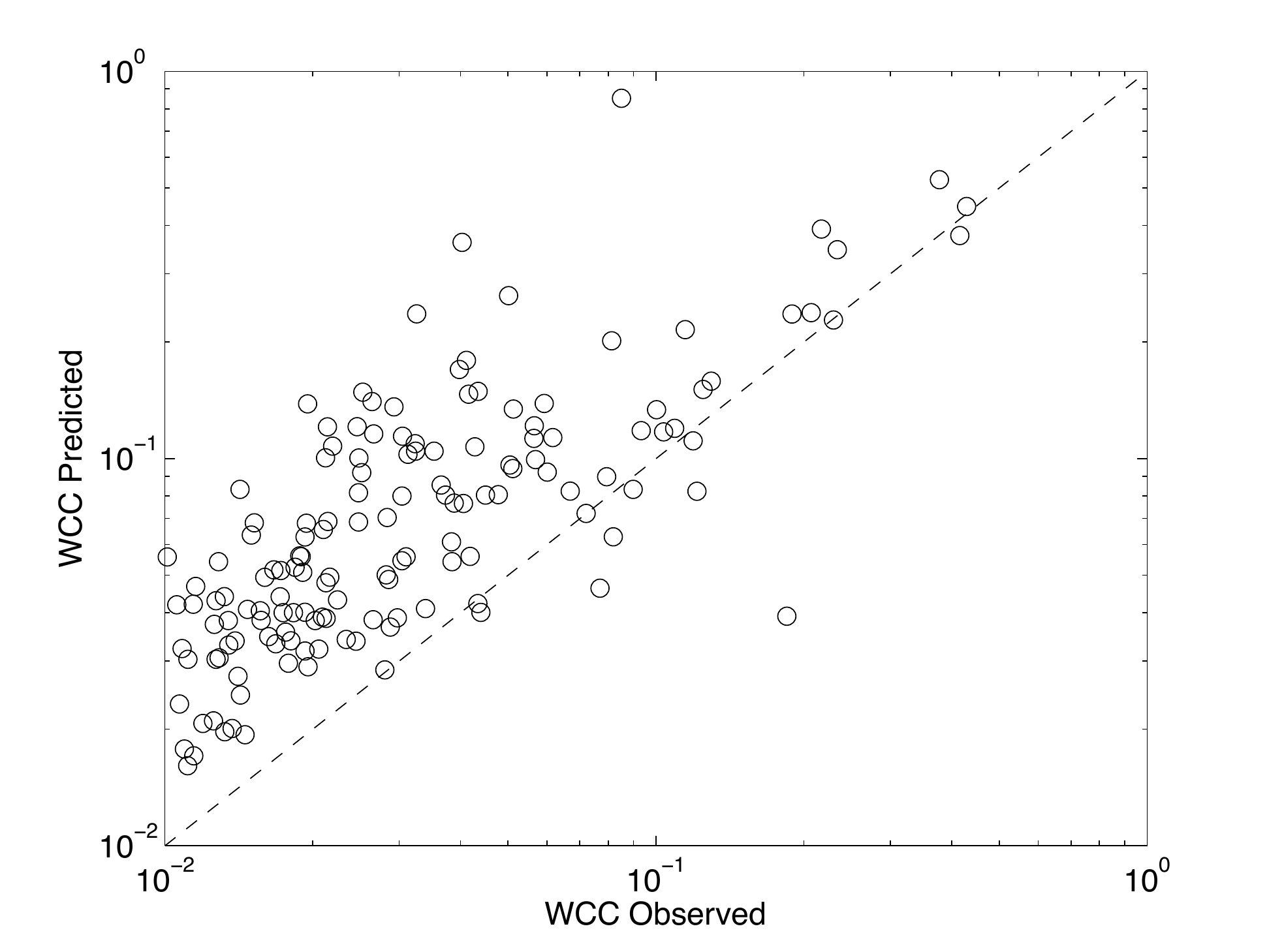}}
        \end{scriptsize}
        \end{minipage}\hfill
        \begin{minipage}[h]{8.5cm}
        \begin{scriptsize}
        \centering {\large{{(d)}}}
        \centering {\includegraphics[height=6cm]{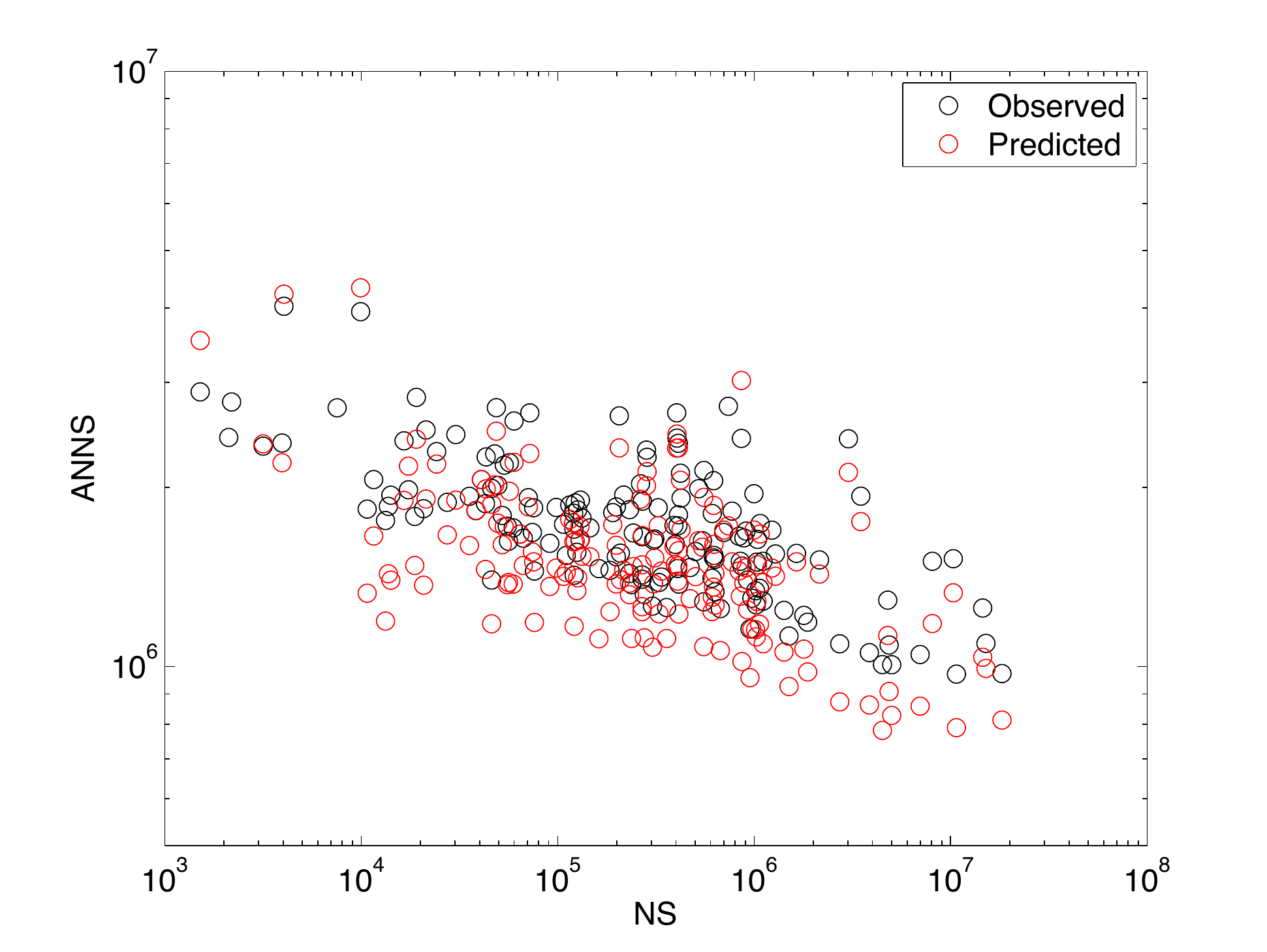}}
        \end{scriptsize}
        \end{minipage}
		\vskip 1cm
        \begin{minipage}[h]{8.5cm}
        \begin{scriptsize}
        \centering {\large{{(e)}}}
        \centering {\includegraphics[height=6cm]{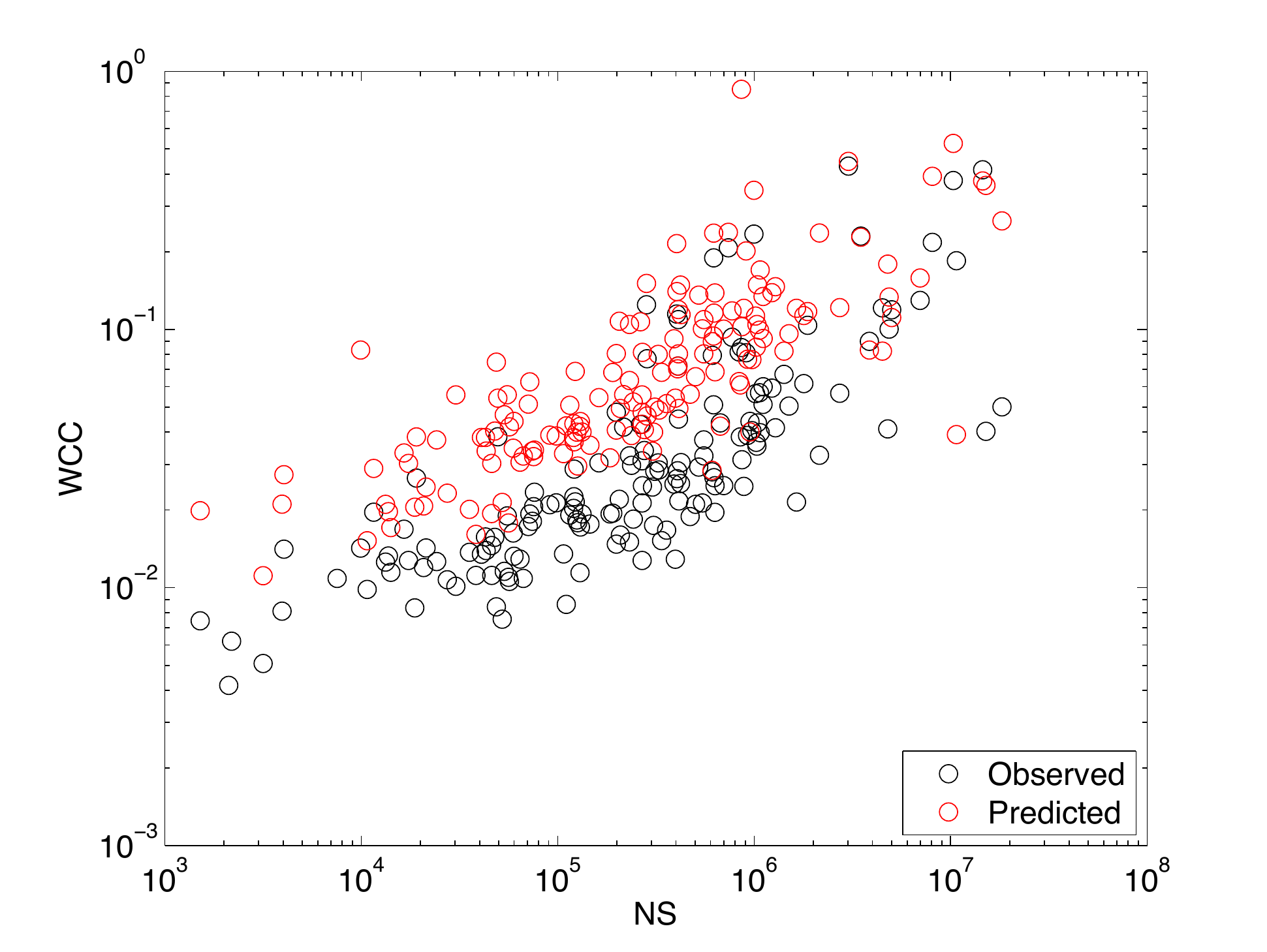}}
        \end{scriptsize}
        \end{minipage}
        \caption{Gravity model estimation of the weighted structure of the IMN. (a) Observed vs. predicted node in-strength (NSin) and out-strength (NSout); (b) Observed vs. predicted average nearest-neighbor strength (ANNS); (c) Observed vs. predicted weighted clustering coefficient (WCC); (d) Observed vs. predicted correlation between ANNS and total node strength (NS); (e) Observed vs. predicted correlation between weighted clustering coefficient (WCC) and total node strength (NS). Each dot corresponds to the average across H=1000 simulated weighted IMN matrices binary structures. Each simulated matrix has a binary structure generated using gravity-based logit predicted link probabilities, on the top of which are superimposed ZIP estimations of the relative migration stock (in levels) using gravity-based pseudo-maximum likelihood (PPML) predicted weights.}\label{fig:grav_weighted}
\end{figure}

Motivated by the success of the gravity model in reproducing the observed binary topology of the IMN, we now turn to the analysis of its weighted structure. Our main findings are reported in Figure \ref{fig:grav_weighted}, where we show observed vs.\ gravity-predicted average weighted IMN statistics. All statistics are here computed on migrant levels (and not logs), as the gravity specification allows to get a better precision, and then plotted in a log-log scale. Panel (a) shows that the gravity model is able to correctly predict in- and out-strength. This is not surprising, as strengths are linear combinations of bilateral migrant stocks, which are relatively nicely reproduced by the gravity model. Thanks to the  good performance in replicating the binary structure in each simulation, also ANND and WCC are well-fitted by the model ---panels (b) and (c). Finally, disassortativity is replicated as well as the positive correlation between weighted clustering and node-strength (emerging when migrant stock levels and not their logs are used to weigh the links).

These results show that a relatively parsimonious specification of a gravity model of migration can not only reproduce the IMN local network structure (degrees and strengths), but with some success also the underlying topology and the correlation structure between higher-order statistics. The goodness of fit of the migration-based gravity employed here is robust to alternative specifications, e.g. using additional variables sometimes employed in gravity studies (employment at origin and destination, institutional quality, etc.) \citep{Letouze_2009} or alternative estimation techniques (panel data approach using time dummies, etc.).

Note that, on the contrary, in the ITN a gravity specification was able to correctly replicate the weighted structure only if the observed binary structure is kept fixed. Indeed, a trade-gravity model is not able to correctly predict the ITN binary topology \cite{Duenas_Fagiolo_2011}. 

In the IMN, instead, country-specific and bilateral interaction terms related to social, economic, geographical and political factors can explain a good deal of details of the IMN. In particular, interacting terms controlling for origin-destination distance, geographical contiguity and common socio-political features can greatly improve the explanation of country-specific variables that, as shown in the previous sub-section, were only weakly significant in affecting local and higher-order node statistics. This has straightforward implications for both control and prediction of the future evolution of the IMN structure. 



\section{Conclusions\label{sec:Conclusions}}
International migration is increasingly recognized as a central issue to understand future trends in our globalized world. Current estimates predict that in 2050 the population of migrants will achieve 405 million, more than twice the figure for 2010 \cite{wmr_2010}. Prospective migrants will progressively face lower barriers to leave their home, facilitated by better traveling technologies and increasing access to global information networks. Moreover, migration flows will be strongly affected by the rapid growth in labor force in less developed countries (as compared to that in more developed ones) coupled with future environmental changes.

Countries will therefore face bigger and more complex issues and will need to develop clever policies to efficiently manage their borders and cope with both legal and illegal migration. A careful management of migration can indeed turn a problem into an opportunity to boost economic growth and innovation at destination, and reduce poverty at origin, while allowing for better life conditions for migrants in better culturally and socially integrated societies.  

Quantifying international migration in a globalized world becomes therefore crucial in order to provide policy makers with the right tools. In this paper, we have made a preliminary step in this direction. We have applied a complex-network approach to international migration using a comprehensive database recently made available to the community of researchers \cite{Ozden_etal_data_2011}. Describing migration data using a network representation allows one to capture the complexity of international-migration linkages between countries and gives one the possibility to study migration from a systemic perspective, where both direct and indirect linkages are taken into consideration. 

Our exercises show that the international migration network (IMN) has remained relatively stable through the years as far as migration corridors are concerned. However, the intensity of migration links has been strongly increasing leading to very skewed (power-law) distributions for weighted link and node-statistics. The IMN is organized around quite a modular structure characterized by a binary small-world pattern displaying disassortativity and high clustering. We have also shown that geographical distance, country size, relative income, together with a host of country-specific and bilateral dummy variables including common language, common religion and south-north linkages, are able to satisfactorily explain in a gravity-like fashion most of the binary and weighted topological structure of the IMN. Interestingly, the local topological structure of the network cannot fully replicate its higher-order structure. This points to a preponderance of socio-economic, geographical and political factors in the process shaping the IMN.

The foregoing analysis can be extended in many ways. First, one could try to fit to the data more sophisticated dynamic models of weighted network formation \cite{caldarelli_etal_2012} where migrant corridors and flows are stochastically added over time, possibly as a function of network (weighted) topology. This could give more hints as to why local network topology is instead unable to replicate higher-order network structure in null models. Second, our community-structure results can be strengthen and expanded both using different community-structure detection algorithms \cite{fortunato2010} and by modifying the null-model employed within the modularity function. A possibility in the latter case is to use gravity-model predictions as expected values for link weights in the modularity function (eq. \ref{eq:modularity}). Third, a comparative-network study of the ITN and IMN can be performed. In addition to directly comparing the topological structure of the ITN and the IMN (given the same set of countries), one can address the migration-trade link \cite{Parsons_2012} to explore whether indirect network effects in migration \cite{Felbermayr_etal_2009}, and more generally the position of a country in the IMN, can explain bilateral trade in the ITN.         

\newpage
\bibliographystyle{apsrev}
\bibliography{fm_migration}

\newpage

\onecolumngrid

\begin{table}[H]
\centering
\begin{tabular}{lccccc} 
\hline \hline
{Years}  & {1960} & {1970} & {1980} & {1990} & {2000}  \\ 
\hline
{Total $\sharp$ Migrants}  & 93.0m &	105.7m &	120.2m	& 141.9m	& 167.1m  \\ 
{Maximum Link Weight}  & 8.7. & 8.1m & 4.8m & 5.2m & 9.4m  \\ 
{Average Link Weight} & 5646 & 5841 & 6221 & 6528 & 7044 \\ 
{Median Link Weight} &   17 & 20 & 22 & 23 & 27 \\ 
{$\sharp$  Links}  &  16485 &  18110 & 19319 & 21731 &  23718 \\  
{Density}  & 0.32	& 0.36 &  0.38 & 0.43 & 0.47 \\ 
{$\sharp$ Bilateral Links}    & 9552 & 10792 & 11688 & 13818 & 15886 \\ 
{Bilateral Density }  & 0.58 & 0.59 & 0.60 &  0.63	& 0.67 \\ 
{Weighted Asymmetry$^\ast$ }  & 0.19 & 0.25 & 0.39 &  0.44	& 0.67 \\ 
{$\sharp$ Strongly Conn. Comp. (SCC)} & 3 & 3 & 2 & 2 & 1 \\ 
{Size of Largest SCC} & 223 & 223 & 225 & 225 & 226 \\ 
{Diameter} & 4 & 4 & 4 & 4 & 3 \\ 
{Average Path Length} & 1.75 & 1.70 & 1.67 & 1.61 & 1.53 \\
\hline \hline
\end{tabular}
\caption{The IMN, Descriptive Statistics. Source: Migration Data from Ref. \cite{Ozden_etal_data_2011}. ($\ast$): Weighted asymmetry index in \citep{Fagiolo2006EcoBull}. }
\label{tab:descriptives}
\end{table}

\begin{table}[htbp]
\begin{tabular}{r|c|cccc|cc}

\hline \hline
 & & \multicolumn{4}{c}{Community Size} & & \\
Year & No. Comm. & Min & Max & Mean & Std Dev & Q & NMI(t,t-10) \\ \hline
1960 & 14 & 3 & 67 & 16.1429 & 15.4913 & 0.6012 &  \\ 
1970 & 13 & 2 & 68 & 20.5455 & 21.5211 & 0.5926 & 0.6242 \\ 
1980 & 11 & 4 & 69 & 20.5455 & 17.9766 & 0.6029 & 0.6312 \\ 
1990 & 11 & 4 & 70 & 17.3846 & 12.3820 & 0.5792 & 0.8067 \\ 
2000 & 7 & 5 & 75 & 32.2857 & 23.4077 & 0.5450 & 0.4470 \\ 
\hline \hline
\end{tabular}
\caption{Community detection in the weighted IMN. Descriptive statistics. Communities have been detected using Newman-Girvan modularity algorithm \cite{newman_girvan2004}. $Q$=optimized modularity using Tabu Search \cite{glover1998}. NMI(t,t-10)=Normalized Mututal Information \citep{nmi} computed between partitions of two consecutive waves.}
\label{tab:commstruct}
\end{table}

\begin{table}[htbp]
\begin{center}
\begin{tabular}{lccccccccc}
\hline \hline
BIN/WEI & \multicolumn{1}{c}{Betweenness} & \multicolumn{1}{c}{In-Closeness} & \multicolumn{1}{c}{Out-Closeness} & \multicolumn{1}{c}{Bonacich} & \multicolumn{1}{c}{Katz} & \multicolumn{1}{c}{In-PageRank} & \multicolumn{1}{c}{Out-PageRank} & \multicolumn{1}{c}{Authorities} & \multicolumn{1}{c}{Hubs} \\ \hline
Betweenness & {$\ast$} & 0.5052 & 0.3683 & 0.3768 & 0.4122 & 0.5148 & 0.4631 & 0.4481 & 0.3597 \\ 
In-Closeness & 0.7386 & {$\ast$} & 0.6690 & 0.7004 & 0.7242 & 0.9387 & 0.7498 & 0.9412 & 0.7136 \\
Out-Closeness & 0.7204 & 0.6096 & {$\ast$} & 0.9430 & 0.9454 & 0.6501 & 0.9363 & 0.6492 & 0.9453 \\
Bonacich & 0.6561 & 0.5697 & 0.9877 & {$\ast$} & 0.9967 & 0.7040 & 0.9788 & 0.7098 & 0.9948 \\ 
Katz & 0.6921 & 0.5930 & 0.9973 & 0.9965 & {$\ast$} & 0.7279 & 0.9922 & 0.7283 & 0.9931 \\ 
In-PageRank & 0.7512 & 0.9981 & 0.6322 & 0.5956 & 0.6172 & {$\ast$} & 0.7560 & 0.9880 & 0.7151 \\ 
Out-PageRank & 0.7375 & 0.6178 & 0.9989 & 0.9799 & 0.9931 & 0.6397 & {$\ast$} & 0.7474 & 0.9772 \\
Authorities & 0.6702 & 0.9903 & 0.5943 & 0.5642 & 0.5824 & 0.9876 & 0.5998 & {$\ast$} & 0.7280 \\
Hubs & 0.6451 & 0.5936 & 0.9894 & 0.9950 & 0.9952 & 0.6161 & 0.9831 & 0.5906 & {$\ast$} \\ \hline \hline
\end{tabular}
\end{center}
\caption{Pearson correlation coefficients between global centrality indicators for the IMN in year 2000. Lower diagonal entries are for the binary directed IMN. Upper diagonal entries are for the weighted directed IMN. Global centrality indicators employed: betweenness centrality \cite{betweenness}); in- and out-closeness centrality \cite{closeness,Opsahl2010b}; Bonacich centrality \cite{bonacich1,bonacich2}; Katz centrality \cite{katz}; in- and out- PageRank centrality \cite{pagerank}; hub/authority centrality \cite{Kleinberg_1999,Perra_Fortunato_2008}. 
}
\label{table:centrality_corr_2000}
\end{table}

\begin{table}[htbp]
\begin{center}
\begin{tabular}{lrrrrrr}

\hline \hline & \multicolumn{1}{c}{NDin} & \multicolumn{1}{c}{NDout} & \multicolumn{1}{c}{ND} & \multicolumn{1}{c}{NSin} & \multicolumn{1}{c}{NSout} & \multicolumn{1}{c}{NS} \\ \hline
Betweenness & 0.7403 & 0.7217 & 0.8143 & 0.5022 & 0.4439 & 0.5094 \\ 
In-Closeness & 0.9999 & 0.6096 & 0.9062 & 0.9402 & 0.7405 & 0.9098 \\ 
Out-Closeness & 0.6125 & 0.9999 & 0.8875 & 0.6484 & 0.9404 & 0.8359 \\ 
Bonacich & 0.5730 & 0.9872 & 0.8578 & 0.7023 & 0.9880 & 0.8905 \\ 
Katz & 0.5961 & 0.9971 & 0.8764 & 0.7259 & 0.9973 & 0.9088 \\ 
In-PageRank & 0.9985 & 0.6323 & 0.9174 & 0.9992 & 0.7449 & 0.9465 \\ 
Out-PageRank & 0.6206 & 0.9991 & 0.8918 & 0.7532 & 0.9985 & 0.9254 \\ 
Authorities & 0.9898 & 0.5941 & 0.8921 & 0.9921 & 0.7399 & 0.9399 \\ 
Hubs & 0.5963 & 0.9890 & 0.8722 & 0.7158 & 0.9851 & 0.8970 \\ \hline \hline
\end{tabular}
\end{center}
\caption{Pearson correlation coefficients between global and local centrality indicators for the IMN in year 2000. Columns labeled as NDin, NDout and ND represent correlation between global centrality indicators computed in the binary directed IMN and node degree in (NDin), out (NDout) and total (ND). Columns labeled as NSin, NSout and NS represent correlation between global centrality indicators computed in the weighted directed IMN and node strength in (NSin), out (NSout) and total (NS). Global centrality indicators employed: betweenness centrality \cite{betweenness}); in- and out-closeness centrality \cite{closeness,Opsahl2010b}; Bonacich centrality \cite{bonacich1,bonacich2}; Katz centrality \cite{katz}; in- and out- PageRank centrality \cite{pagerank}; hub/authority centrality \cite{Kleinberg_1999,Perra_Fortunato_2008}.
}
\label{table:centrality_corr_ND_NS_2000}
\end{table}

\begin{table}[htbp]
\begin{tabular}{lrrrrr}

\hline \hline
 & 1960 & 1970 & 1980 & 1990 & 2000 \\ 
\hline
NDin & -0,1358 & -0,1021 & -0,1576 & -0,1447 & -0,1888 \\ 
NDout & -0,3608 & -0,3720 & -0,3755 & -0,3603 & -0,3685 \\ 
ND & -0,2683 & -0,2593 & -0,2965 & -0,2819 & -0,3056 \\ 
ANND & -0,0855 & -0,0372 & -0,0905 & -0,0521 & 0,0896 \\ 
BCC & 0,0540 & 0,1067 & 0,0886 & 0,0906 & 0,1700 \\ 
NSin & -0,1406 & -0,1406 & -0,1820 & -0,1953 & -0,2238 \\ 
NSout & -0,3717 & -0,3711 & -0,3614 & -0,3449 & -0,3423 \\ 
NS & -0,2687 & -0,2698 & -0,2876 & -0,2868 & -0,2971 \\ 
ANNS & -0,0211 & -0,0160 & -0,0316 & 0,0215 & 0,1337 \\ 
WCC & -0,3433 & -0,3533 & -0,3615 & -0,5085 & -0,4428 \\ 
\hline \hline
\end{tabular}
\caption{Pearson correlation coefficients between country remoteness, defined as the average geographical (log of) distance of a country with respect to all the others, and node topological properties.}
\label{table:corr_distance_nodevars}
\end{table}

\begin{table}[htbp]
\begin{tabular}{lrrrr}
\hline \hline
 & GDP & Growth & pcGDP & Population \\ \hline
Ndin & \textbf{0,2931} & -0,0117 & \textbf{0,3563} & \textbf{0,2187} \\ 
Ndout & \textbf{0,3511} & 0,0611 & \textbf{0,2209} & \textbf{0,3616} \\ 
ND   & \textbf{0,3599} & 0,0242 & \textbf{0,2214} & \textbf{0,3210} \\ 
ANND & \textbf{-0,3306} & -0,0071 & \textbf{-0,1099} & \textbf{-0,3041} \\ 
BCC & \textbf{-0,3504} & -0,0393 & \textbf{-0,1688} & \textbf{-0,3262} \\ 
Nsin & \textbf{0,5070} & 0,0079 & \textbf{0,4064} & \textbf{0,2166} \\ 
Nsout & \textbf{0,4588} & 0,0507 & \textbf{0,1899} & \textbf{0,4548} \\ 
NS & \textbf{0,5222} & 0,0287 & \textbf{0,2581} & \textbf{0,3470} \\ 
ANNS & \textbf{-0,3147} & -0,0079 & \textbf{-0,1139} & \textbf{-0,2954} \\ 
WCC & 0,0863 & 0,0457 & 0,0086 & -0,0545 \\ 
\hline\hline
\end{tabular}
\caption{Pearson correlation coefficients between country-network statistics and continuous country-specific variables in year 2000. GDP: Gross domestic product. Growth: yearly country growth rate of GDP. pcGDP: Per-capita GDP. In boldface correlation values that are significantly different from zero.}
\label{table:corr_netvars_contvars}
\end{table}

\newpage \clearpage

\begin{table}[htbp]
\begin{center}
\begin{tabular}{r|r@{.}lr@{.}lr@{.}lr@{.}l|r@{.}lr@{.}lr@{.}lr@{.}l}
\hline \hline
 & \multicolumn{ 14}{c}{Dependent Variable (Year: 2000)} \\ 
\multicolumn{1}{l|}{Covariates} & \multicolumn{2}{c}{NDin} & \multicolumn{2}{c}{NDout} & \multicolumn{2}{c}{ANND} & \multicolumn{2}{c}{BCC} & \multicolumn{2}{c}{NSin} & \multicolumn{2}{c}{NSout} & \multicolumn{2}{c}{ANNS} & \multicolumn{2}{c}{WCC} \\ \hline 
\multicolumn{1}{l|}{log(Population)} & 13&17*** & 18&07*** & -7&39*** & -0&03*** & 100&69*** & 110&03*** & -50&13*** & 0&00 \\ 
\multicolumn{1}{l|}{GDP Growth} & 0&40 & -0&10 & -0&04 & 0&00 & 0&27 & -0&71 & -0&36 & 0&00 \\ 
\multicolumn{1}{l|}{North-South Dummy} & 22&44** & 8&10 & 0&61 & 0&00 & 172&67*** & 2&95 & 0&52 & 0&00 \\  \hline
\multicolumn{14}{l}{Sub-Region Dummies}\\ 
\textit{ANZ} & 65&83* & 51&25*** & -16&26 & -0&09** & 567&64*** & 249&95** & -114&76 & 0&02 \\ 
\textit{Caribbean} & 4&64 & 24&71** & -2&02 & 0&02 & 68&22 & 169&75*** & -27&91 & 0&03*** \\ 
\textit{C America} & -2&27 & 13&90 & 17&70** & 0&08** & -39&43 & 48&41 & 73&26 & 0&03** \\ 
\textit{C Asia} & -72&58** & 10&52 & 5&92 & 0&04 & -143&60 & 123&02 & 49&01 & 0&06*** \\ 
\textit{E Africa} & -16&36 & 8&24 & 17&71*** & 0&08** & -75&29 & 25&99 & 88&01* & 0&03*** \\ 
\textit{E Asia} & -46&57 & 10&98 & 18&36** & 0&07* & -120&65 & 129&07 & 114&43* & 0&05*** \\ 
\textit{E Europe} & -13&09 & 30&27** & 10&56 & 0&04 & -33&92 & 212&34*** & 49&32 & 0&05*** \\ 
\textit{Micronesia} & -0&88 & 25&57* & -4&05 & -0&02 & 103&14 & 193&79** & -18&89 & 0&00 \\ 
\textit{Mid Africa} & -29&73 & 14&41 & 18&43** & 0&09*** & -96&39 & 83&41 & 103&62** & 0&04*** \\ 
\textit{N Africa} & -21&16 & 36&67** & 11&22 & 0&04 & -105&64 & 181&80* & 42&29 & 0&04*** \\ 
\textit{N America} & 29&93 & 34&52** & -0&63 & -0&03 & 544&45*** & 356&78*** & -22&25 & 0&03** \\ 
\textit{N Europe} & 32&97 & 43&75*** & 0&81 & -0&01 & 273&87** & 276&07*** & -16&83 & 0&04*** \\ 
\textit{Polynesia} & 19&31 & 32&18** & -16&93** & -0&06* & 198&69 & 244&90*** & -107&46** & 0&00 \\ 
\textit{S America} & 4&51 & 22&03* & 10&72 & 0&04 & 3&48 & 106&58 & 36&96 & 0&03*** \\ 
\textit{S-E Asia} & -29&92 & 13&00 & 19&99** & 0&08** & -57&73 & 82&76 & 112&06** & 0&04*** \\ 
\textit{S Africa} & 10&85 & 1&85 & 21&43*** & 0&08** & 51&99 & -16&69 & 102&43** & 0&03*** \\ 
\textit{S Asia} & -44&92 & 11&74 & 25&85*** & 0&09** & -161&85 & 152&23* & 164&05*** & 0&04*** \\ 
\textit{S Europe} & 7&02 & 43&77*** & 5&57 & 0&00 & 145&51 & 287&70*** & 17&76 & 0&04*** \\ 
\textit{W Africa} & -35&68 & 17&43 & 17&23** & 0&08*** & -83&75 & 100&78 & 99&21** & 0&04*** \\ 
\textit{W Asia} & -16&84 & 37&96*** & 4&83 & 0&01 & 45&54 & 212&11*** & 6&67 & 0&05*** \\ 
\textit{W Europe} & 49&21* & 53&46*** & -4&26 & -0&04 & 472&44*** & 397&70*** & -49&89 & 0&04*** \\  \hline
\multicolumn{14}{l}{Religion Dummies}\\ 
\textit{Buddhism} & -24&62 & -17&36 & 9&94 & 0&05 & 41&13 & -126&02 & 64&97 & 0&02* \\ 
\textit{Christian} & -45&46 & -14&88 & 7&88 & 0&05 & 9&84 & -59&44 & 60&94 & 0&03*** \\ 
\textit{Hindu} & -13&11 & -14&72 & 3&02 & 0&03 & 64&73 & -53&35 & 20&22 & 0&02 \\ 
\textit{Islam} & -42&18 & -21&24 & 14&27 & 0&08** & -44&92 & -130&28 & 93&69 & 0&03** \\ 
\textit{Jews} & 35&10 & 12&49 & -8&51 & -0&03 & 361&95 & 43&16 & -44&51 & 0&00 \\ 
\textit{Trad Beliefs} & -32&54 & -27&03* & 16&05* & 0&08* & 1&12 & -152&98* & 102&16* & 0&03** \\  \hline
\multicolumn{14}{l}{Language Dummies}\\ 
\textit{English} & 14&61 & 8&37 & -3&33 & -0&02 & 136&70** & 92&19*** & -21&85 & 0&00 \\ 
\textit{French} & 19&64 & -5&15 & -2&67 & -0&02 & 94&18 & -2&74 & -23&97 & 0&00 \\ 
\textit{Arabic} & 10&06 & -2&17 & 2&28 & 0&01 & 105&81 & 49&30 & 23&28 & 0&00 \\ 
\textit{Spanish} & 18&22 & -7&75 & 2&76 & 0&00 & 91&54 & -38&16 & 19&29 & 0&01 \\ 
\hline \hline
\multicolumn{14}{l}{Diagnostics} \\ 
\multicolumn{1}{r|}{N} & 193 & 00 & 193 & 00 & 193 & 00 & 193 & 00 & 193 & 00 & 193 & 00 & 193 & 00 & 193 & 00\\ 
\multicolumn{1}{r|}{$R^2$ } & 0&55 & 0&85 & 0&68 & 0&73 & 0&72 & 0&84 & 0&69 & 0&37 \\ 
\multicolumn{1}{r|}{Adj $R^2$ } & 0&45 & 0&82 & 0&61 & 0&67 & 0&66 & 0&81 & 0&62 & 0&23 \\ 
\multicolumn{1}{r|}{df(R)} & 158 & 00 & 158 & 00 & 158 & 00 & 158 & 00 & 158 & 00 & 158 & 00 & 158 & 00 & 158  & 00\\ 
\multicolumn{1}{r|}{df(M)} & 34 & 00 & 34 & 00 & 34 & 00 & 34 & 00 & 34 & 00 & 34 & 00 & 34 & 00 & 34  & 00\\ 
\multicolumn{1}{r|}{F} & 5&60*** & 27&00*** & 9&67*** & 12&32*** & 12&10*** & 25&20*** & 10&25*** & 2&69*** \\ 
\hline \hline
\end{tabular}
\end{center}
\caption{OLS regressions of node networks statistics vs. continuous and categorical migration dterminants. Year: 2000. Population: Country population, World Bank (http://data.worldbank.org/country). GDP growth: Yearly GDP growth (http://data.worldbank.org/country). North-South dummy: CIA Factbook, https://www.cia.gov. Religion: country official religion (Cia Factbook, www.cia.go). Language: country official language (Cia Factbook, www.cia.go). Notes: Constant included. Significance levels: * p$<$ 0.10; ** p$<$ 0.05; *** p$<$ 0.01.}
\label{table:country_vars_regressions}
\end{table}

\begin{table}[htbp]
\begin{tabular}{lrrrrrr}
\hline \hline
 & \multicolumn{1}{r}{Coef.} & \multicolumn{1}{r}{Std. Err.} & \multicolumn{1}{c}{z} & \multicolumn{1}{l}{$P>z$} & \multicolumn{ 2}{c}{[95\% Conf. Interval]} \\ 
\hline
$\log P_i$  & 0.6648 & 0.0709 & 9.3800 & 0.0000 & 0.52585 & 0.80365 \\ 
$\log P_j$  & 0.6595 & 0.0837 & 7.8800 & 0.0000 & 0.49545 & 0.82348 \\ 
$\log d_{ij}$  & -0.7259 & 0.1502 & -4.8300 & 0.0000 & -1.02027 & -0.43161 \\ 
$rY_{ij}$  & -0.3662 & 0.0836 & -4.3800 & 0.0000 & -0.53017 & -0.20233 \\ 
CONT & 1.5572 & 0.2428 & 6.4100 & 0.0000 & 1.08128 & 2.03321 \\ 
COMLANG & 0.8447 & 0.2611 & 3.2300 & 0.0010 & 0.33291 & 1.35658 \\ 
COMCOL & 1.8411 & 0.2411 & 7.6400 & 0.0000 & 1.36853 & 2.31364 \\ 
COMRELIG & 0.6163 & 0.2981 & 2.0700 & 0.0390 & 0.03211 & 1.20054 \\ 
SS & 1.2945 & 0.3696 & 3.5000 & 0.0000 & 0.57008 & 2.01895 \\ 
SN & 1.5010 & 0.4602 & 3.2600 & 0.0010 & 0.59908 & 2.40300 \\ 
NN & 0.4440 & 0.3997 & 1.1100 & 0.2670 & -0.33931 & 1.22739 \\ 
\hline\hline
\end{tabular}
\caption{Gravity-model estimation. Dependent variable: bilateral migrant stocks in year 2000. Full-sample Poisson pseudo-maximum likelihhod (PPML) fit. Country dummy variables included. Independent variables: $(P_i,P_j)$ country population. $d_{ij}$: geographical distance. $rY_{ij}$: relative origin-destination per-capita GDP. CONT: contiguity dummy. COMLANG: common official language dummy. COMCOL: common colony relationship dummy. COMRELIG: common official religion dummy. SS: south-south dummy. SN: south-north dummy. NN: north-north dummy. Data are from Ref. \citep{Bhavnani2002} and the World Bank. Due to data limitations, we employ here $N=154$ countries, accounting for 93\% of total migration stocks. Diagnostics: Pseudo $R^2=0.6477$ (model with constant and country dummy variables excluded). Wald test statistic: 800.51 (0.0000). Log pseudo-likelihood = -1.232e+08.}
\label{Table:Gravity}
\end{table}

\newpage \clearpage

\newpage

%

\clearpage

\newpage \clearpage

\newpage \clearpage

\twocolumngrid


\end{document}